\documentclass[preprint,12pt]{elsarticle}
\nopreprintlinetrue

\usepackage[letterpaper,top=1in,bottom=1.5in,left=1.0in,right=1.0in]{geometry}%

\usepackage{amssymb,mathrsfs,amsmath}
\usepackage{MnSymbol,bbding,pifont}
\usepackage{extarrows}
\usepackage{soul}
\usepackage{hyperref}
\usepackage{tocloft}
\hypersetup{hidelinks,
	colorlinks=true,
	allcolors=black,
	pdfstartview=Fit,
	breaklinks=true}

\usepackage{color}
\usepackage{courier}
\usepackage{booktabs}
\usepackage{makecell}
\usepackage{multirow}
\usepackage{siunitx}
\usepackage{bm}
\usepackage[figuresright]{rotating}
\usepackage[dvipsnames]{xcolor}
\usepackage[most]{tcolorbox}
\usepackage[toc,page,header]{appendix}
\usepackage{minitoc}
\usepackage{algorithm}
\usepackage{algpseudocode}
\usepackage{adjustbox}
\usepackage[switch]{lineno}

\usepackage{pifont}
\usepackage{graphicx}

\usepackage{colortbl}   
\usepackage{array}

\renewcommand\appendix{\par
    \setcounter{section}{0}
    \setcounter{subsection}{0}
    \gdef\thesection{\Alph{section}}}

{}
\newtheorem{proposition}{\bf Proposition}{}
{}
{}

\usepackage{lineno}

\begin{document}
\doparttoc 
\faketableofcontents 
\parttoc 
\begin{frontmatter}



\title{An Efficient Graph-Transformer Operator for Learning Physical Dynamics with Manifolds Embedding}


\author[control]{Pengwei Liu\corref{cor1}}
\author[control]{Xingyu Ren\corref{cor1}}
\author[control]{Pengkai Wang\corref{cor1}}
\author[cst]{Hangjie Yuan}
\author[thu]{Zhongkai Hao}
\author[ic]{Guanyu Chen}
\author[control]{Chao Xu}
\author[ict]{Dong Ni\corref{cor2}}
\author[control]{Shengze Cai\corref{cor2}}

\cortext[cor1]{These authors contributed equally.}
\cortext[cor2]{Corresponding author: \href{dni@zju.edu.cn}{dni@zju.edu.cn}, \href{shengze_cai@zju.edu.cn}{shengze\_cai@zju.edu.cn}.}

\affiliation[control]{organization={College of Control Science and Engineering},
            addressline={Zhejiang University}, 
            city={Hangzhou},
            state={Zhejiang},
            country={China},
            }
\affiliation[ic]{organization={College of Integrated Circuits},
            addressline={Zhejiang University}, 
            city={Hangzhou},
            state={Zhejiang},
            country={China},
            }
\affiliation[cst]{organization={College of Computer Science and Technology},
            addressline={Zhejiang University}, 
            city={Hangzhou},
            state={Zhejiang},
            country={China},
            }
\affiliation[thu]{organization={College of Computer Science and Technology},
            addressline={Tsinghua University}, 
            city={Beijing},
            country={China},
            }
\affiliation[ict]{organization={State Key Laboratory of Industrial Control Technology},
            addressline={Zhejiang University}, 
            city={Hangzhou},
            state={Zhejiang},
            country={China}
            }

\begin{abstract}
Accurate and efficient physical simulations are essential in science and engineering, yet traditional numerical solvers face significant challenges in computational cost when handling  simulations across dynamic scenarios involving complex geometries, varying boundary/initial conditions, and diverse physical parameters.
While deep learning offers promising alternatives, existing methods often struggle with flexibility and generalization, particularly on unstructured meshes, which significantly limits their practical applicability.
To address these challenges, we propose PhysGTO, an efficient Graph-Transformer Operator for learning physical dynamics through explicit manifold embeddings in both physical and latent spaces.
In the physical space, the proposed Unified Graph Embedding module aligns node-level conditions and constructs sparse yet structure-preserving graph connectivity to process heterogeneous inputs.
In the latent space, PhysGTO integrates a lightweight flux-oriented message-passing scheme with projection-inspired attention to capture local and global dependencies, facilitating multilevel interactions among complex physical correlations. 
This design ensures linear complexity relative to the number of mesh points, reducing both the number of trainable parameters and computational costs in terms of floating-point operations (FLOPs), and thereby allowing efficient inference in real-time applications.
We introduce a comprehensive benchmark spanning eleven datasets, covering problems with unstructured meshes, transient flow dynamics, and large-scale 3D geometries.  
PhysGTO consistently achieves state-of-the-art accuracy while significantly reducing computational costs, demonstrating superior flexibility, scalability, and generalization in a wide range of simulation tasks.
\end{abstract}


\end{frontmatter}
\section{Introduction}
Physical spatiotemporal simulations are essential in a wide range of scientific domains and engineering applications, providing a foundation for downstream tasks involving physical systems and offering a cost-effective alternative to resource-intensive real-world experiments \cite{wang2023scientific,joseph2021physics,ma2021deep,hatfield2021data,kochkov2024neural,bauer2023quantum}. 
High-fidelity modeling strategies, such as Direct Numerical Simulation (DNS), are widely employed to capture complex physical dynamics.
These models rely on numerical discretization methods, such as the finite element method (FEM) \cite{huebner2001finite} and finite volume method (FVM) \cite{eymard2000finite}, to solve governing equations over complex domains.
Mesh-based solvers, particularly those that utilize adaptive mesh refinement, play a central role in maintaining spatial resolution while improving computational efficiency \cite{berger1984adaptive}. 
Such strategies form the basis for modern industrial simulation platforms, such as COMSOL \cite{pryor2009multiphysics} and ANSYS \cite{stolarski2018engineering}.

However, traditional numerical solvers typically require numerous degrees of freedom to discretize partial differential equations (PDEs) based on first principles \cite{strikwerda2004finite, huynh2014high}, resulting in prohibitive computational costs, especially for large-scale complex systems \cite{anandkumar2020neural, karniadakis2021physics, brunton2024promising}. 
This computational burden becomes even more pronounced when simulating systems with varying domain geometries, boundary and initial conditions, or input parameters—factors that are essential for real-world applications.
The worse is that DNS-driven approaches are inherently problem-specific and struggle to generalize, particularly when transitioning between irregular computational grids, thereby limiting their adaptability across diverse scenarios.
In industrial settings, where addressing families of related problems is a common practice, this lack of flexibility presents a critical bottleneck for large-scale deployment.

Recently, deep learning has emerged as a promising tool in computational physics, particularly for overcoming the computational complexity and case-specific limitations of traditional numerical solvers in solving complex problems \cite{reichstein2019deep, sanchez2020learning, karniadakis2021physics, lu2021learning, chen2021physics, azizzadenesheli2024neural, kontolati2024learning, liu2024papm, liu2024deep, renuncertainty}. 
Leveraging their remarkable ability to model nonlinear relationships, deep-learning models can be trained to learn mappings from input data to output solutions for PDE-governed tasks, enabling significantly faster inference compared to conventional numerical solvers. 
This makes them particularly well-suited for handling high-dimensional parameter spaces and solving computationally intensive tasks in engineering. 

An important family of approaches in this direction is physics-informed neural networks (PINNs) \cite{raissi2019physics, mao2020physics, cai2021physics}, which integrate domain-specific knowledge to enhance generalization in specialized domains.
However, classic PINNs face two main limitations: 
(1) they require explicit governing equations to be pre-defined and embedded into the loss function, which becomes challenging in facing multi-physics scenarios and complex geometries where such equations are often unavailable;
and (2) optimizing parameters in high-dimensional feature spaces is computationally inefficient, particularly when dealing with highly complex geometric topologies. 
To efficiently model complex nonlinear behaviors in an end-to-end function space, neural operators, such as DeepONet \cite{lu2021learning}, Fourier Neural Operator (FNO) \cite{li2020fourier}, and their variants \cite{jin2022mionet, kovachki2023neural,li2023fourier,cao2024laplace,kontolati2024learning}, have emerged as a promising alternative paradigm for learning such functional dependencies.

Although vanilla neural operators have shown significant potential in learning nonlinear function mappings, they still face several fundamental limitations like traditional paradigms. 
The fixed input–output structures inherently constrain their flexibility across diverse dynamical systems, and their applicability is predominantly limited to simple geometric domains, thereby posing significant obstacles to addressing real-world physical problems on complex topologies.
To address the difficulties associated with complex geometries, deep geometric learning (DGL) has emerged as a powerful framework~\cite{atz2021geometric, dunn2021geometric, pineda2023geometric}, offering geometry-aware representations of physical fields. 
A central idea is how to project high-dimensional physical domains into latent computational spaces that are more easy for representation and learning. 
For instance, DIMON~\cite{yin2024scalable} utilises diffeomorphic transformations based on precomputed deformation matrices to map complex physical domains onto predefined canonical ones. 
However, this approach relies on the assumption of geometric diffeomorphism and requires a fixed correspondence between input and output points, significantly limiting scalability in dynamic scenarios with varying geometries across cases or in large-scale settings where such mappings become impractical to compute.
Alternatively, methods such as Geo-FNO~\cite{li2023fourier} and GINO~\cite{li2024geometry} aim to learn these projections without reliance on precomputation matrices. 
Nevertheless, they remain closely tied to the FNO architecture which presents scalability challenges for extensive three-dimensional datasets.
Collectively, these limitations highlight a fundamental bottleneck in current methods: the lack of an efficient and learnable representation space that generalizes across irregular geometries and scales with problem complexity.

In parallel, graph neural networks (GNNs) \cite{pfaff2020learning, scherbela2022solving, 
schuetz2022combinatorial, witman2023defect, liang2023recurrent, zhong2024accelerating} and transformer-based models \cite{hao2023gnot, wu2024transolver, casert2024learning, brunton2024promising, bauer2024digital, liu2024automated} have also emerged as powerful tools for surrogate modeling, facilitating the prediction of physical simulations over transient and steady-state problems. 
Their ability to handle complex geometries makes them particularly attractive for modeling physical dynamics.
However, both architectures face some limitations on irregular geometries and large-scale interactions, restricting their expressive capacity.
GNNs, which rely on node-edge message passing, suffer from two major challenges.
First, the oversmoothing problem arises due to graph convolutions acting as low-pass filters, suppressing high-frequency signals and homogenizing node features, which reduces their capacity to capture local variations~\cite{chen2020measuring}.
Second, as the number of nodes and message-passing iterations grows, the computational complexity scales quadratically, creating a severe bottleneck for large-scale simulations~\cite{chiang2019cluster}.
Transformers, which process data by projecting points into latent spaces and applying self-attention, also face scalability challenges.
First, self-attention mechanisms exhibit quadratic complexity with respect to input tokens, leading to high memory and computational costs in large-scale applications.
Second, their reliance on multilayer perceptrons (MLPs) to model relationships in unstructured domains lacks explicit topological constraints, making it difficult to preserve geometric and physical correlations.
Therefore, addressing these issues requires integrating topology-aware mechanisms to improve both modeling capability and computational efficiency for large-scale scenarios.

Moreover, integrating GNNs with Transformers has shown promise in mitigating their respective limitations~\cite{yuan2025survey, liu2025aerogto}. 
Representative hybrids, e.g., GTN~\cite{yun2019graph}, Graphormer~\cite{ying2021transformers}, GraphGPS~\cite{rampavsek2022recipe}, GraphViT~\cite{he2023generalization}, and recent task-specific variants~\cite{buterez2025end}, combine local message passing for structural inductive bias with global attention for long-range dependency modeling. 
Complementary lines of work on multi-scale GNNs~\cite{gao2019graph, cao2022bi, fortunato2022multiscale, suk2023se, barwey2023multiscale, wu2025mpg} further improve scalability via coarsen–refine schemes. 
However, directly applying these approaches to large-scale, mesh-based PDE surrogate modeling faces two key challenges: (i) many of these general-purpose graph transformers still employ standard self-attention mechanisms, which retain the quadratic $O(N^2)$ computational complexity, rendering them intractable for high-resolution industrial meshes; and (ii) these architectures are often designed for general graph-level or node-level classification tasks and thus may lack the specific physics-aware inductive biases needed for field regression, such as topology-preserving connectivity, flux directionality, and boundary conditioning. 
While some methods achieve scalability by operating in a mesh-reduced space~\cite{han2022predicting, janny2023eagle}, these approaches often involve complex, multi-stage training pipelines~\cite{han2022predicting} and have primarily been validated on 2D problems. 
A significant gap therefore remains in developing a hybrid operator that is simultaneously topology-aware, computationally scalable (ideally with linear complexity), and demonstrates broad generality across a diverse range of both steady-state and transient dynamics in large-scale, 3D physical systems.

To overcome the difficulties of representing complex physical correlations arising from varying geometries, initial/boundary conditions and input parameters while maintaining computational efficiency, we propose PhysGTO, an efficient Graph-Transformer Operator for learning physical dynamics. 
The key innovation of PhysGTO lies in its explicit manifold embeddings in both physical and latent spaces across diverse physical conditions.
In the physical space, PhysGTO employs a Unified Graph Embedding (UGE) module to process heterogeneous inputs.
It aligns node-level conditions with a Multi-Condition Aligner and constructs efficient, structure-preserving graph connectivity using a Topology-Aware Mesh Sampler.
In the latent space, PhysGTO integrates a lightweight message-passing mechanism with a linear-complexity attention mechanism to capture both local and global dependencies while significantly reducing computational overhead.
This manifold embedding strategy enhances geometric adaptability and representation consistency across complex simulation scenarios, enabling accurate velocity field prediction for temporal problems and aerodynamic coefficient estimation for forward problems, while ensuring scalability and efficiency in large-scale physical applications.

Overall, our key contributions are as follows. 
(1) We introduce PhysGTO, an efficient Graph-Transformer Operator for learning physical dynamics across diverse conditions. By integrating manifold embeddings, PhysGTO enables scalable learning and accurate predictions for both transient and steady-state dynamics while preserving computational efficiency.  
(2) A lightweight message-passing mechanism is proposed, that can capture local dependencies while preserving structural consistency. By incorporating a directional flux-oriented message-passing scheme and a topology-preserving sampling strategy, this framework significantly demonstrates strong scalability on large-scale unstructured meshes with a relatively low computational cost.  
(3) A projection-inspired attention is designed to integrate global and local information through a learnable consistency query. This mechanism enhances the modeling of long-range dependencies with linear computational complexity while reducing the information fragmentation introduced by topology-aware sampling.
(4) A comprehensive benchmark is established to evaluate model generalization and scalability in various physical simulation scenarios in this field. PhysGTO consistently achieves state-of-the-art performance across these challenging tasks, demonstrating its effectiveness in maintaining high accuracy with practical computational efficiency.
\section{Results} \label{results}
In this section, we first present a formal definition of operator learning tasks within unstructured mesh spaces, encompassing both steady-state and transient physical dynamics under a unified framework.
An overview of the proposed PhysGTO framework is then presented, highlighting its architecture and key design principles. 
A comprehensive evaluation of the proposed benchmark is conducted, which consists of three classes of data covering various physical simulation scenarios. All notations are summarized in Supplementary Information (SI) Sec.~\ref{supp notations}.

\subsection{Problem setup} \label{problem setup}
In this work, we first consider a class of parametric partial differential equations defined in various domains.
We assume that the problem domain is bounded and orientable manifolds embedded in some background Euclidean space $\Omega \subset \mathbb{R}^d$. 
The objective of a neural operator is to learn a mapping function $\mathcal{G}: \mathcal{A} \mapsto \mathcal{U}$, where $\mathcal{U}$ is the solution function space over $\Omega$, and $\mathcal{A}$ is the input function space which may include various types of input information, such as initial/boundary condition, geometry, coefficient, and source terms.

For multiphysics problems, the output extends to multiple physical fields, such as $\mathbf{s} := (U, V, P, T)$, where $U$ and $V$ denote velocity components, $P$ represents pressure, and $T$ corresponds to temperature in certain scenarios. 
Moreover, our operator framework is designed to address not only steady-state problems in physical simulations but also transient phenomena. 
For the latter, we adopt an autoregressive formulation to achieve accurate fitting.  
Further details are provided in the specific cases in the following.

In practice, however, operator approximation in industrial-scale simulations is typically performed on discretized Riemannian manifolds. 
Specifically, the continuous manifold $ \Omega \subset \mathbb{R}^d $ is discretized into a mesh structure $ \mathcal{M} = (\mathcal{V}, \mathcal{E}) $, where $ \mathcal{V} = \{x_i\}_{i=1}^N \subset \Omega $ denotes the set of sampled points on the manifold, and $ \mathcal{E} \subseteq \mathcal{V} \times \mathcal{V} $ encodes the local geometric or topological relationships between these points.
Building on this foundation, we formulate the problem as an approximate operator learning task:
\begin{equation}
    \mathcal{G} \approx \mathcal{G}_\mathcal{M}: \mathcal{A}_\mathcal{M} \rightarrow \mathcal{U}_\mathcal{M},
\end{equation}
where $\mathcal{A}_\mathcal{M} = \{f_i = f(x_i) \mid x_i \in \mathcal{V} \} $ denotes the input functions defined on the mesh nodes, and $\mathcal{U}_\mathcal{M} = \{u_i = u(x_i) \mid x_i \in \mathcal{V} \}$ represents the corresponding discrete target solutions. 
Thus, we can obtain a training set $\{(\boldsymbol{a}^k,\boldsymbol{u}^k)\}_{i=1}^{D}$, where each input $\boldsymbol{a}^k \in \mathcal{A}_\mathcal{M}$ and corresponding output $\boldsymbol{u}^k \in \mathcal{U}_\mathcal{M}$ are discretized representations of physical simulations, and $D$ denotes the total number of training samples.
We can introduce a parameterized neural network $\mathcal{G}_\theta$ to approximate such a solution mapping, where the parameters of the neural network are optimized using the training dataset. 

Additionally, for long-term prediction of physical dynamics, we adopt an autoregressive formulation, where the model recursively uses its previously predicted state $\hat{\mathbf{u}}^k$ to generate the next-step prediction $\hat{\mathbf{u}}^{k+1}$. 
Formally, given the input $\mathbf{a}^k \in \mathcal{A}_{\mathcal{M}}$ at time step $k$, the model first predicts the state $\hat{\mathbf{u}}^k = \mathcal{G}_\theta(\mathbf{a}^k)$. 
Then, the predicted state $\hat{\mathbf{u}}^k$ is incorporated into the subsequent input $\mathbf{a}^{k+1}$ to obtain $\hat{\mathbf{u}}^{k+1} = \mathcal{G}_\theta(\mathbf{a}^{k+1})$, and so forth, iterating until the desired future horizon is reached. 
This iterative prediction scheme enables the model to simulate the evolution of multi-physics fields across extended temporal horizons and large spatial domains, achieving fast inference even under high spatial complexity \cite{chen2020autoreservoir, antunes2024crystal}.

\begin{figure}[]
\centering
\includegraphics[width=\textwidth]{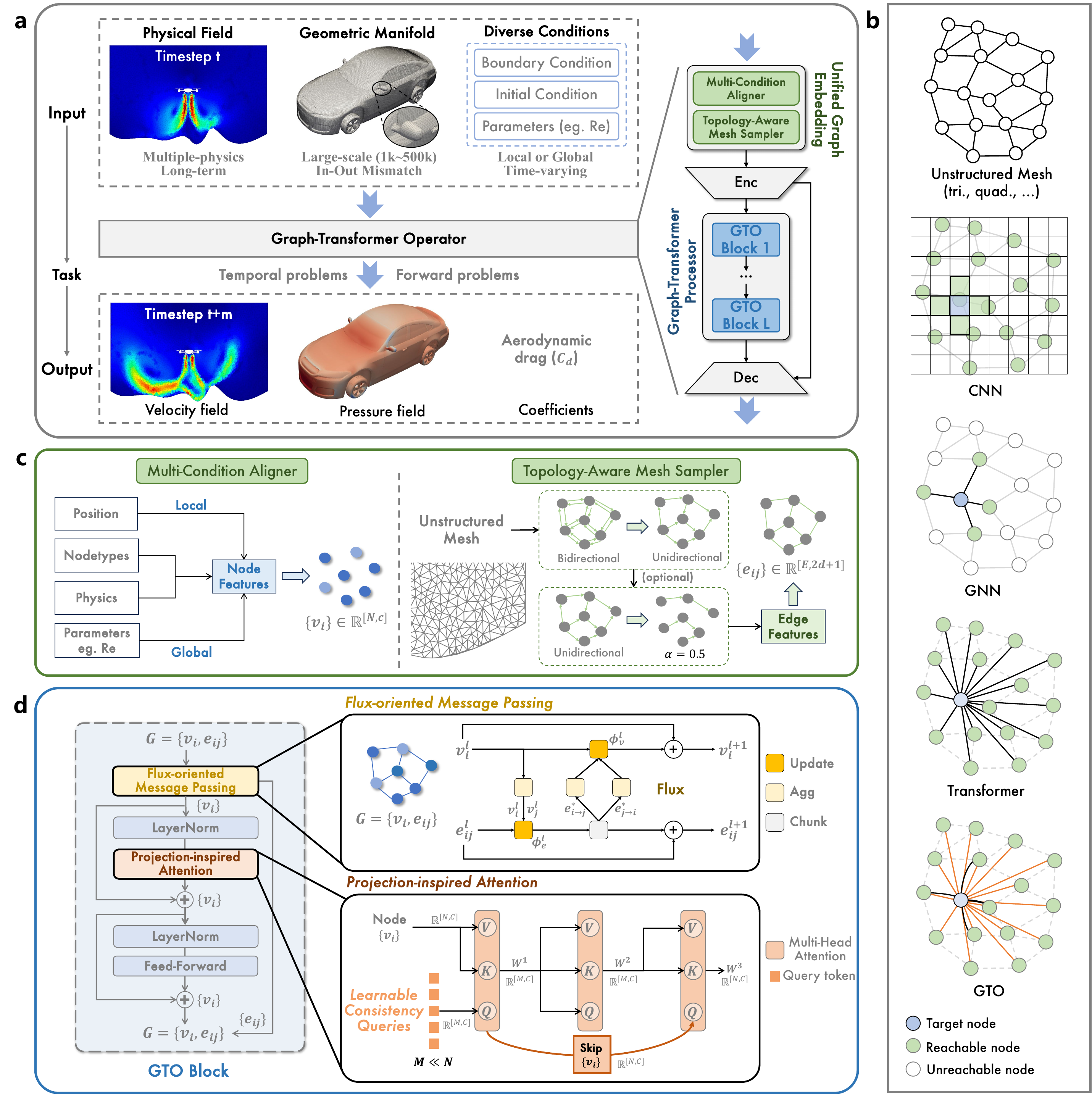}
\vspace{-6mm}
\caption{The overall pipeline of PhysGTO, an efficient Graph-Transformer Operator for learning physical dynamics. 
(a) Taking multiple conditions as inputs, PhysGTO can  predict velocity fields for temporal problems and estimate aerodynamic coefficients for forward problems.
(b) Differences between base blocks, including CNNs, GNNs, Transformers, and our proposed GTO. GTO integrates local and global interactions enabling effective information flow across unstructured domains. 
(c) Detailed structure of the Unified Graph Embedding (UGE) module, which encodes diverse input conditions into a unified graph representation. UGE consists of two components: Multi-Condition Aligner, and Topology-Aware Mesh Sampler.
(d) Internal architecture of a GTO block. Each block includes two core components: Flux-Oriented Message Passing for local interactions, and Projection-Inspired Attention for long-range dependencies. 
}
\label{pipline}
\end{figure}

\subsection{Overview of PhysGTO}
As illustrated in Fig.~\ref{pipline}a, PhysGTO is a unified framework designed to handle different physical simulation tasks, including both temporal predictions and forward inference. 
Fig.~\ref{pipline}b provides an intuitive comparison between base neural network blocks, including convolutional neural networks (CNNs), graph neural networks (GNNs), Transformers, and our proposed graph-transformer operator (GTO), which integrates local and global interactions to enable effective information flow over unstructured domains.
The model begins by embedding heterogeneous input conditions—such as geometry, boundary/initial conditions, and physical parameters—into a graph representation using the Unified Graph Embedding (UGE) module.
As shown in Fig.~\ref{pipline}c, this module comprises a Multi-Condition Aligner, which fuses node-level information through condition-wise alignment, and a Topology-Aware Mesh Sampler, which constructs a sparse yet structure-preserving edge set. 
This design effectively reduces computational overhead while preserving critical geometric structure.
The aligned node and sampled edge features are then projected into the latent space and passed through a stack of Graph Transformer Operator (GTO) blocks.

As shown in Fig.~\ref{pipline}d, each GTO block alternates between two complementary components.
The first is a Flux-Oriented Message Passing scheme, which mitigates the computational and memory burdens commonly associated with conventional GNNs.
By combining a directional flux-oriented update scheme with a topology-preserving sampling strategy, it helps maintain structural consistency while significantly reducing resource requirements for representing complex systems.
The second component is a Projection-Inspired Attention module, which enhances the model's capacity to capture long-range dependencies with linear complexity. 
By introducing a learnable spatial consistency query, it embeds global graph-level context into the node-edge message-passing process, ensuring effective information propagation and facilitating multilevel interactions among intricate physical correlations.
In addition, this design mitigates potential fragmentation caused by the sampling strategy, seamlessly restoring information flow across disconnected subgraphs that may arise during topology-preserving sampling.

Finally, the decoder maps latent features to outputs, followed by a BC-aware correction that enforces boundary conditions without additional learning complexity, enhancing fidelity in complex geometries. 
Together, this unified and modular framework enables PhysGTO to flexibly and efficiently model complex physical dynamics, while maintaining strong scalability and generalization across a wide range of simulation scales and problem types.
More details about the proposed model can be found in Methods (Sec.~\ref{methods}). 

To evaluate the performance, efficiency, and scalability of PhysGTO, we establish a comprehensive benchmark comprising three classes of computational physics experiments: unstructured mesh problems, complex transient flow prediction, and large-scale 3D geometry problems. 
These tasks are designed to systematically assess both static and time-dependent PDEs across diverse geometries and parameterized conditions. 
The PhysGTO’s ability to generalize across complex spatial domains and time-dependent systems is also assessed by comparing with strong baselines under appropriate settings.
All details on the evaluation protocol, hyperparameter configurations, and baseline implementations are provided in SI Sec.~\ref{method detaials}.

\subsection{PhysGTO excels in learning complex physical patterns on unstructured meshes}\label{sec_example_1}
The real-world problems generally involve complex computational domains, such as curved surfaces and uneven regions, which are often geometrically intricate and typically represented using unstructured meshes. 
To evaluate the modeling performance of PhysGTO under various parameterized conditions with complex domains, we analyze benchmark simulation problems from~\cite{chen2024learning}, treating them as a class of unstructured mesh-based challenges. 

Following the experimental settings established in \cite{chen2024learning}, we evaluate PhysGTO on five PDE operator learning tasks spanning diverse physical dynamics and geometric complexities. 
These tasks include three canonical benchmarks (Darcy flow, Pipe turbulence, Heat transfer) and two engineering applications involving multi-physics coupling (Composite workpiece deformation) and transient dynamics (Blood flow dynamics). 
They cover a wide range of scenarios, from static mappings on 2D manifolds to time-dependent predictions in 3D domains. For baseline comparison, we benchmark PhysGTO against several popular neural operators, including DeepONet \cite{lu2021learning}, POD-DeepONet \cite{lu2022comprehensive}, Fourier Neural Operator (FNO) \cite{li2020fourier}, Geo-FNO \cite{li2023fourier}, GraphSAGE \cite{hamilton2017inductive}, and the state-of-the-art model NORM \cite{chen2024learning}. Due to architectural limitations—such as FNO’s requirement for structured grids and GNNs' difficulty handling heterogeneous input-output formats—not all baselines are evaluated across all datasets. 

\begin{table}[t]
\centering
\vspace{-5mm}
\caption{Performance comparison ($L_2$ error, lower is better) on the unstructured mesh problems adopted from~\cite{chen2024learning}. \textbf{Bold} and \underline{underline} numbers indicate the best and second-best performance, respectively.
}\label{task_1_result_1}
\resizebox{\textwidth}{!}{%
\begin{tabular}{l|c|c|c|c|c}
\toprule
Model& Irregular Darcy & Pipe Turbulence & Heat Transfer & Composite & Blood Flow \\
\midrule
GraphSAGE\cite{hamilton2017inductive} 
& 6.73e-2
& 2.36e-1
& -       
& 2.09e-1
& -       \\
DeepONet\cite{lu2021learning}         
& 1.36e-2
& 9.36e-2 
& 7.20e-4
& 1.88e-2
& 8.93e-1\\
POD-DeepONet\cite{lu2022comprehensive}
& 1.30e-2 
& 2.59e-2
& 5.70e-4 
& 1.44e-2
& 3.74e-1\\
FNO\cite{li2020fourier}               
& 3.83e-2
& 3.80e-2
& -       & -       & -       \\
Geo-FNO\cite{li2023fourier}
&2.96e-2
&1.64e-2
&1.20e-2
&1.21e-2
&6.67e-2\\
NORM\cite{chen2024learning}           
& \underline{1.05e-2}
& \underline{1.01e-2}
& \underline{2.70e-4} 
& \underline{9.99e-3} 
& \underline{4.82e-2}\\
\midrule
\textbf{PhysGTO} (ours)               
&\textbf{6.65e-3}    
&\textbf{6.30e-3}
&\textbf{1.56e-4} 
&\textbf{9.27e-3}
&\textbf{2.40e-2}\\
\bottomrule
\end{tabular}
}
\end{table}


\begin{figure}[]
\centering
\vspace{-4mm}
\includegraphics[width=\textwidth]{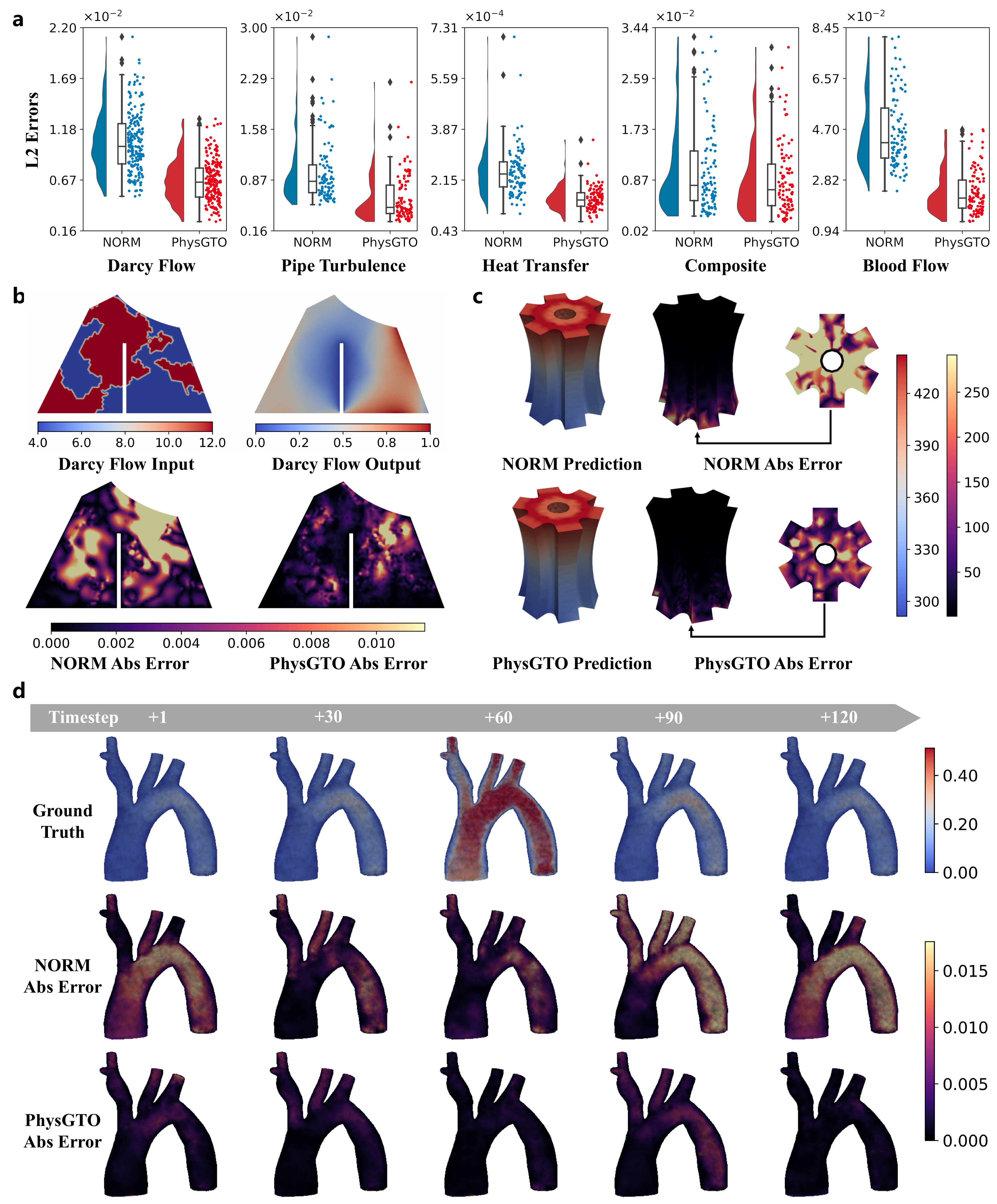}
\vspace{-1cm}
\caption{PhysGTO excels in learning complex physical patterns on unstructured meshes: a comparison with a state-of-the-art model, NORM~\cite{chen2024learning}. 
(a) Raincloud plots depicting the relative $L_2$ errors of the current SOTA model and the proposed PhysGTO across five datasets. 
(b-d) Visualization of some representative cases: (b) Darcy problem, (c) Heat transfer, and (d) Blood flow. 
Here, \textbf{Abs Error} refers to $|\hat{u}(x) - u(x)|$, where $\hat{u}(x)$ and $u(x)$ denote the predicted and ground-truth values, respectively. More visualization results can be found in SI Sec.~\ref{supp visulaization}.
}
\label{Unstructured Mesh Problems}
\end{figure}

The results presented in Tab.~\ref{task_1_result_1} demonstrate that PhysGTO consistently achieves high predictive accuracy in all evaluated test cases, showing its strong ability in approximating different physical phenomena under complex geometric and parametric variations. 
Meanwhile, NORM consistently ranks as the second-best model across all five datasets. 
To further understand the modeling advantages of PhysGTO, we perform a detailed comparative analysis against NORM, focusing on both prediction accuracy and error distribution patterns.
First, we compare the test $L_2$ errors for PhysGTO and NORM across five benchmark datasets.
As Fig.~\ref{Unstructured Mesh Problems}a shows, PhysGTO consistently yields lower error magnitudes compared to NORM, indicating that it achieves not only higher predictive accuracy but also improved generalization. 
The more compact and left-shifted distribution of errors suggests that PhysGTO exhibits greater stability and robustness.
Then we visualize representative test samples to further highlight PhysGTO's superior physical simulation capabilities on irregular meshes.
As shown in Fig.~\ref{Unstructured Mesh Problems}b and Fig.~\ref{Unstructured Mesh Problems}c, PhysGTO produces lower pointwise prediction errors across irregular domains and demonstrates superior performance near complex and composite boundary regions when compared to NORM.
Furthermore, we present the modeling results of blood flow dynamics at multiple time instances in Fig.~\ref{Unstructured Mesh Problems}d. 
PhysGTO consistently achieves high simulation accuracy along the temporal trajectory, showcasing its superior capability in capturing dynamic physical processes.
For a more comprehensive comparison, additional visualizations across various datasets and models can be found in SI Sec.~\ref{supp visulaization}.

\subsection{PhysGTO enables reliable dynamics forecasting over long time horizons}\label{sec_example_2}
In the second category, we focus on transient flow prediction problems with unstructured meshes. Accurate long-term forecasting is crucial for time-dependent PDEs, as many engineering applications, ranging from thermal regulation to aerodynamic optimization, requires high-fidelity dynamic simulations. Evaluating model performance in these scenarios provides critical insight into its ability to maintain physical consistency over extended temporal evolutions.

For long-horizon transient flow prediction tasks, we adopt the experimental setup introduced in EAGLE \cite{janny2023eagle}, evaluating PhysGTO on four challenging time-dependent simulations: Cylinder Flow \cite{pfaff2020learning}, EAGLE \cite{janny2023eagle}, and two industrial-scale simulations (Heat Flow and ICP Plasma). These problems involve diverse geometries, time-varying boundary/initial conditions, and complex parameter spaces, reflecting real-world physical scenarios. 
Each task is formulated as predicting future multi-physics fields (e.g., velocity, pressure, temperature, electron density) at time $t+h$, given the full simulation state $\mathcal{M}^t$ and future mesh structure. 
We compare PhysGTO against strong baselines designed for irregular spatio-temporal domains, including: MeshGraphNet \cite{pfaff2020learning}, a classical GNN for irregular spatio-temporal domains; 
Geo-FNO~\cite{li2023fourier}, a variant of FNO \cite{li2020fourier} designed to extend operator learning capabilities to irregular meshes; GraphViT \cite{janny2023eagle}, a hybrid model that incorporates node clustering and global attention to improve scalability and long-range reasoning; GNOT \cite{hao2023gnot} and Transolver \cite{wu2024transolver} are two transformer-based geometric operators. 
To further assess the model’s dynamic prediction capabilities, we adopt a short-to-long horizon training and testing protocol in EAGLE \cite{janny2023eagle}.

\begin{figure}[]
\centering
\includegraphics[width=\textwidth]{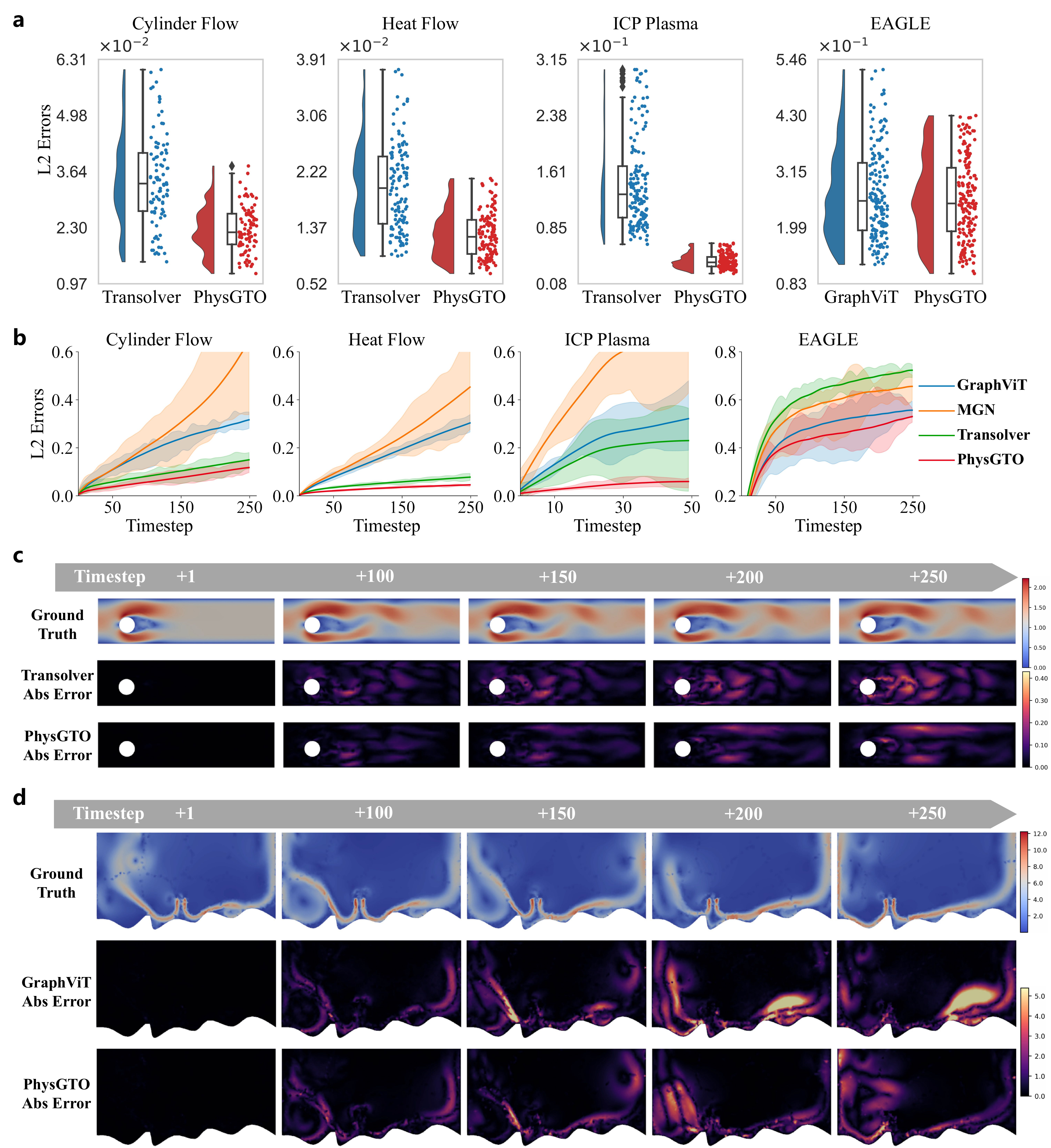}
\vspace{-0.8cm}
\caption{
PhysGTO enables reliable dynamics forecasting over long time horizons. 
(a) Raincloud plots depicting the relative $L_2$ errors of SOTA models and the proposed PhysGTO across four datasets. 
(b) Relative $L_2$ errors at each time step over long temporal horizons across four datasets, comparing four deep learning models.
(c-d) Visualization of two representative cases: (c) Cylinder Flow, (d) EAGLE dataset. 
The first row shows the ground-truth velocity magnitude ($\sqrt{u^2+v^2}$), while the second and third rows present the prediction errors of the state-of-the-art baseline and the proposed PhysGTO, respectively. Here, \textbf{Abs Error} refers to $|\hat{u}(x) - u(x)|$, where $\hat{u}(x)$ and $u(x)$ denote the predicted and ground-truth values, respectively.
}
\vspace{-2mm}
\label{Complex Transient Flow Prediction}
\end{figure}

\begin{table}[t]
\centering
\caption{
Performance comparison ($L_2$ error, lower is better) on complex transient flow cases. \textbf{Bold} and \underline{underline} numbers indicate the best and second-best performance, respectively. “-” indicates failure due to error explosion during autoregressive rollout, reflecting instability in long-horizon prediction.
}\label{result_2}
\resizebox{\textwidth}{!}{%
\begin{tabular}{l|c|cc|c|cc|c|cc|c|c}
\toprule
\multirow{3}{*}{Model} 
&\multicolumn{3}{c|}{\textbf{Cylinder Flow} \cite{pfaff2020learning}}  
&\multicolumn{3}{c|}{\textbf{EAGLE} \cite{janny2023eagle}}  
&\multicolumn{3}{c|}{\textbf{Heat Transfer}}  
&\multicolumn{2}{c}{\textbf{ICP Plasma}}\\ 
&in&\multicolumn{2}{c|}{out}&in&\multicolumn{2}{c|}{out}&in&\multicolumn{2}{c|}{out}&in&\multicolumn{1}{c}{out}\\
&+5&+50&+250&+5&+50&+250&+5&+50&+250&+5&+50\\
\midrule
MeshGraphNet\cite{pfaff2020learning}
&1.39e-2 &6.23e-2 &2.76e-1 
&6.37e-2 &3.06e-1 &3.06e-1 
&8.22e-3 &4.99e-2 &2.15e-1 
&9.46e-2 &4.77e-1\\ 
GeoFNO\cite{li2023fourier} 
&6.17e-2 &2.45e-1 &-
&1.83e-1 &- &-
&4.74e-2 &2.84e-1 &-
&2.40e-1 &-\\
GNOT \cite{hao2023gnot}               
&6.05e-2 &2.21e-1 &- 
&1.86e-1 &6.27e-1 &- 
&3.21e-2 &1.59e-1 &- 
&5.58e-2 &2.84e-1\\ 
GraphViT\cite{janny2023eagle}  
&\underline{1.06e-2} &5.88e-2 &1.95e-1 
&\underline{6.12e-2} &\underline{2.65e-1} &\underline{4.59e-1} 
&7.00e-3 &4.44e-2 &1.64e-1 
&4.10e-2 &2.16e-1\\
Transolver \cite{wu2024transolver} 
&1.09e-2 &\underline{3.56e-2} &\underline{8.86e-2} 
&7.45e-2 &3.35e-1 &5.94e-1 
&\underline{5.24e-3} &\underline{2.24e-2} &\underline{5.06e-2} 
&\underline{2.69e-2} &\underline{1.65e-1}\\
\midrule
\textbf{PhysGTO} (ours)                  
&\textbf{8.34e-3} &\textbf{2.41e-2} &\textbf{6.67e-2} 
&\textbf{5.98e-2} &\textbf{2.60e-1} &\textbf{4.32e-1} 
&\textbf{5.21e-3} &\textbf{1.40e-2} &\textbf{3.07e-2} 
&\textbf{1.29e-2} &\textbf{4.11e-2}\\
\bottomrule
\end{tabular}    
}    
\end{table}


As shown in Tab.~\ref{result_2}, PhysGTO consistently outperforms all baselines across four complex transient flow benchmarks, under both short- and long-horizon prediction settings. We adopt a two-stage evaluation protocol where models are trained on short-step sequences (step = 5), and then tested on both in-distribution (+5 step) and out-of-distribution rollout settings (+50 and +250 steps). 
This setup reflects the practical learning challenges faced by operators in dynamic multi-physics systems, where models must remain predictive over extended time horizons despite limited training information.

PhysGTO demonstrates not only strong accuracy in short-term predictions but also maintains remarkable stability during long-horizon rollout. For example, in the ICP Plasma case, PhysGTO reduces the prediction error by nearly an order of magnitude compared to the best baseline, effectively capturing both localized dynamics and coherent global evolution. In contrast, Geo-FNO and GNOT fail to produce meaningful results at step = 250. The “-” entries in Tab.~\ref{result_2} denote numerical divergence or error explosion during autoregressive rollout, which stems from their architectural reliance on static one-shot mappings and aligned input-output domains. 

As Fig.~\ref{Complex Transient Flow Prediction}a depicts, PhysGTO outperforms the competitive Transolver with tighter, smaller error distributions on Cylinder Flow, Heat Flow, and ICP Plasma. Even in the EAGLE dataset, PhysGTO still achieves comparable performance to the SOTA GraphViT, optimized for this dataset, with more stable, concentrated error distributions.
Moreover, a major challenge in long-horizon prediction lies in the accumulation of errors at each intermediate time step, and we present the $L_2$ errors across different forecasting horizons. 
Remarkably, Fig.~\ref{Complex Transient Flow Prediction}b demonstrates that PhysGTO exhibits marginal increases in error as the prediction length grows, significantly outperforming existing baselines in long-term forecasting accuracy and stability.
Additionally, we also visualize some sample predictions along a unified temporal trajectory in Fig.~\ref{Complex Transient Flow Prediction}c and Fig.~\ref{Complex Transient Flow Prediction}d. 

These results highlight PhysGTO’s key advantage: it enables stable and geometry-adaptive operator learning on unstructured meshes over long time horizons. By explicitly embedding both physical and latent manifolds, PhysGTO remains robust to error accumulation and generalizes well in dynamic systems where geometry, boundary conditions, and field variables evolve independently.
\subsection{PhysGTO scales effectively to large-scale 3D geometry problems}\label{sec_example_3}
The third category evaluates large-scale industrial simulations, with a particular focus on real-world automotive aerodynamic applications. 
This setting is motivated by a critical practical concern: in high-fidelity CFD workflows, real-world simulations typically require meshes with millions of elements to capture fine-grained geometric and flow-field details. 
Consequently, a learning-based model must demonstrate not only accuracy but also scalability in order to be practical as a surrogate in such situations.
To this end, we consider two datasets—Ahmed-Body \cite{li2024geometry} and DrivAerNet \cite{elrefaie2024drivaernet}—which involve predicting pressure fields and aerodynamic coefficients over highly detailed and irregular 3D meshes. 
These simulations significantly surpass the scale of conventional benchmarks, posing challenges that go beyond canonical geometries such as cylinders. 
Ahmed-Body includes parametric car designs with 551 samples, while DrivAerNet provides more than 4000 3D car shapes with full-field aerodynamic quantities such as pressure and the drag coefficient.
In particular, the underlying automotive bodies exhibit complex structural variations, including non-uniform chassis, curved surfaces, and intricate window geometries, which increase the difficulty of accurate prediction.
In addition, the wide-ranging Reynolds number (from $4.35 \times 10^5$ to $6.82 \times 10^6$) is also a big modeling challenge in this dataset.
As a result, this benchmark serves as a comprehensive evaluation of PhysGTO’s generalization capability and scalability.

To benchmark PhysGTO under these settings, we compare it against five strong baselines: MeshGraphNet \cite{pfaff2020learning}, a classic GNN; GNOT \cite{hao2023gnot}, a scalable linear-transformer with mixture-of-experts; Transolver \cite{wu2024transolver} and IPOT \cite{lee2024inducing}, two transformer-based geometric operators; and GINO \cite{li2024geometry}, a neural operator framework built for large-scale 3D flow using Fourier-based backbones. Following prior works, we adopt the same task formulations as in Li et al.~\cite{li2024geometry} for pressure field prediction, and Elrefaie et al.~\cite{elrefaie2024drivaernet} for aerodynamic coefficient estimation.

\begin{figure}[]
\centering
\includegraphics[width=\textwidth]{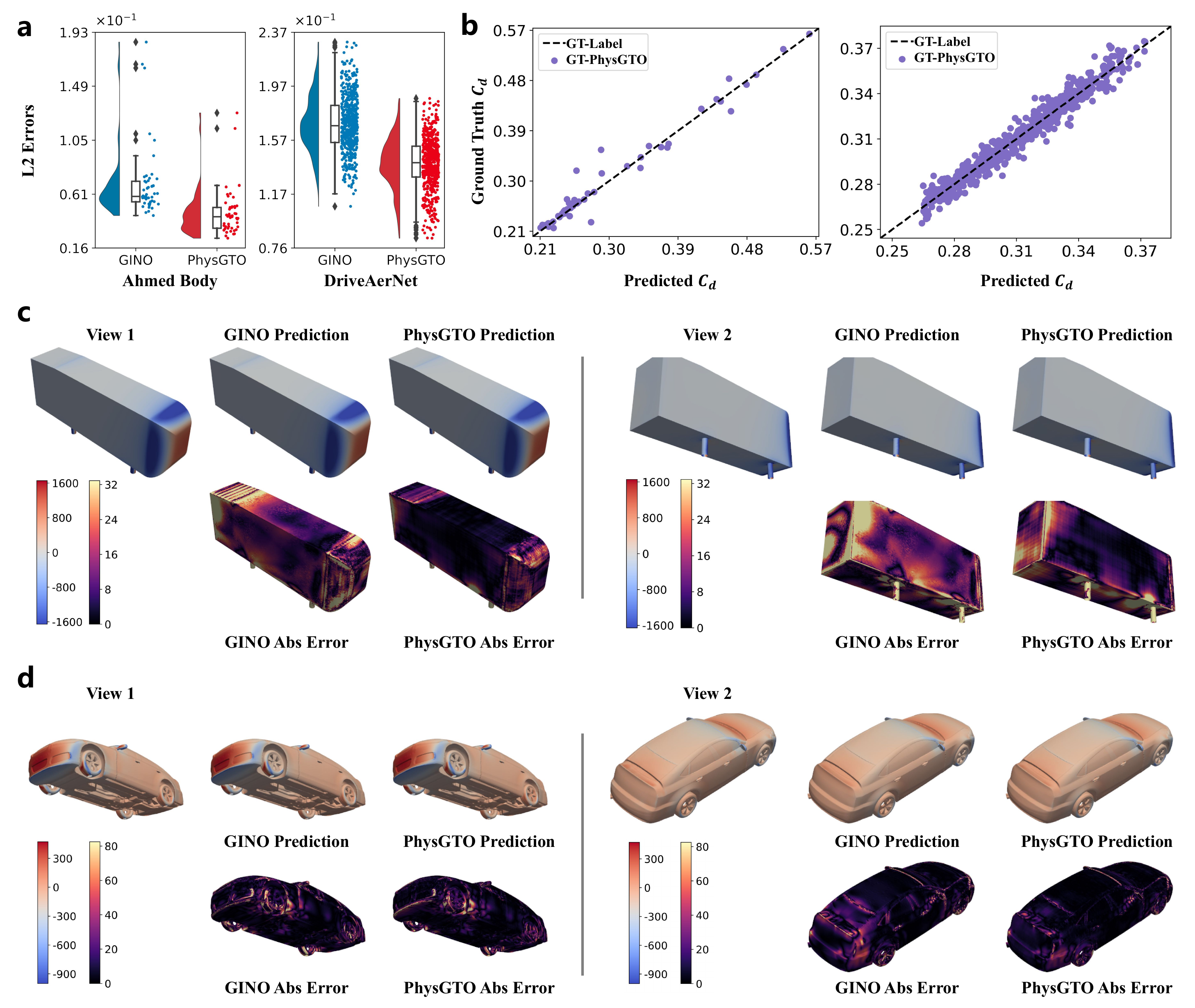}
\vspace{-8mm}
\caption{
PhysGTO scales effectively to large-scale 3D geometry problems. (a) Raincloud plots depicting the relative L2 errors of the current SOTA GINO and our proposed PhysGTO for Ahmed-Body (left) and DrivAerNet (right) test datasets. (b) Correlation plot of the drag coefficient $C_d$ predicted by our PhysGTO model against the ground truth for Ahmed-Body (left) and DrivAerNet (right) test datasets, achieving an R2 score of 0.9250 for Ahmed-Body and 0.9595 for DrivAerNet. The dotted line denotes the line of ideal correlation. 
(c-d) Visualization of a ground-truth pressure and corresponding prediction by the current SOTA, GINO, and our proposed model, PhysGTO for Ahmed-Body (c) and DrivAerNet (d).
}
\label{Large-Scale 3D Geometry Problems}
\end{figure}



\begin{table}[t]
\centering
\begin{minipage}[t]{0.47\textwidth}
\centering
\caption{Performance on surface pressure estimation for Ahmed-Body and DrivAerNet datasets ($L_2$ error). \textbf{Bold} and \underline{underline} numbers indicate the best and second-best performance, respectively.}
\label{result_car_1}
\resizebox{\textwidth}{!}{%
\begin{tabular}{l|c|c}
\toprule
\textbf{Model} & Ahmed-Body $\downarrow$& DrivAerNet $\downarrow$\\
\midrule  
MGN \cite{pfaff2020learning} & 1.39e-1 & 2.31e-1 \\
GNOT \cite{hao2023gnot}      & 1.29e-1 & 2.51e-1 \\
IPOT \cite{lee2024inducing}  & 1.00e-1 & 1.99e-1 \\
Transolver \cite{wu2024transolver} & 9.27e-2 & 1.77e-1 \\
GINO \cite{li2024geometry}         & \underline{8.31e-2} & \underline{1.65e-1} \\
\midrule
\textbf{PhysGTO} (Ours)            & \textbf{5.70e-2}    & \textbf{1.45e-1} \\
\bottomrule
\end{tabular}
}
\end{minipage}
\hfill
\begin{minipage}[t]{0.48\textwidth}
\centering
\caption{Performance on drag coefficient ($c_d$) prediction. \textbf{Bold} and \underline{underline} numbers indicate the best and second-best performance, respectively.\vspace{4mm} }
\label{result_3_2}
\resizebox{\textwidth}{!}{%
\begin{tabular}{l|c|c|c}
\toprule
\textbf{Model} & $c_d$ \textbf{MSE} $\downarrow$   & $c_d$ \textbf{MAE} $\downarrow$  & $c_d$ $R^2$ $\uparrow$\\ 
\midrule
PointNet++\cite{qi2017pointnet++}     & 7.81e-5 & 6.76e-3  & 0.896\\
MGN\cite{pfaff2020learning}           & 6.00e-5 & 6.08e-3  & 0.917\\
PointNeXt\cite{qian2022pointnext}     & 4.59e-5 & 5.20e-3  & 0.939\\
RegDGCNN\cite{elrefaie2024drivaernet} & 8.00e-5 & 6.91e-3  & 0.901\\
FIGConvNet\cite{choy2025factorized}   & \underline{3.23e-5} & \underline{4.42e-3}  & \underline{0.957}\\
\midrule
\textbf{PhysGTO} (Ours)               & \textbf{2.94e-5} & \textbf{4.15e-3} & \textbf{0.960}\\
\bottomrule
\end{tabular}
}
\end{minipage}
\end{table}


We first evaluate PhysGTO’s capability in pressure field reconstruction, a fundamental task in computational fluid dynamics. 
As shown in Tab.~\ref{result_car_1}, PhysGTO achieves SOTA accuracy, significantly outperforming previous methods. 
Beyond pressure prediction, another crucial metric in aerodynamic analysis is the drag coefficient $C_d$, which plays a central role in vehicle performance evaluation.
As summarized in Tab.~\ref{result_3_2}, PhysGTO consistently achieves the lowest MAE, lowest MSE, and highest $R_2$ score, further confirming its robustness across a variety of vehicle geometries.
Furthermore, to provide deeper insight into performance, we compare the error distributions of PhysGTO and the current leading model, GINO. 
As illustrated in Fig.~\ref{Large-Scale 3D Geometry Problems}a, PhysGTO exhibits not only lower average errors but also a more compact and concentrated error distribution, highlighting its improved generalization across complex geometries.
Moreover, we visualize the predicted aerodynamic drag coefficients ($C_d$) against ground-truth values, which exhibit a strong alignment in Fig.~\ref{Large-Scale 3D Geometry Problems}b. 
To further illustrate model performance, we visualize the predicted pressure fields of PhysGTO and GINO on both the Ahmed-Body and DrivAerNet datasets. 
As shown in Fig.~\ref{Large-Scale 3D Geometry Problems}c and Fig.~\ref{Large-Scale 3D Geometry Problems}d, despite the substantial increase in mesh resolution (Ahmed-Body) and the presence of more intricate boundary structures (DrivAerNet), PhysGTO consistently delivers accurate predictions. 
These results underscore the model’s strong generalization capacity and highlight its potential as a surrogate for industrial-scale simulations. 
Additional visualization of PhysGTO’s prediction errors compared to the current state-of-the-art method, GINO can be seen in SI Sec.~\ref{supp visulaization}.

\subsection{Analysis on the Scalability and Complexity of PhysGTO}
While the previous sections focus on benchmark-level comparisons, where PhysGTO  achieves top performance across a diverse set of mechanics tasks, it remains crucial to move beyond black-box metrics and probe the underlying mechanisms that enable such strong generalization.  
In this section, we present a more detailed investigation of PhysGTO’s internal structure and learning ability, aiming to better understand why and how the model works.
We begin by analyzing the scalability of PhysGTO, revealing how its architectural design naturally accommodates data at different spatial resolutions and mesh complexities. 
Next, we investigate the Graph-Transformer Operator (GTO) block through the kernel learning theory, showing that it can be understood as a compositional kernel over both local and global neighborhoods, which offers insights into its ability to capture spatial correlations. 
A formal theoretical analysis, including its formulation as an integral neural operator and discussion on discretization over unstructured manifolds, is provided in SI Sec.~\ref{supp_gto}.
Eventually, we provide a comprehensive complexity analysis, quantifying the computational behavior as problem dimension scales. 
We note that further experimental configurations, including sampling strategies, model scaling setups in SI Sec.~\ref{method detaials}, and ablation results in Sec.~\ref{Ablation Studies}.

\begin{figure}[]
\centering
\includegraphics[width=\textwidth]{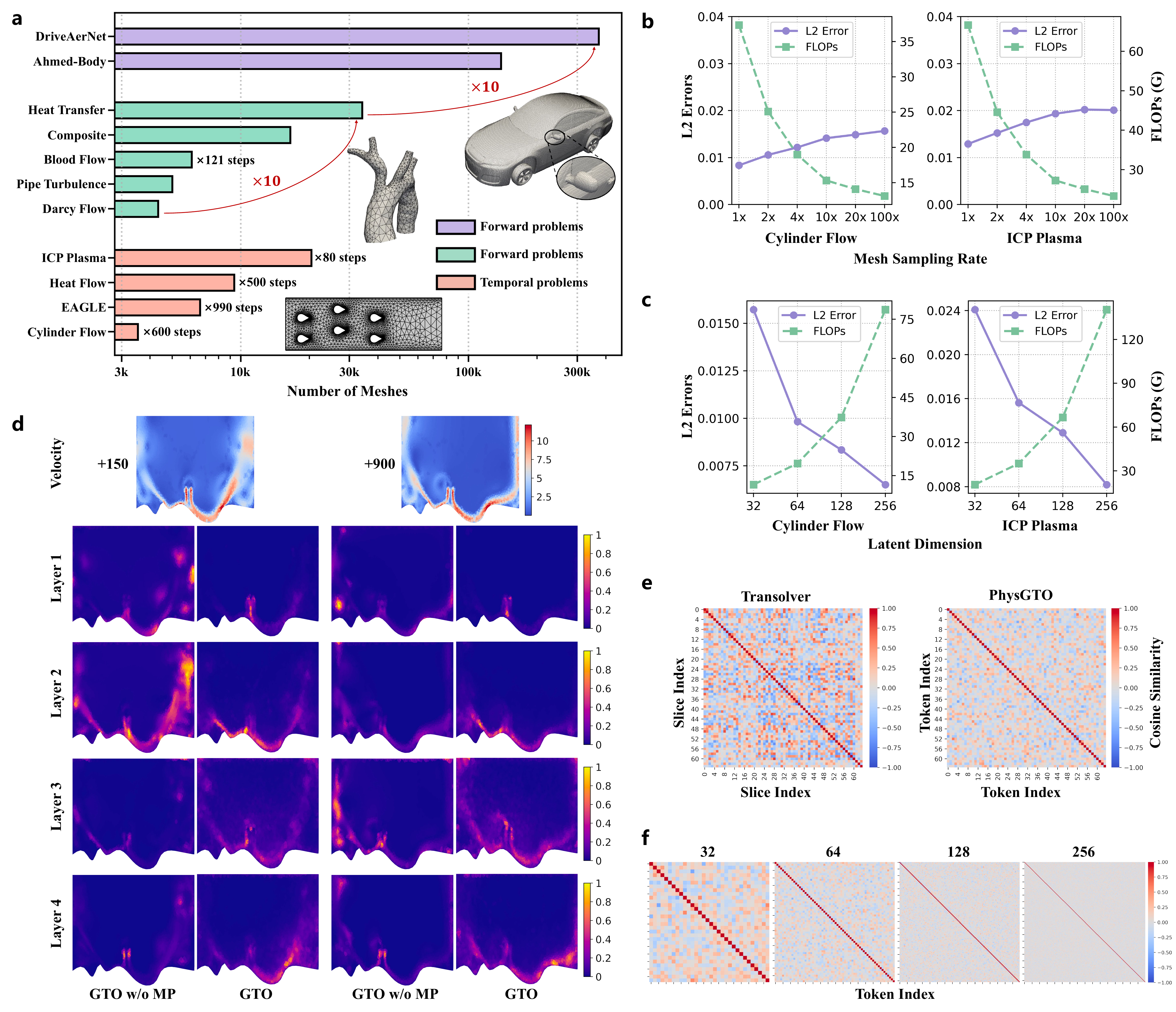}
\vspace{-8mm}
\caption{
Comprehensive analysis of physGTO: scalability and adaptability.
(a) Comparison of all benchmarks. PhysGTO is evaluated across eleven representative datasets spanning three major benchmark categories with mesh resolutions ranging from 3k to over 300k. Notably, PhysGTO achieves state-of-the-art performance across all datasets, demonstrating strong scalability and adaptability to high-fidelity tasks under limited computational resources.
(b) Varying the sampling rates of the input-output mesh, and testing on the full mesh.
(c) Model scalability on the cylinder flow and plasma dataset for latent dimension.
(d) Visualization of the attention weights associated with the learnable queries across four different layers (rows 2–5) of PhysGTO on the EAGLE dataset, evaluated at two time steps. 
Each adjacent column pair presents a side-by-side comparison between GTO w/o MP (left), which removes local graph message passing, and the full PhysGTO (right).
(e) Comparison of the attention token correlation matrices between the strong baseline Transolver and the proposed PhysGTO.
(f) Correlation matrix of the learnable consistency queries of the last layer in PhysGTO, showing the correlation coefficients from $M=32$ to $M=256$.
}
\label{model analyze}
\end{figure}
\subsubsection{The scalability of PhysGTO} \label{the scalability of PhysGTO}
Large-scale simulation applications often involve processing massive, complex meshes, posing significant computational challenges. 
As shown in Fig.~\ref{model analyze}a, both the dataset scales and individual mesh resolutions vary substantially. 
As mesh resolution increases, particularly in large-scale datasets, each sample may contain hundreds of thousands to millions of irregular cells. 
This rapid growth in complexity raises a key question: how can we effectively represent the full physical domain without exhaustively processing every mesh point during training? Thus, we investigate two complementary directions: (1) whether reduced and structured sampling strategies over large meshes can preserve prediction accuracy, and (2) how increasing the model capacity of PhysGTO impacts its ability to capture fine-grained physical patterns. These analyses provide a mechanistic understanding of PhysGTO’s strong generalization across diverse physical scenarios and highlight its potential as a foundational surrogate model for large-scale, real-world applications.

First, we reinterpret large-scale mesh learning as an operator approximation task under constrained topological observations, where training is performed on a reduced subset of mesh elements while aiming to recover the full spatial distribution.
To this end, we define a representative subset $\mathcal{M}' \subset \mathcal{M}$ and implement a point-edge consistent sampling strategy. 
Here, we first perform edge-based sampling that prioritizes structural connectivity, then select the covered nodes, and randomly add nodes from uncovered regions to ensure adequate spatial coverage. 
This method, termed the Topology-Aware Mesh Sampler (see Sec.~\ref{methods}), allows us to vary the sampling ratio $\alpha$, ensuring $N(\mathcal{M}') \approx \alpha \cdot N(\mathcal{M})$ to reduce computational costs during training, while inference is always performed on the full mesh.
As visualized in Fig.~\ref{model analyze}b, PhysGTO maintains stable predictive accuracy across a wide range of sampling ratios.
This indicates that even with significantly fewer topological observations, PhysGTO retains the ability to generalize across the whole physical domain, highlighting its suitability for scalable simulation surrogates in complex environments.

On the other hand, we investigate the impact of model capacity on predictive performance. 
As shown in Fig.~\ref{model analyze}c, we observe a consistent improvement in accuracy as the dimensionality of hidden tokens increases. 
This finding suggests that PhysGTO possesses strong potential to scale favorably as a physics foundation model, offering both architectural flexibility and high-fidelity performance under increasing task complexity. 

\subsubsection{Learned Attention for Physics}
With the adoption of the Unified Graph Embedding (UGE) strategy, PhysGTO efficiently handles datasets of varying scales. The processed data is subsequently processed by the GTO block, which comprises two key components: a message-passing module and a projection-inspired attention mechanism, both contributing to capture the underlying physical characteristics of the data. 
Here, the following proposition can be obtained:
\begin{proposition}
The GTO block of PhysGTO can be seen as the sum of a local learnable kernel operator and a global learnable operator.\label{GTO operator}
\end{proposition}
The proof of Proposition~\ref{GTO operator} is presented in SI Sec.~\ref{supp_gto}. 
On the one hand, the local learnable kernel operator in PhysGTO is designed to preserve critical topological information in the discretized mesh domain $\mathcal{M}$.  
By introducing edge-based sampling, it effectively complements the global attention component, which tends to overlook local structures.

As shown in Fig.~\ref{model analyze}d, we visualize the attention weights learned by a four-layer GTO architecture on the EAGLE dataset from Sec.~\ref{sec_example_2}. 
Specifically, we extract the attention maps from each block, defined as $\text{softmax}\left(\frac{Q_{ls} W_0^T}{\sqrt{C}}\right)$, with further details provided in Sec.~\ref{methods}.
These are compared with those obtained from a model that retains only the global attention component (GTO w/o MP). 
The results clearly show that incorporating local kernels enables PhysGTO to better capture neighborhood-level physical features, enhance attention on adjacent regions, and preserve topological detail. 
This local sensitivity proves critical for improving physical simulation accuracy. Additional supporting studies are provided in Sec.~\ref{Ablation Studies}. 

On the other hand, the core of global attention in PhysGTO lies in the learned subspace queries $Q_{ls} =\{q_i\}\in \mathbb{R}^{[M, C]}$, where $q_i\in \mathbb{R}^{[C]}$, $M$ denotes the subspace dimension and $C$ the latent feature size. 
Each token $q_i$ in $Q_{ls}$ can be interpreted as a basis vector of a learned subspace, encoding representative patterns or features within the physical domain. 
By projecting the original high-dimensional tokens into this compact, learnable subspace, PhysGTO effectively captures long-range dependencies across the mesh. 
Additionally, to further validate this basis-oriented interpretation, we compare PhysGTO with DeepONet-based methods that rely on handcrafted basis functions in SI Sec.~\ref{supplementary: compare with deeponet} for details.

We further investigate the expressiveness and diversity of the learned queries in PhysGTO by analyzing their attention token correlations. As shown in Fig.~\ref{model analyze}e, we compare the attention token correlation matrices between the strong baseline Transolver~\cite{wu2024transolver} and the proposed PhysGTO. PhysGTO exhibits more decorrelated and diverse query patterns, indicating improved capacity for capturing complex spatial dependencies. 
To further quantify this, we visualize the correlation matrices of $Q_{ls}$ across $M$ query vectors in the final layer, with $M$ varying from 32 to 256, as illustrated in Fig.~\ref{model analyze}f. As the number of queries increases, PhysGTO maintains low inter-token correlation, highlighting its ability to flexibly model multiscale physical patterns while preserving high representational capacity.
The results show that the subspace queries consistently span a set of nearly uncorrelated latent directions, confirming that they form a robust basis with high expressivity—serving as an effective global kernel operator. 
Beyond the EAGLE dataset, we also conduct more experiments on different configurations of $M$ and $C$, and some of them are summarized in Fig.~\ref{model analyze}c.
The results indicate that the model’s representation power (measured via the $L_2$ error) is tightly coupled with the rank of $Q_{ls}$, where $\text{rank}(Q_{ls}) \leq \min(M, C)$. Based on this observation, we adopt the configuration $M = C$ to ensure an efficient balance between model capacity and computational cost. More comprehensive results and discussions on the effect of varying $M$ and $C$ can be found in Sec.~\ref{sec_ablation_scalab}.
\subsubsection{Model Complexity Analysis}
In analyzing the overall computational complexity of the PhysGTO model, we focus on two main components: message-passing and projection-inspired attention. 
The message-passing layer has a complexity of $O(N \times d \times \alpha/2)$, where $d$ is the maximum graph degree, and $\alpha$ is a sampling factor. 
The projection-inspired attention has a complexity of $O(2 \times N \times M \times C + M^2 \times C)$, where $N$ is the number of mesh points, $M$ is the number of tokens, and $C$ is the feature dimensionality. 
Since $M \leq N$, the complexity of global attention scales linearly with the number of mesh points $N$. 
Therefore, the total complexity of PhysGTO is $O(N \times \beta)$, where $\beta = d \times \alpha/2 + 2 \times M \times C + M^2 \times C/N$.
Given that $M \ll N$, the overall complexity for large-scale physical simulations remains manageable and efficient.

\begin{table}[t]
\centering
\vspace{-5mm}
\caption{
Comparison of performance (average relative $L_2$ error, as $\bar{\varepsilon}_{L_2}$), the number of trainable parameters ($N_P/$M) and computational cost (FLOPs/GFLOPs) between PhysGTO and the state-of-the-art (SOTA) baselines across three tasks: unstructured mesh problems (Case 1), complex transient flow problems with 5-step rollout (Case 2), and large-scale 3D geometry problems (Case 3). \textbf{Bold} numbers indicate the best performance. $\Delta$ denotes the relative improvement between PhysGTO and the SOTA baseline.}\label{task_efficiency}
\resizebox{\textwidth}{!}{%
\begin{tabular}{l|ccc@{\hspace{0.8cm}}|ccc@{\hspace{0.8cm}}|ccc}
\toprule
Dataset & \multicolumn{3}{c@{\hspace{0.8cm}}|}{$\bar{\varepsilon}_{L_2}$ $\downarrow$} & \multicolumn{3}{c@{\hspace{0.8cm}}|}{$N_P$/M$\downarrow$} & \multicolumn{3}{c}{FLOPs/GFLOPs$\downarrow$}\\
\midrule
\rowcolor{gray!15}
\multicolumn{10}{c}{\textbf{Case 1: Unstructured Mesh Problems\cite{chen2024learning}}}\\
\cmidrule(lr){2-4} \cmidrule(lr){5-7} \cmidrule(lr){8-10}
& NORM\cite{chen2024learning} & PhysGTO & $\Delta$ & NORM\cite{chen2024learning} & PhysGTO & $\Delta$ & NORM\cite{chen2024learning} & PhysGTO & $\Delta$\\
\midrule 
Irregular Darcy 
&1.05e-2&\textbf{6.65e-3}&\textbf{36.67\%}
&0.533  &\textbf{0.436}  &\textbf{18.20\%}
& 0.677 &\textbf{0.306}  &\textbf{54.80\%}\\
Pipe Turbulence 
&1.01e-2&\textbf{6.30e-3}&\textbf{37.62\%} 
&0.533  &\textbf{0.440}  &\textbf{17.45\%} 
& 0.790 &\textbf{0.664}  &\textbf{15.95\%}\\
Heat Transfer   
&2.70e-4&\textbf{1.56e-4}&\textbf{42.22\%} 
&2.122  &\textbf{0.472}  &\textbf{77.76\%} 
&1.205  &\textbf{1.006}  &\textbf{16.51\%}\\
Composite       
&9.99e-3&\textbf{9.27e-3}&\textbf{7.21\%}
&0.533  &\textbf{0.497}  &\textbf{6.75\%}
&2.431  &\textbf{1.360}  &\textbf{44.06\%}\\
Blood Flow      
&4.82e-2&\textbf{2.40e-2}&\textbf{50.21\%} 
&43.963 &\textbf{1.197}  &\textbf{97.28\%} 
&59.17  &\textbf{3.744}  &\textbf{93.67\%}\\
\midrule
\rowcolor{gray!15}
\multicolumn{10}{c}{\textbf{Case 2: Complex Transient Flow Problems (5-step rollout)}}\\
\cmidrule(lr){2-4} \cmidrule(lr){5-7} \cmidrule(lr){8-10}
& GraphViT\cite{janny2023eagle} & PhysGTO & $\Delta$ & GraphViT\cite{janny2023eagle} & PhysGTO & $\Delta$ & GraphViT\cite{janny2023eagle} & PhysGTO & $\Delta$\\
\midrule
Cylinder Flow\cite{pfaff2020learning} 
& 1.06e-2 & \textbf{8.34e-3} & \textbf{21.32\%} 
& 10.318 & \textbf{2.379} & \textbf{76.94\%} 
& 52.500 & \textbf{25.094} & \textbf{52.20\%}\\
EAGLE\cite{janny2023eagle} 
& 6.12e-2 & \textbf{5.98e-2} & \textbf{2.29\%} 
& 10.318 & \textbf{2.379} & \textbf{76.94\%} 
& 101.353 & \textbf{48.302} & \textbf{52.34\%}\\
\midrule
& Transolver \cite{wu2024transolver} & PhysGTO & $\Delta$ & Transolver \cite{wu2024transolver} & PhysGTO & $\Delta$ & Transolver \cite{wu2024transolver} & PhysGTO & $\Delta$\\
\midrule
Heat Flow 
& 5.24e-3 & \textbf{5.21e-3} & \textbf{0.57\%} 
& 2.950 & \textbf{2.379} & \textbf{19.36\%} 
& 63.382 & \textbf{54.507} & \textbf{14.00\%}\\
ICP Plasma 
& 2.69e-2 & \textbf{1.29e-2} & \textbf{52.04\%} 
& 2.950 & \textbf{2.379} & \textbf{19.36\%} 
& 52.270 & \textbf{44.753} & \textbf{14.38\%}\\
\midrule
\rowcolor{gray!15}
\multicolumn{10}{c}{\textbf{Case 3: Large-scale 3D Geometry Problems}}\\
\cmidrule(lr){2-4} \cmidrule(lr){5-7} \cmidrule(lr){8-10}
& GINO\cite{li2024geometry} & PhysGTO & $\Delta$ & GINO\cite{li2024geometry} & PhysGTO & $\Delta$ & GINO\cite{li2024geometry} & PhysGTO & $\Delta$\\
\midrule
Ahmed-body\cite{li2024geometry} 
& 8.31e-2 & \textbf{5.70e-2} & \textbf{31.41\%} 
& 230.690 & \textbf{1.791} & \textbf{99.22\%} 
& 30839.283 & \textbf{194.113} & \textbf{99.37\%}\\
DriveAerNet\cite{elrefaie2024drivaernet} 
& 1.65e-1 & \textbf{1.45e-1} & \textbf{12.12\%} 
& 230.690 & \textbf{1.746} & \textbf{99.24\%} 
& 97501.312 & \textbf{613.790} & \textbf{99.37\%}\\
\midrule
\rowcolor{blue!10}
Mean $\Delta$ & - & - & \textbf{26.70\%} & - & - & \textbf{55.32\%} & - & - & \textbf{50.60\%}\\
\bottomrule
\end{tabular}
}
\end{table}
We evaluate PhysGTO’s scalability and computational complexity with respect to critical architectural factors, as discussed in Sec.~\ref{the scalability of PhysGTO}. These analyses demonstrate how internal design choices—such as token dimensionality and network depth—affect computational efficiency. 
Furthermore, Tab.~\ref{task_efficiency} benchmarks PhysGTO against state-of-the-art (SOTA) baselines across three representative tasks: unstructured mesh problems (Sec.~\ref{sec_example_1}), complex transient flows (Sec.~\ref{sec_example_2}), and large-scale 3D geometry simulations (Sec.~\ref{sec_example_3}). 
Notably, PhysGTO achieves superior accuracy on all datasets while using significantly fewer trainable parameters and floating-point operations (FLOPs).
In particular, on industrial-scale 3D datasets (Ahmed-Body and DrivAerNet), PhysGTO reduces computational cost (FLOPs) by more than two orders of magnitude compared to GINO~\cite{li2024geometry}, while maintaining high predictive fidelity. \textbf{Across all 11 benchmark tasks}, PhysGTO achieves an average reduction of \textbf{26.70\% in relative $L_2$ error}, with \textbf{55.32\% fewer parameters} and \textbf{50.60\% lower computational cost}, underscoring its architectural efficiency, scalability, and practical readiness for real-world scientific computing workloads.
\section{Discussion}
In this work, we present PhysGTO, an efficient Graph-Transformer Operator for learning physical dynamics on unstructured manifolds. PhysGTO integrates manifold embeddings to enhance geometric adaptability across diverse physical conditions. By combining a lightweight message-passing mechanism with a linear-complexity attention mechanism, PhysGTO effectively captures both local and global dependencies while significantly reducing resource demands. This design ensures rapid and accurate inference in diverse physical scenarios while maintaining flexibility and scalability for large-scale applications. Extensive evaluations across a wide range of PDE-governed benchmarks underscore the performance of PhysGTO. It not only achieves state-of-the-art accuracy but also significantly reduces computational costs, demonstrating superior efficiency even in two large-scale scenarios. Furthermore, in time-series applications, PhysGTO effectively mitigates error accumulation over time, making it particularly well suited for downstream tasks that demand high-precision operations.

Specifically, PhysGTO introduces several key innovations that enhance its efficiency and effectiveness in learning physical dynamics on complex geometries.
First, compared to traditional GNN-based models, a lightweight message-passing layer is used to capture local dependencies while preserving structural consistency. By incorporating a directional, flux-oriented update scheme and a topology-aware sampling strategy, the model ensures stable local information propagation with reduced computational and memory overhead, enabling it to scale effectively on large unstructured meshes.
Second, while attention-based architectures—such as those used in GNOT \cite{hao2023gnot}, IPOT \cite{lee2024inducing}, and Transolver \cite{wu2024transolver}—are effective for capturing global dependencies and modeling solution operators as integral kernels, they often lack the inductive bias necessary for accurate local approximation, particularly in regions with sharp variations. 
To address this, PhysGTO incorporates message-passing layers that explicitly encode local topological relations. These layers serve as discrete approximations of local differential neighborhoods, reinforcing the model’s ability to learn fine-grained physical behaviors.
This hybrid design complements the projection-inspired attention mechanism, which introduces a learnable spatial consistency query to capture long-range dependencies efficiently while maintaining linear complexity. By combining the strengths of local message passing and global attention, PhysGTO achieves improved accuracy and generalization across non-Euclidean domains.
Finally, although message-passing layers enhance local modeling, their application to large-scale meshes can lead to increased cost due to the rapid growth in edge connections. 
To address this, PhysGTO employs a sparsified edge sampling strategy, demonstrating that a randomly selected subset of edges is sufficient to retain the benefits of local connectivity. 
With the attention mechanism’s global receptive field supporting feature propagation, this design maintains the necessary local-global balance while significantly reducing computational overhead.

Beyond these demonstrated performance, two key insights emerge from our study.  
First, the computational efficiency of PhysGTO’s topology-aware message passing and projection-inspired attention provides insights into improving physics-informed learning models.  
By ensuring structural consistency through topology-preserving sampling and compressing local physical interactions with linear-complexity feature aggregation, PhysGTO enables efficient information propagation across spatially and temporally complex domains.  
Unlike traditional numerical solvers, whose complexity scales nonlinearly with mesh resolution, PhysGTO exhibits resolution-invariant generalization, making it a potential candidate for large-scale physics pre-training. These properties highlight its potential for broader applications in scalable physics-based learning frameworks. 
Second, while PhysGTO effectively models physical dynamics under diverse conditions, it remains a task-specific operator rather than a general-purpose foundation model. 
In contrast to large-scale pre-trained models in natural language processing, where tasks are conditionally aligned through unified representations, current physics-driven learning methods lack the ability to generalize across multiple PDE classes.  
A promising direction is to explore cross-task conditional alignment, leveraging generative models to build a pre-trained foundation model for scientific computing.  
By incorporating unified physical priors and multi-task training, such a model could enhance generalization across diverse PDE-driven applications, offering an accessible direction toward more flexible physics-based learning. 

Additionally, diffusion models~\cite{ho2020denoising,song2020denoising,song2020score} have recently emerged as a promising tool for modeling physical fields, particularly in flow field prediction.
They are typically classified into conditional models, which generate scenario-specific predictions based on explicit conditioning inputs~\cite{gao2024bayesian, du2024conditional}, and unconditional models, which enhance generalization across diverse conditions~\cite{li2024learning}.
While their generative nature helps mitigate uncertainty by improving distribution alignment between training and test sets, their applicability to complex physical simulations remains limited.
One important reason for this limitation is that current diffusion models for physical systems largely rely on UNet-based architectures~\cite{ronneberger2015u} or simple multilayer perceptrons (MLPs) for noise generation and denoising, both of which have restricted expressive capacity in modeling intricate physical phenomena.
In this context, employing the GTO blocks from PhysGTO as the backbone for noise generation and removal within diffusion models offers a highly promising direction, potentially enabling significantly improved expressivity and fidelity in complex physics-driven tasks.

Developing generalizable physics-based models is not only a matter of architectural design, but also hinges on the availability of high-quality benchmark datasets.
A major bottleneck in current research is the lack of comprehensive benchmarks that reflect the diversity and complexity of real-world physical systems.
To address this, we introduce three representative datasets covering steady-state problems, temporal dynamics, and large-scale computational fluid dynamics.
These datasets are designed to support standardized, task-diverse evaluation, thereby encouraging the development of models that generalize beyond narrowly defined scenarios.
Overcoming the limitations of current benchmarks calls for the development of more diverse and scalable datasets that can faithfully capture the complexity inherent in real-world engineering applications.

Lastly, we delineate the scope of the present design choices regarding boundary conditions, symmetry, and physical consistency. The BC-aware step in PhysGTO employs a standard hard Dirichlet enforcement, representing a conventional but reliable approach for maintaining boundary fidelity. Similarly, the current architecture does not explicitly encode physical symmetries (e.g., equivariance \cite{brandstetter2022lie}) or incorporate PDE-residual constraints \cite{yu2022gradient, lippe2023pde}. These aspects are omitted by design to emphasize the primary contribution of this work—the hybrid operator kernel that integrates flux-oriented message passing with projection-inspired attention, while preserving linear computational complexity. Potential extensions, such as symmetry-preserving flux kernels or residual-based regularization, are considered orthogonal to the present focus and may be integrated within the same framework in future studies. The demonstrated stability over long-horizon forecasts and the strong generalization to large-scale 3D industrial meshes (e.g., Ahmed-Body and DrivAerNet) collectively substantiate the efficacy of the proposed architectural formulation.

\section{Methods}\label{methods}
This section presents a comprehensive description of the proposed PhysGTO framework. It first introduces the overall model architecture and its key components. Subsequently, the benchmark datasets and baseline methods used in the experiments are described. Then the error metrics employed for model evaluation are outlined. Finally, we also provide detailed ablation studies to further evaluate the effectiveness of its components.
\subsection{Model Architecture of PhysGTO}
As shown in Fig.\ref{pipline}a, PhysGTO is a unified framework designed to handle diverse physical simulation tasks, including both temporal predictions and forward inference. The model begins by embedding heterogeneous physical inputs into a graph representation via the Unified Graph Embedding (UGE) module. The resulting aligned node features and sparsified edge structures are then projected into a latent space and processed through a stack of Graph Transformer Operator (GTO) blocks. Finally, the decoder maps the latent representations back to the physical domain, followed by a BC-aware post-processing step that ensures strict enforcement of boundary conditions. The SI Sec.~\ref{supp_algorithms} provides the pseudo-code for the entire training process, offering a comprehensive understanding of our approach.
\subsubsection{Unified Graph Embedding}

\textbf{Multi-Condition Aligner.} As shown in Fig.\ref{pipline}c (left), at the node level, the input features consist of local information—such as position coordinates $\boldsymbol{x}_i\in\mathcal{R}^{d}$, relative positions $\boldsymbol{n}_i\in\mathcal{R}^{1}$, and physical fields $\boldsymbol{u}_i\in\mathcal{R}^{c}$—as well as global context, including input parameters $\boldsymbol{a}\in\mathcal{R}^{l}$ and time $t\in\mathcal{R}^{1}$ for transient prediction tasks. In the subsequent encoder, we perform dedicated fusion of these local and global components to ensure effective condition alignment.

\textbf{Topology-Aware Mesh Sampler.} As shown in Fig.\ref{pipline}c (right), to reduce the memory and computational overhead associated with dense graph connectivity, we propose a hierarchical, topology-preserving sampling strategy that adapts to mesh scale and resource constraints. 
Given a physical mesh graph $\mathcal{G} = (\mathcal{V}, \mathcal{E})$, where $\mathcal{V}$ denotes the set of nodes and $\mathcal{E}$ the set of edges, we first construct a directed graph based on a flux-oriented criterion, in which edge directions are aligned with the dominant physical flow (e.g., velocity or gradient fields) to reflect the underlying transport dynamics. 
Specifically, for each edge connecting node $i$ to node $j$, we use the sign of the inner product between the physical vector (e.g., velocity $\vec{u}_i$) and the edge vector $\vec{x}_j - \vec{x}_i$ to determine the direction: the edge is directed from $i$ to $j$ if the product is positive, and from $j$ to $i$ otherwise.

For small-scale meshes ($|\mathcal{V}| < 10^5$), bidirectional edge pairs $(i, j)$ and $(j, i)$ are replaced with a single directed edge $(i, j)$ if the local projected flux $\phi_{ij} > 0$, yielding a reduced edge set $\tilde{\mathcal{E}}$ as $\tilde{\mathcal{E}} = \{ (i, j) \in \mathcal{E} \;|\; \phi_{ij} > 0 \}$. This reduction preserves physical directionality while halving the edge count, enabling efficient message propagation without compromising structural fidelity. 
For medium-scale graphs ($10^5 \leq |\mathcal{V}| \leq 10^6$), further sparsification is achieved via uniform random edge sampling. A fixed ratio $\rho \in (0,1]$ determines the probability of retaining each edge from $\tilde{\mathcal{E}}$ as $\hat{\mathcal{E}} = \{ e \in \tilde{\mathcal{E}} \;|\; \xi_e < \rho, \xi_e \sim \mathcal{U}(0,1)\}$. The resulting graph $\hat{\mathcal{G}} = (\hat{\mathcal{V}}, \hat{\mathcal{E}})$ retains only the nodes participating in sampled edges. Despite the sparsification, long-range dependencies are effectively recovered downstream via the projection-inspired attention mechanism. 
For large-scale scenarios ($|\mathcal{V}| > 10^6$), we build upon the previous sparsification steps by applying a point–edge consistent sampling scheme. 
Specifically, a subset of edges is first selected from $\tilde{\mathcal{E}}$ to preserve topological structure. The nodes covered by these edges are then included, and additional nodes are randomly sampled from uncovered regions to ensure adequate spatial coverage during training.
During inference, we partition the full mesh into $K$ spatially uniform subgraphs $\{\hat{\mathcal{G}}_k\}_{k=1}^{K}$ via edge-based partitioning, and process each subgraph independently in batch mode. Predictions from overlapping regions are aggregated via mean pooling to reconstruct the full-field output. This procedure enables efficient large-scale inference while preserving consistency across subdomains.

This hierarchical graph sampling and aggregation framework ensures scalability across a wide range of spatial resolutions, supporting efficient inference under constrained hardware conditions without sacrificing physical accuracy.
\subsubsection{Encoder}
\textbf{Nodes.} Each node contains two types of information: the \textit{local features} $\boldsymbol{v}_i = (\boldsymbol{x}_i, \boldsymbol{n}_i, \boldsymbol{u}_i)$, which include spatial coordinates, relative positions, and physical field values; and the \textit{global condition} $\boldsymbol{v}_g = (\boldsymbol{a}, t)$, representing the input parameters and the time step for transient prediction. To project node features from the physical space into the latent space, we adopt two separate MLPs, $\Phi^V_1$ and $\Phi^V_2$, and fuse the encoded representations as $\mathbf{v}_i = \Phi^V_1(\boldsymbol{v}_i) + \Phi^V_2(\boldsymbol{v}_g)$. Moreover, to capture rich spatial and frequency-aware patterns from coordinates, we incorporate a sinusoidal positional encoding (SPE) \cite{janny2023eagle} defined as $F(\boldsymbol{X}) = \left[\cos \left(2^i \pi \boldsymbol{X}\right), \sin \left(2^i \pi \boldsymbol{X}\right)\right]_{i=-\delta, \ldots, \delta}$, where $\delta \in \mathbb{Z}^+$. 
The encoding $F(\boldsymbol{X})$ injects multi-scale positional context into the node features and is explicitly integrated into the processor, enabling more expressive message passing over complex geometric domains.

\textbf{Edges} Edges encode the geometric relationships between nodes and are constructed based on the sampled mesh connectivity from the topology-preserving sampling stage. 
For each edge $(i,j)$, we extract relative positional features including the displacement vector $\mathbf{x}_{ij} = \boldsymbol{x}_i - \boldsymbol{x}_j$, its inverse $\mathbf{x}_{ji}$, and its magnitude $|\mathbf{x}_{ij}|$. 
These components are concatenated and passed through a learnable MLP $\Phi^E$ to obtain the edge-level latent representation as $\mathbf{e}_{ij} = \Phi^E(\mathbf{x}_{ij}, \mathbf{x}_{ji}, |\mathbf{x}_{ij}|)$. 

The final output of the encoder consists of the latent node representations $\boldsymbol{V} = \{\mathbf{v}_i\} \in \mathbb{R}^{N \times C}$ and latent edge representations $\boldsymbol{E} = \{\mathbf{e}_{ij}\} \in \mathbb{R}^{E \times C}$, which are subsequently processed by the Graph-Transformer module.
\subsubsection{Processor}
As shown in Fig.\ref{pipline}d, the processor is designed to capture both local and global dependencies through integrating flux-oriented message passing with a projection-inspired global attention mechanism. This combination enables PhysGTO to model intricate physical correlations across multiple spatial and temporal scales while maintaining computational efficiency.

\textbf{Flux-Oriented Message Passing} 
At each GTO block, local message passing is performed over the sampled mesh-based edges $\boldsymbol{E}^M$ and graph $\boldsymbol{G} = (\boldsymbol{V}, \boldsymbol{E})$, where $\boldsymbol{V}=[\boldsymbol{V}, F(\boldsymbol{X}))]$ is enhanced via the positional embedding. First, the message passing updates edge features via an MLP $\Phi_P^E$ using the features of connected nodes as $\mathbf{e}_{ij}^M = \mathbf{e}_{ij}^M + \Phi_P^E(\mathbf{e}_{ij}^M, \mathbf{v}_i, \mathbf{v}_j)$. Here $\mathbf{v}_i$ and $\mathbf{v}_j$ represent the features of sender and receiver nodes respectively. 
The edge embedding $\mathbf{e}_{ij}^M \in \mathbb{R}^{C}$ encodes directional flux between nodes $i$ and $j$. 
To represent bidirectional interactions, it is partitioned into two components as $\mathbf{e}_{i \rightarrow j}, \mathbf{e}_{j \rightarrow i} \in \mathbb{R}^{C/2}$, capturing the influence from $i$ to $j$ and from $j$ to $i$, respectively. Then, node features are updated via bidirectional flux aggregation. 
For a given node $k$, we separately aggregate the outgoing and incoming directional flux embeddings using mean aggregation as follows, 
\begin{equation}
\mathbf{m}_k = \left[ 
\frac{1}{|\mathcal{N}_k^{\text{out}}|} \sum_{l \in \mathcal{N}_k^{\text{out}}} \mathbf{e}_{k \rightarrow l},\quad 
\frac{1}{|\mathcal{N}_k^{\text{in}}|} \sum_{m \in \mathcal{N}_k^{\text{in}}} \mathbf{e}_{m \rightarrow k} 
\right] \in \mathbb{R}^{C},
\end{equation}
where $\mathcal{N}_k^{\text{out}}$ and $\mathcal{N}_k^{\text{in}}$ denote the sets of target and source neighbors of node $k$, respectively.
Finally, a node-wise MLP $\Phi^V$ updates each node's representation by integrating its current feature $\mathbf{v}_k$ with the aggregated bidirectional flux message $\mathbf{m}_k$ as $\mathbf{v}_k = \mathbf{v}_k+\Phi^V(\mathbf{v}_k, \mathbf{m}_k)$. 
This message-passing scheme explicitly leverages directional flux information derived from edge interactions to refine node representations.

\textbf{Projection-inspired Attention} To enhance long-range reasoning under linear complexity, we propose a projection-inspired attention mechanism. Let $W_0 := \{\mathbf{v}_i\} \in \mathbb{R}^{N \times C}$ denote the local node representations. A set of learnable queries $Q_{ls} \in \mathbb{R}^{M \times C}$ with $M \ll N$ is used to define a latent projection space. The attention mechanism proceeds as:
\begin{equation}
\left\{
\begin{aligned}
&W_1 = \text{softmax}\left(\frac{Q_{ls} W_0^T}{\sqrt{C}}\right) W_0, \\
&W_2 = \text{softmax}\left(\frac{W_1 W_1^T}{\sqrt{C}}\right) W_1, \\
&W_3 = \text{softmax}\left(\frac{W_0 W_2^T}{\sqrt{C}}\right) W_2,
\end{aligned}
\right.
\end{equation}\label{attention_eq}
where $Q_{ls} \sim \mathcal{N}(0,1)$ is randomly initialized and jointly optimized. $W_{ls} = \text{softmax}\left(\frac{Q_{ls} W_0^T}{\sqrt{C}}\right) \in \mathbb{R}^{M \times N}$ represents the attention weight matrix (i.e., the attention map), where each of the $M$ learnable query tokens attends to $N$ spatial points through dot-product attention over $C$-dimensional features.
This mechanism allows node features to be projected into a learned global subspace ($W_1$), refined ($W_2$), and reprojected back to the original node space ($W_3$), yielding global representations enriched with structural context.

\textbf{Pre-Norm Structure and Multi-Head Attention} To ensure stable training, we adopt a pre-normalization structure with residual connections:
\begin{equation}
\left\{
\begin{aligned}
&V^{1}_{G} = W_0 + W_3, \\
&V^{2}_{G} = V^{1}_{G} + \text{FFN}(\text{LN}(V^{1}_{G})),
\end{aligned}
\right.
\end{equation}
where $\text{FFN}$ is a feed-forward network and $\text{LN}$ denotes layer normalization. Furthermore, we employ multi-head attention to model diverse physical interactions and enhance representational capacity. 
By stacking $L$ such processor blocks, PhysGTO captures hierarchical relationships between local and global physical patterns.

\subsubsection{Decoder} The decoder maps latent node representations back to the physical output space. Depending on the task, two decoding strategies are used: one for steady-state problems and another for transient scenarios. The former problems such as aerodynamic coefficient prediction or pressure field estimation, the decoder directly transforms the latent node feature $\mathbf{v}_i$ into the target physical variable using an MLP as $\boldsymbol{p}_i = \Phi_D^V(\mathbf{v}_i, \mathbf{V}_g)$, where $\mathbf{V}_g$ is the global-conditional encoding, and $\Phi_D^V$ is the decoding MLP. The output $\boldsymbol{p}_i$ corresponds to the steady-state physical value at node $i$. For the latter, we adopt the first-order Euler format as $\boldsymbol{\hat{u}}_i^t = \boldsymbol{\hat{u}}_i^{t-1}+\Phi_D^V(\mathbf{v}_i, \mathbf{V}_g)$, $t \in [1, T]$, where $\boldsymbol{\hat{u}}_i^0=\boldsymbol{u}_i^0$, the global-conditional encoding $\mathbf{V}_g$ also contains the embedding of time $t$.

\textbf{Post-processor}. 
To ensure strict compliance with prescribed physical constraints, we introduce a post-processing module that explicitly enforces boundary conditions on the predicted physical fields. 
As a representative example, we consider the no-slip boundary condition, which is commonly used in fluid dynamics to impose zero velocity at solid boundaries. 
Let $\mathcal{B} \subset \mathcal{V}$ denote the set of boundary nodes subject to this condition. The post-processor corrects the predicted values at these nodes by enforcing $\tilde{\boldsymbol{u}}_i^t = 0$ for all $i \in \mathcal{B}$. 
This simple yet effective correction ensures that the final output strictly adheres to the no-slip condition. 
The same framework can be extended to other boundary types (e.g., Neumann or periodic) by adjusting the correction rules accordingly. 
By decoupling constraint enforcement from the learning process, this design preserves physical fidelity without increasing model complexity, which is especially beneficial in high-resolution simulations with complex geometries.

\subsubsection{Training of PhysGTO} \label{training of PhysGTO}

In practice, we can obtain a training set $\{(\boldsymbol{a}^k,\boldsymbol{u}^k)\}_{i=1}^{D}$, where each input $\boldsymbol{a}^k \in \mathcal{A}_\mathcal{M}$ and corresponding output $\boldsymbol{u}^k \in \mathcal{U}_\mathcal{M}$ are discretized representations of physical simulations, and $D$ denotes the total number of training samples.
To approximate the operator $\mathcal{G}_{\mathcal{M}}$,  we introduce a parameterized neural network $\mathcal{G}_{\boldsymbol{\theta}}$ to approximate this solution set, and output $\mathcal{G}_{\boldsymbol{\theta}}(\boldsymbol{a}^k)=\hat{\boldsymbol{u}}^k$, where $1\leq k \leq D$, $\boldsymbol{\theta}$ is a set of the network parameters.

To approximate the target solution operator $\mathcal{G}_{\mathcal{M}}$ introduced in Sec.~\ref{problem setup}, we employ a parameterized neural network $\mathcal{G}_{\boldsymbol{\theta}}$ to learn the mapping from $\boldsymbol{a}^k$ to $\boldsymbol{u}^k$, yielding the prediction $\hat{\boldsymbol{u}}^k = \mathcal{G}_{\boldsymbol{\theta}}(\boldsymbol{a}^k)$ for $1 \leq k \leq D$, where $\boldsymbol{\theta}$ denotes the set of trainable parameters. Therefore, our goal is to minimize the $L_2$ relative error loss between the prediction $\hat{\boldsymbol{u}}^k$ and real data $\boldsymbol{u}^k$ in the training dataset as
\begin{equation}
\min_{\boldsymbol{\theta} \in \boldsymbol{\Theta}} 
\frac{1}{D} \sum_{k=1}^{D} 
\mathcal{L}_k(\boldsymbol{\theta})
=\min_{\boldsymbol{\theta} \in \boldsymbol{\Theta}}
\frac{1}{D} \sum_{k=1}^{D}
\frac{\parallel \boldsymbol{u}^k - \hat{\boldsymbol{u}}^k \parallel_2}{\parallel \boldsymbol{u}^k \parallel_2},
\label{eq: all_lossfunction}
\end{equation}
where $\mathcal{L}_k(\boldsymbol{\theta})$ denotes the relative $L_2$ loss for the $k$-th input while $\boldsymbol{\theta}$ is the whole parameter space.

\subsection{Benchmark Datastes} \label{dataset_details}

\textbf{Unstructured mesh problems.} The involved cases encompass a diverse range of problem settings and input-output structures in PDE operator learning from ~\cite{chen2024learning}. Specifically, Darcy flow benchmark targets the mapping of the diffusion coefficient field to the pressure field defined on an irregular 2D domain; Pipe turbulence task predicts the velocity field’s evolution from its current state on an irregular pipe domain with unstructured meshes; Heat transfer task models the solution mapping from boundary temperature conditions to the temperature field within an irregular 3D domain; Composite workpiece deformation is a more complex task to predict the deformation field from the temperature field in an irregular discretized 3D composite domain governed by multi-physics processes beyond PDEs; Blood flow dynamics case aims to  predict the spatiotemporal velocity field in an irregular 3D aortic domain from time-varying boundary conditions with cyclic fluctuations, and also driven by multi-physics interactions.

\textbf{Transient flow dynamics.} We tested PhysGTO on four complex transient flows prediction problems in open-access benchmarks, Cylinder Flow \cite{pfaff2020learning}, EAGLE \cite{janny2023eagle}, and two industrial-level tasks (Heat Flow and ICP Plasma). These tasks involve complex parameter spaces and diverse geometries, along with varying initial conditions (ICs), boundary conditions (BCs), and other parameters. These characteristics commonly found in real-world industrial applications.

\textbf{Large-scale 3D geometries.} We applied PhysGTO to an industrial design task using two high-level automotive datasets: Ahmed-Body Car \cite{li2024geometry} and DrivAerNet \cite{elrefaie2024drivaernet}. The Ahmed-Body dataset consists of six design variables with inlet velocities ranging from $10\,\text{m/s}$ to $70\,\text{m/s}$, corresponding to Reynolds numbers between $4.35 \times 10^5$ and $6.82 \times 10^6$. Simulated using the SST $k-\omega$ turbulence model, the dataset includes 551 shapes, 500 of which are used for training and 51 for validation. DrivAerNet, on the other hand, features 4000 detailed 3D car meshes with 0.5 million surface mesh faces, along with comprehensive aerodynamic data, including 3D pressure, velocity fields, and wall shear stresses. This dataset addresses the pressing need for large-scale datasets in deep learning-based engineering applications. Together, these datasets represent diverse 3D vehicle geometries with Reynolds numbers up to five million and mesh sizes ranging from 100k to 500k faces, making them valuable benchmarks for evaluating large-scale 3D physical simulations.

More details about the datasets can be found in SI Sec.~\ref{supp datasets details}.

\subsection{Baseline methods}
\textbf{Unstructured mesh problems.} PhysGTO is compared against several popular neural operators, including DeepONet \cite{lu2021learning}, POD-DeepONet \cite{lu2022comprehensive}, Fourier Neural Operator (FNO) \cite{li2020fourier}, and Geo-FNO~\cite{li2023fourier}, which introduces a geometry transformation network that maps the original function defined on an irregular physical domain to a regular computational domain, where fast Fourier transforms (FFT) can be efficiently executed. Additionally, we evaluate GraphSAGE \cite{hamilton2017inductive} as the representative of Graph Neural Networks, and the current SOTA model for these test cases, NORM \cite{chen2024learning}. For fair comparisons, irregular 2D domains were interpolated onto structured grids for FNO.However, FNO was excluded from 3D cases due to its reliance on structured grids, making spatial interpolation infeasible in highly irregular domains. Additionally, GNNs were not evaluated for heat transfer and blood flow dynamics, as these tasks involve heterogeneous input-output structures that are difficult to represent using a single graph topology.

\textbf{Transient flow dynamics.} We compare PhysGTO with several strong graph-based baselines. MeshGraphNet \cite{pfaff2020learning} is a classical GNN model that uses stacked message-passing layers to capture spatio-temporal dynamics on irregular meshes. GNOT \cite{hao2023gnot} is a scalable transformer framework utilizing linear attention and a mixture-of-experts approach \cite{fedus2022switch}, which demonstrates the advantages of linear-attention methods. Geo-FNO~\cite{li2023fourier} is a variant of FNO \cite{li2020fourier} designed to extend operator learning capabilities to irregular meshes. GraphViT \cite{janny2023eagle} is a hybrid graph-transformer that combines clustering, pooling, and global attention to capture long-range dependencies with fewer message-passing steps, enhancing scalability. Transolver \cite{wu2024transolver} is a transformer-based model tailored for geometric learning operators, offering low computational complexity.

\textbf{Large-scale 3D geometries.} Large-scale physical simulations require models with high scalability and efficiency. We evaluate PhysGTO against five competitive baselines on two large-scale car datasets. In addition to previously introduced methods (MGN, GNOT, and Transolver), we include two strong baseline methods. IPOT \cite{lee2024inducing} is another variant of the transformer-based and has been further customized for geometric learning. GINO \cite{li2024geometry} is a neural operator leveraging Fourier-based architectures (FNO \cite{li2020fourier}) for modeling of complex large-scale simulations.

More details about the implementation of these baseline methods can be found in SI Sec.~\ref{method detaials}.

\subsection{Error metric}
Across all three different classes of experiments, we adopt the mean $L_2$ relative error, abbreviated as $\epsilon$, as our evaluation metric for validation and testing datasets to assess the performance of PhysGTO.
$\epsilon$ is formulated as follows:
\begin{equation}
\epsilon = \frac{1}{D\times T\times m} \sum_{k=1}^{D} \sum_{t=1}^{T} \sum_{i=1}^{m} \frac{\|\boldsymbol{u}^{t,k,i} - \hat{\boldsymbol{u}}^{t,k,i}\|_2}{\|\boldsymbol{u}^{t,k,i}\|_2},
\end{equation}
where $D$ is the number of data sets, $T$ is the number of time steps, $m$ is the number of physical fields. $\boldsymbol{u}^{t,k,i}\in\mathcal{R}^{[N]}$ is the $i$-th physical field in set $k$ at time step $t$, $\hat{\boldsymbol{u}}^{t,k,i}\in\mathcal{R}^{[N]}$ is the predicted, $\|\cdot\|_2$ is the Euclidean norm, and $N$ is the number of points.

\subsection{Ablation Studies}\label{Ablation Studies}
The ablation and complexity analyses summarize the impact of PhysGTO’s key components and confirm its efficiency and scalability. The flux-oriented message passing and projection-inspired attention function cooperatively to capture both local and global physical correlations, enabling accurate and consistent multi-scale representations. Overall, PhysGTO achieves a balanced trade-off between accuracy and computational cost while maintaining stable generalization across complex three-dimensional physical systems.

\subsubsection{Impacts of different components}\label{sec_ablation_1}
PhysGTO is built from a Graph Topology Operator (GTO) that couples a flux-oriented message passing (MP) module with a projection-inspired global attention, and a UGE module that improves adaptability to multi-scale and multi-condition regimes. To quantify the contribution of each part, we perform ablations along three axes: (i) removing local or global modules and the BC-aware post-processing, (ii) replacing the proposed single-sided flux-oriented MP with a standard bidirectional variant, and (iii) comparing our projection-inspired attention with self-attention and Transolver-style designs. Unless otherwise stated, UGE is kept fixed across variants.

\begin{table}[t]
\centering
\caption{Ablation study of PhysGTO architecture. We compare the full model with variants using only GNN (w/o Atten), only attention (w/o MP), or no post-processing (w/o BC-aware). \textbf{Bold} and \underline{underline} numbers indicate the best and second-best performance, respectively.}\label{tab: example 2 ablation}
\resizebox{\textwidth}{!}{%
\begin{tabular}{l|c|c|c|c|c}
\toprule
\multirow{2}{*}{Config} &
\multirow{2}{*}{$N_P$/M} &
\multicolumn{2}{c|}{FLOPS/GFLOPs}&
\multicolumn{2}{c}{$L_2$ $\downarrow$} \\
&&Cylinder Flow&ICP Plasma&Cylinder Flow&ICP Plasma\\
\midrule
\textbf{PhysGTO}             
&2.379 & 25.094 & 44.753 & \textbf{8.3413e-3} & \textbf{1.2893e-2} \\
\textbf{w/o Atten}   
&\underline{1.973} & \underline{23.051} & \underline{40.944} & 1.3206e-2 & 2.7243e-2 \\
\textbf{w/o MP}     
&\textbf{0.971} & \textbf{8.443}  & \textbf{15.025} & 4.5963e-2 & 3.9447e-2 \\
\textbf{w/o BC-aware}
&2.379 & 25.094 & 44.753 & \underline{8.9917e-3} & \underline{1.2902e-2} \\
\bottomrule
\end{tabular}
}
\end{table}

\textbf{Overall architecture.} We first assess the importance of each component by removing it from the full model. The results are summarized in Table~\ref{tab: example 2 ablation}. Removing the global attention module (\textbf{w/o Atten}) forces the model to rely solely on local message passing. This resulted in a dramatic surge in $L_2$ error, which increased by \textbf{58\%} for Cylinder Flow (8.34e-3 vs 1.32e-2) and \textbf{111\%} for ICP Plasma (1.29e-2 vs 2.72e-2). This result strongly indicates that local information is insufficient for modeling large-scale physical systems and that a global attention mechanism is indispensable for capturing long-range dependencies. Conversely, removing the message passing module (\textbf{w/o MP}) had a detrimental impact on performance. The error skyrocketed by \textbf{451\%} for Cylinder Flow (8.34e-3 vs 4.60e-2) and \textbf{206\%} for ICP Plasma (1.29e-2 vs 3.94e-2). Although this pure-attention variant is the most efficient in terms of parameters ($N_P$) and FLOPs, its poor performance demonstrates that without explicit modeling of local physical interactions based on topological structure, the model fails to learn fine-grained physical dynamics. Finally, omitting the BC-aware module (\textbf{w/o BC-aware}) led to a slight but consistent performance degradation, confirming its role as an effective post-processing calibration step that enhances the model's physical fidelity. In summary, these findings clearly establish that the core architecture of PhysGTO—a synergy between local message passing and global attention—is fundamental to achieving high accuracy.

These ablation results demonstrate that both local and global modules are essential for modeling large-scale physical systems, particularly in capturing long-range dependencies and resolving multi-scale physical correlations. The superior performance of PhysGTO highlights the importance of its design strategy for multi-level representation and collaborative modeling between topological structure and projection-based global attention.

\textbf{Efficiency Trade-off of Flux-Oriented Message Passing.} Having established the necessity of the MP module, we now justify our choice of a \textbf{Flux-Oriented (single-sided)} design. The primary objective of this design is not to surpass the raw accuracy of standard bidirectional message passing, but rather to achieve a more favorable \textbf{efficiency-accuracy balance}, thereby enhancing scalability.

\begin{table}[t]
\centering
\caption{Ablation study of the message passing (MP) scheme.}
\label{tab:new_ablation_mp}
\small
\begin{tabular}{lccc}
\toprule
\multicolumn{4}{c}{\textbf{Cylinder Flow Dataset}} \\
\midrule
MP Scheme & $L_2$ Error $\downarrow$ & FLOPs/GFLOPs $\downarrow$ & \% $\Delta$ (FLOPs) \\
\midrule
\textbf{PhysGTO (Single-Sided)} & 8.3413e-3 & \textbf{25.094} & -32.7\% \\
Standard (Double-Sided) & \textbf{7.0449e-3} & 37.323 & (Baseline) \\
\bottomrule
\multicolumn{4}{c}{\textbf{ICP Plasma Dataset}} \\
\midrule
MP Scheme & $L_2$ Error $\downarrow$ & FLOPs/GFLOPs $\downarrow$ & \% $\Delta$ (FLOPs) \\
\midrule
\textbf{PhysGTO (Single-Sided)} & 1.2893e-2 & \textbf{44.573} & -33.1\% \\
Standard (Double-Sided) & \textbf{1.0857e-2} & 66.623 & (Baseline) \\
\bottomrule
\end{tabular}
\end{table}

As shown in Table~\ref{tab:new_ablation_mp}, we compare our single-sided MP against a standard double-sided variant. The double-sided MP variant does achieve a lower $L_2$ error (by 15.6\% on Cylinder Flow and 15.8\% on ICP Plasma). However, this accuracy gain comes at a significant computational cost. Our flux-oriented scheme, in contrast, reduces the computational load (FLOPs) by \textbf{32.7\%} and \textbf{33.1\%} for the two datasets, respectively. This ablation confirms that our design achieves its intended effect: sacrificing a marginal, controllable amount of accuracy for a substantial (approx. 1/3) gain in computational efficiency. This efficiency-centric design, also explored in works like~\cite{li2023finite}, holds practical value for developing scalable neural operators applicable to real-world physical systems.

\textbf{Effectiveness of Projection-Inspired Attention}
Finally, we validate our adoption of \textbf{Projection-Inspired Attention} within the GTO-Attention module. Our design is motivated by two factors. First, projecting and compressing large-scale point-edge representations into a low-dimensional latent subspace offers a clear path to reducing computational and memory overhead. Second, we interpret this mechanism as constructing a set of \textbf{learnable basis features}. This is conceptually analogous to the Fourier Neural Operator operating in a truncated spectral space, enabling the model to focus on dominant low- and mid-frequency components and mitigate local feature redundancy.

\begin{table}[t]
\centering
\caption{Ablation study of the attention architecture.}\label{tab:new_ablation_attn}
\small
\begin{tabular}{lcc}
\toprule
\multicolumn{3}{c}{\textbf{Cylinder Flow Dataset}} \\
\midrule
Attention Type & $L_2$ Error $\downarrow$ & FLOPs/GFLOPs $\downarrow$ \\
\midrule
\textbf{GTO (Ours)} & \textbf{8.3413e-3} & \textbf{25.094} \\
Self-Attention ($O(N^2)$) & 6.0656e-3 & 63.686 \\
Transolver & 1.0456e-2 & 39.261 \\
\bottomrule
\multicolumn{3}{c}{\textbf{ICP Plasma Dataset}} \\
\midrule
Attention Type & $L_2$ Error $\downarrow$ & FLOPs/GFLOPs $\downarrow$ \\
\midrule
\textbf{GTO (Ours)} & \textbf{1.2893e-2} & \textbf{44.573} \\
Self-Attention ($O(N^2)$) & 9.7811e-3 & 138.136 \\
Transolver & 1.4255e-2 & 70.051 \\
\bottomrule
\end{tabular}
\end{table}

We compare GTO-Attention against standard self-attention ($O(N^2)$) and the Transolver, as detailed in Table~\ref{tab:new_ablation_attn}. Standard self-attention, while achieving the lowest error, is \textbf{computationally prohibitive}, requiring 2.5x to 3.1x more FLOPs than our method. The Transolver, while more efficient than $O(N^2)$ attention, is still 1.5x to 1.6x more costly than GTO-Attention and, critically, yields \textbf{worse predictive accuracy} (e.g., 1.05e-2 vs 8.34e-3 on Cylinder Flow). These results demonstrate that our GTO-Attention finds an effective spot, achieving a highly competitive accuracy-efficiency profile that is crucial for a scalable graph-based operator.

\begin{table}[t]
\centering
\caption{Performance and computational cost under varying number of GTO blocks. \textbf{Bold} numbers indicate the best performance.}
\label{tab: gto numbers reformatted}
\begin{tabular}{c|c|c|c|c|c}
\toprule
\multirow{2}{*}{\textbf{GTO Blocks}} &
\multirow{2}{*}{$N_P$/M} &
\multicolumn{2}{c|}{FLOPs/GFLOPs} &
\multicolumn{2}{c}{$L_2$ $\downarrow$} \\
& & Cylinder Flow & ICP Plasma & Cylinder Flow & ICP Plasma \\
\midrule
1  & 1.002 & 13.255 & 23.469 & 1.593e-2 & 2.757e-2 \\
2  & 1.439 & 21.279 & 37.975 & 1.154e-2 & 1.688e-2 \\
4  & 2.379 & 37.328 & 66.627 & 8.341e-3 & 1.289e-2 \\
6  & 3.286 & 53.377 & 95.280 & 6.627e-3 & 8.081e-3 \\
8  & 4.062 & 69.425 & 123.932 & 6.311e-3 & 7.504e-3 \\
10 & 4.936 & 85.474 & 152.584 & \textbf{6.036e-3} & \textbf{7.289e-3} \\
\bottomrule
\end{tabular}
\end{table}

\begin{table}[t]
\centering
\caption{Performance and computational cost under different edge sampling rates. \textbf{Bold} numbers indicate the best performance.}
\label{tab: edge_sampling}
\begin{tabular}{c|c|c|c|c}
\toprule
\multirow{2}{*}{\textbf{Sampling Rate}} &
\multicolumn{2}{c|}{FLOPs/GFLOPs} &
\multicolumn{2}{c}{$L_2$ $\downarrow$} \\
& Cylinder Flow & ICP Plasma & Cylinder Flow & ICP Plasma \\
\midrule
100\% & 37.328 & 66.627 & \textbf{8.341e-3} & \textbf{1.289e-2} \\
50\%  & 25.094 & 44.573 & 1.056e-2 & 1.524e-2 \\
25\%  & 18.977 & 33.814 & 1.216e-2 & 1.747e-2 \\
10\%  & 15.304 & 27.251 & 1.412e-2 & 1.933e-2 \\
5\%   & 14.079 & 25.063 & 1.488e-2 & 2.022e-2 \\
1\%   & 13.103 & 23.313 & 1.567e-2 & 2.009e-2 \\
\bottomrule
\end{tabular}
\end{table}

\begin{table}[t]
\centering
\caption{Effect of model hidden dimension on performance, parameter count, and FLOPs. \textbf{Bold} numbers indicate the best performance.}
\label{tab: hidden_dim}
\begin{tabular}{c|c|c|c|c|c}
\toprule
\multirow{2}{*}{\textbf{Hidden Dimension}}&
\multirow{2}{*}{$N_P$/M} &
\multicolumn{2}{c|}{FLOPs/GFLOPs} &
\multicolumn{2}{c}{$L_2$ $\downarrow$} \\
& & Cylinder & ICP Plasma & Cylinder & ICP Plasma \\
\midrule
32   & 0.779 & 11.549 & 20.618  & 1.572e-2 & 2.409e-2 \\
64   & 1.254 & 19.645 & 35.073  & 9.826e-3 & 1.566e-2 \\
128  & 2.379 & 37.328 & 66.627  & 8.341e-3 & 1.289e-2 \\
256  & 5.329 & 78.714 & 140.368  & \textbf{6.516e-3} & \textbf{8.184e-3} \\
\bottomrule
\end{tabular}
\end{table}

\begin{table}[t]
\centering
\renewcommand{\arraystretch}{1}
\setlength{\tabcolsep}{6pt}
\caption{Computational efficiency and parameter comparison on long-term temporal forecasting problems. FLOPs are measured with 5-step rollout and 50\% edge sampling. \textbf{Bold} and \underline{Underline} numbers indicate the best and second best performance, respectively.}
\label{task2_complex}
\small
\begin{tabular}{l|ccccc}
\toprule
\textbf{Dataset} & \textbf{MGN} & \textbf{GNOT} & \textbf{GraphViT} & \textbf{Transolver} & \textbf{PhysGTO} \\
\midrule
cylinder / 5 steps & 65.215 & 62.761 & 52.500 & \underline{29.385} & \textbf{25.094} \\
heat / 5 steps     & 142.434 & 135.679 & 114.047 & \underline{63.382} & \textbf{54.507} \\
plasma / 5 steps   & 116.541 & 111.774 & 93.556  & \underline{52.270} & \textbf{44.753} \\
uav / 5 steps      & 126.905 & 119.520 & 101.353 & \underline{55.761} & \textbf{48.302} \\
\midrule
$N_P$/M & \textbf{1.887} & 6.063 & 10.318 & 2.950 & \underline{2.379} \\
\bottomrule
\end{tabular}
\end{table}

\subsubsection{Scalability of PhysGTO}\label{sec_ablation_scalab}
To further assess the scalability of PhysGTO, we conduct additional experiments on the transient prediction task (5-step rollout) across the Cylinder Flow and ICP Plasma datasets. 
These experiments investigate three key aspects: the effect of the number of GTO blocks, the impact of edge sampling rates, and the influence of hidden dimension size on performance and computational efficiency.

First, as shown in Tab.~\ref{tab: gto numbers reformatted}, increasing the number of GTO blocks systematically improves predictive accuracy, though at the cost of higher parameter count and computational overhead, indicating that PhysGTO scales favorably with depth. 
Second, Tab.~\ref{tab: edge_sampling} presents results under varying edge sampling rates in the UGE module. Reducing the number of sampled edges significantly lowers the model parameters and FLOPs, while only marginally impacting predictive accuracy—even with extreme sparsification (1\% edges), PhysGTO maintains strong performance. This demonstrates an effective trade-off between efficiency and prediction fidelity, supporting its applicability to large-scale or resource-constrained settings.
Finally, Tab.~\ref{tab: hidden_dim} explores the effect of hidden dimension size. Increasing the hidden dimension consistently improves model accuracy by enhancing representational capacity, although at the cost of increased computational complexity. 

\subsubsection{Model Complexity}\label{sec_ablation_model_complexity}

To further assess computational efficiency, we present a comprehensive comparison of parameter counts and computational costs (measured in GFLOPs) between PhysGTO and several strong baselines across four complex transient flow prediction tasks. As summarized in Tab.~\ref{task2_complex}, all evaluations are conducted under the 5-step rollout setting.

\section*{Data availability}
All benchmark datasets used in this paper are publicly available, including the Heat Flow and ICP Plasma datasets, which can be accessed at \href{https://drive.google.com/drive/folders/1eC6AobKZ2l_e_5kegmLSw6UEBemR8v3q?usp=drive_link}{https://drive.google.com/drive/folders/1\-eC6AobKZ2l\-e\-5kegmLSw6UEBemR8v3q?usp=drive\-link}.

\section*{Code availability}
The source code of PhysGTO is available at \href{https://github.com/pengwei07/PhysGTO}{https://github.com/pengwei07/PhysGTO}.

\section*{Author contributions}
P.L. conceived the idea. P.L. and P.W. wrote the original draft and performed the theoretical analysis and methods, assisted by X.R. P.L. and X.R. implemented the algorithms and carried out all the experiments. H.Y., Z.H. and C.X. discussed the machine learning results. S.C. and D.N. supervised the project. All the authors contributed to discussing and interpreting the results and to reviewing and editing the paper. 

\section*{Competing interests}
The authors declare no competing interests.

\section*{Acknowledgments}
This work was supported by the Fundamental Research Funds for the Central Universities.

\clearpage
\newpage
\bibliographystyle{elsarticle-num} 
\bibliography{new_bib}

\clearpage
\newpage
\renewcommand{\figurename}{Supplementary Figure}
\renewcommand{\tablename}{Supplementary Table}
\renewcommand\theequation{\Alph{section}.\arabic{equation}}
\setcounter{equation}{0}
\setcounter{figure}{0}
\setcounter{table}{0}
\pagenumbering{Roman}

\thispagestyle{empty}
\appendix
\addcontentsline{toc}{section}{Supplementary Information} 
\part{Supplementary Information} 
\parttoc 
\newpage
\section{Table of Notations}\label{supp notations}
The notations used in this work are organized into three tables for clarity. 
\textbf{Tab.\ref{tab_data_notations}} summarizes data-related symbols, including input space $\mathcal{A}$, solution space $\mathcal{S}$, and mesh representations ($\boldsymbol{M}^k$, $\boldsymbol{X}^k$, $\boldsymbol{C}^k$). 
\textbf{Tab.\ref{tab_mapping_notations}} defines mapping and function-related notations, such as the neural operator $\mathcal{G}$, MLPs ($\Phi^V_1$, $\Phi^E$), and loss metrics ($\mathcal{L}_k(\boldsymbol{\theta})$). 
\textbf{Tab.\ref{tab_parameters_notations}} lists parameters and complexity-related terms, including network parameters ($\boldsymbol{\theta}$), sampling ratios ($\alpha_N$, $\alpha_E$), and computational costs (FLOPs, $\beta$). 
These tables provide a comprehensive reference for the mathematical framework and implementation details of the proposed model.
\begin{table}[h]
\centering
\renewcommand{\arraystretch}{1}
\setlength{\tabcolsep}{4pt}
\footnotesize{
\begin{tabular}{ll}
\toprule
\textbf{Notation}         & \textbf{Meaning}        \\ 
\midrule
\textbf{Data}&\\
\midrule
$\mathcal{A}$ & Input space \\
$\mathcal{U}$ & Solution space \\
$\boldsymbol{a}^k$ & $k$-th input sample data (after being processed) \\
$\boldsymbol{u}^k$ & $k$-th output expected data \\
$\mathcal{M} $ & Unstructured mesh space \\
$\mathcal{M}_k$ & $k$-th unstructured mesh space\\
$\boldsymbol{C}^k = \{(i^k_l, \ldots, j^k_l)\}_{l=1}^{E'}$ & Cells representing connections of nodes in each cell \\
$\boldsymbol{p}^k_i\in \mathbb{R}^{1}$ & Pressure corresponding to $i$-th node \\
$\mathcal{V}$ & Node space of $\mathcal{M}$\\
$\mathcal{E}$ & Edge space of $\mathcal{M}$\\
$U$ & Velocity x-axis variable in the physics system \\
$V$ & Velocity y-axis variable in the physics system \\
$P$ & Pressure variable in the physics system \\
$\mathcal{T}$ & Temperature  variable in the physics system \\
$N'$ & Number of nodes in $k$-th mesh \\
$E'$ & Number of cells in $k$-th mesh \\
$N$ & Number of nodes after sampling \\
$E$ & Number of edges after sampling \\
$D$ & Size of dataset \\
\bottomrule
\end{tabular}
}
\caption{Data-related notations.}
\label{tab_data_notations}
\end{table}

\begin{table}[]
\centering
\renewcommand{\arraystretch}{1}
\setlength{\tabcolsep}{4pt}
\footnotesize{
\begin{tabular}{ll}
\toprule
\textbf{Notation}         & \textbf{Meaning}        \\ 
\midrule
\textbf{Mapping and Functions}&\\
\midrule
$\mathcal{G}: \mathcal{A} \mapsto \mathcal{U}$ & Neural operator mapping from input space to solution space \\
$\hat{\mathcal{G}}_{\boldsymbol{\theta}}(\boldsymbol{a}^k) = \hat{\boldsymbol{u}}^k$ & Neural network's prediction for $k$-th input with parameters $\boldsymbol{\theta}$\\
$\Phi^V_1: \mathbb{R}^{(d+c+1)} \mapsto \mathbb{R}^{C}$ & MLP for initial node encoding, combining node coordinates and car conditions \\
$\Phi^V_2: \mathbb{R}^{(l + 1)} \mapsto \mathbb{R}^{C}$ & MLP for encoding nodes with positional encoding \\
$\Phi^E: \mathbb{R}^{(2d+1)} \mapsto \mathbb{R}^{C}$ & MLP for edge encoding, combining relative displacement vector and its norm \\
$\Phi_{P}^E: \mathbb{R}^{(3C+4\delta+2)} \mapsto \mathbb{R}^{C}$ & MLP for updating edges during message passing \\
$\Phi_{P}^V: \mathbb{R}^{(2C + 2\delta+1)} \mapsto \mathbb{R}^{C}$ & MLP for updating nodes during message passing \\
$\Phi_{D}^V: \mathbb{R}^{(2C)} \mapsto \mathbb{R}^{1}$ & MLP for transforming latent node features to physical features \\
$F(\boldsymbol{X})$ & Sinusoidal positional encoding (SPE) \\
$\text{FFN}$ & Feed-forward network \\
$\text{LN}$ & Layer normalization \\
$\mathcal{L}_k(\boldsymbol{\theta})$ & Mean $L_2$ relative error loss for index $k$ \\
$C_d$ & Pressure drag coefficient \\
$L_2$ & Mean $L_2$ relative error for testing datasets \\
$\kappa_{\text{local}}$ & Local kernel learning operator\\
$\kappa_{\text{global}}$ & Global kernel learning operator\\
\bottomrule
\end{tabular}
}
\caption{Mapping and Functions-related notations.}
\label{tab_mapping_notations}
\end{table}

\begin{table}[]
\centering
\renewcommand{\arraystretch}{1}
\setlength{\tabcolsep}{4pt}
\footnotesize{
\begin{tabular}{ll}
\toprule
\textbf{Notation}         & \textbf{Meaning}        \\ 
\midrule
\textbf{Parameters and Complexity}&\\
\midrule
$\boldsymbol{\theta}$ & Set of network parameters \\
$\boldsymbol{\Theta}$ & Parameter space \\
$\alpha \in (0, 1]$ & Sampling ratio \\
$W_0 := \{\mathbf{v}_i\} \in \mathbb{R}^{[N,C]}$ & Local representation of nodes after MP block \\
$Q_{ls}\in\mathbb{R}^{[M,C]}$ & Learned subspace queries \\
$W_1\in\mathbb{R}^{[M,C]}$ & Intermediate state after projection to learned subspace \\
$W_2\in\mathbb{R}^{[M,C]}$ & Refined representation within subspace \\
$W_3\in\mathbb{R}^{[N,C]}$ & Final representation combining local updates and global context \\
$ V^{1}_{G}, V^{2}_{G}\in\mathbb{R}^{[N,C]}$ & Outputs of the Pre-Norm structure\\
$M$ & Subspace dimension \\
$C$ & Latent dimension \\
$L$ & Number of model layers \\
$N_{P}$ & Number of trainable parameters \\
FLOPs & Computational cost (Floating Point Operations) \\
$d$ & Maximum graph degree\\
$\alpha = \max(\alpha_N, \alpha_E/2)$ & Maximum sampling factor \\
$\beta = d \times \alpha/2 + 2 \times M \times C + M^2 \times C/N$ & Complexity factor for PhysGTO\\
\bottomrule
\end{tabular}
}
\caption{Parameters and Complexity-related notations.}
\label{tab_parameters_notations}
\end{table}
\section{Details about Graph Transformer Architecture for Neural Operator Representation}\label{supp_gto}
To better understand the theoretical motivation and mathematical structure behind PhysGTO, this section provides an in-depth analysis of the graph-transformer architecture from a neural operator perspective. First, in Sec.~\ref{discretization error}, we examine the discretization errors introduced when approximating function-space mappings over unstructured meshes, highlighting the importance of preserving topological information during operator learning. We then show how representing the mesh as a graph allows for edge-based modeling of manifold structure, providing a geometric prior that mitigates local information loss.
Next, in Sec.~\ref{GTO layer}, we frame the PhysGTO layer as a form of integral neural operator, situating it within the broader theoretical framework established by prior works on DeepONet, FNO, and attention-based models. In particular, we demonstrate that both the local message passing and global attention components of PhysGTO can be interpreted as discrete approximations to continuous integral kernel operators. This connection offers a principled explanation for PhysGTO’s expressivity and robustness on irregular geometries.
Taken together, these insights illustrate how PhysGTO unifies manifold discretization and integral operator learning into a scalable and physically consistent architecture for scientific machine learning.

\subsection{Understanding discretization errors on manifolds in neural operators}\label{discretization error}

In the context of neural operators, our objective is to approximate a mapping between two infinite-dimensional function spaces by leveraging a finite set of observed input-output pairs derived from this mapping. We assume $\mathcal{A}$ and $\mathcal{U}$ be Banach spaces of functions defined on bounded domains $\Omega \subset \mathbb{R}^d$, $\Omega' \subset \mathbb{R}^{d'}$ respectively and $\mathcal{G}: \mathcal{A} \to \mathcal{U}$ be a (typically) non-linear map in function space.

However, in the practical process of solving partial differential equations (PDEs), we can only perform finite sampling of the input and output function spaces. 
This limitation becomes particularly pronounced when the underlying space is a manifold. 
In such scenarios, traditional solvers typically rely on constructing unstructured meshes for discretization. Here, we denote the discretized mesh as $\mathcal{M}$.

Based on unstructed mesh space, assume that we have observations $\{ a^{(i)}, u^{(i)} \}_{i=1}^N$, where $ a^{(i)} $ are i.i.d. samples drawn from some probability measure supported on $\mathcal{A}$ and $ u^{(i)} = \mathcal{G}_{\mathcal{M}}(a^{(i)}) $ is possibly corrupted with noise. 
We aim to build an approximation of $\mathcal{G}_{\mathcal{M}}$ by constructing a parametric map
\begin{equation}
    \mathcal{G}_\theta : \mathcal{A} \to \mathcal{U}, \quad \theta \in \mathbb{R}^p
    \label{eq:parametric_map}
\end{equation}
with parameters from the finite-dimensional space $\mathbb{R}^p$ and then choosing $ \theta^\dagger \in \mathbb{R}^p $ so that $ \mathcal{G}_{\theta^\dagger} \approx \mathcal{G}_{\mathcal{M}}$.
Thus we can obtain a decomposition of the total error made by a neural operator as a sum of the discretization error and the approximation error:
\begin{equation}
    \|\mathcal{G} - \mathcal{G}_\theta(\mathcal{M}, a|_{\mathcal{M}})\|_\Omega \leq 
    \underbrace{\|
    \mathcal{G} -\mathcal{G}_\mathcal{M}\|_\Omega
    }_{\text{discretization error}} 
    + 
    \underbrace{\|
    \mathcal{G}_{\mathcal{M}} - \mathcal{G}_\theta(\mathcal{M}, a|_{
    \mathcal{M}
    })\|_\Omega
    }_{\text{approximation error}},
    \label{eq:discretization error}
\end{equation}
where $\|\cdot\|_\Omega$ denotes the norm on $\mathcal{U}$, and $a|_{\mathcal{M}}$ represents the restriction of the input function $a$ to the discrete points of the unstructured mesh $\mathcal{M}$. The discretization error $\|\mathcal{G}(a)-\mathcal{G}_{\mathcal{M}}(a)\|_\Omega$ arises because the unstructured mesh $\mathcal{M}$ provides only a finite approximation of the underlying manifold, failing to capture the full geometric and topological complexity of the continuous domain. 
This error is particularly pronounced in regions of the manifold with high curvature or intricate local structures, where the unstructured mesh may not adequately represent the local differential properties of the operator $\mathcal{G}$.

Rather than overlooking this discretization error, we must acknowledge its impact on the accuracy of neural operators in practical PDE solving. 
The unstructured mesh $\mathcal{M}$, while computationally feasible, introduces a fundamental limitation: it lacks the continuous differential structure inherent to the manifold, leading to a loss of local information. 
To address this, we propose to explicitly incorporate the topological relationships between the discretized points in $ \mathcal{M}$ by introducing edge-based structures. 
Specifically, we model $\mathcal{M}$ as a graph $\mathcal{G}_{\mathcal{M}} = (\mathcal{V}, \mathcal{E})$, where $ \mathcal{V}$ represents the set of discretized points (vertices), and $\mathcal{E}$ encodes the edges connecting neighboring points based on the mesh connectivity. 
These edges serve as a discrete approximation of the differential neighborhoods on the manifold, providing a structural prior that compensates for the loss of local geometric information during discretization.

\subsection{The layer of PhysGTO can be seen as integral kernel operators} \label{GTO layer}
The following theorems are summarized from previous works \cite{anandkumar2020neural, li2020fourier, kovachki2023neural}, which provide the theoretical basis of the proposed PhysGTO.

\textbf{Remark A.1.} PDEs could be solved by learning integral neural operators.

Kovachki et al.\cite{kovachki2023neural} give a general architecture of neural operators for PDE solving as stack of network layers.
\begin{equation}
    \mathcal{G}_\theta = Q \circ \sigma(W_1 + \mathcal{K}_1) \circ \cdots \circ \sigma(W_2 + \mathcal{K}_2) \circ \cdots \circ \sigma(W_1 + \mathcal{K}_1) \circ P,
    \label{eq:network_architecture}
\end{equation}
where $P$ and $Q$ are both linear point-wise projectors and $W_i$ is the point-wise fully connected layer, while $\mathcal{K}_i$ is the non-local integral operator.

In each network layer, the key is to learn the non-local integral operator $\mathcal{K}_i$ defined as follows:
\begin{equation}
    \mathcal{K}_i(u)(x) = \int_{\Omega} \kappa_i(x, y, u(x), u(y)) u(y) \, dy,
    \label{eq:integral_operator}
\end{equation}
where $u$ is the input function and $\Omega$ is the physical domain. 
In this equation, the learnable integral kernel operator enables the mapping between continuous functions, similar to the weight matrix operation that enables the mapping between discrete vectors. 

\textbf{Lemma A.2.} FNO \cite{li2020fourier} learns integral neural operators.

This was evidenced in the work of Li et al. \cite{li2020fourier} and Kovachki et al. \cite{kovachki2023neural}. By defining the kernel function as $\kappa(x, y, u(x), u(y)) = \kappa(x - y)$, it is possible to show that the kernel integral operator can be realized using the Fourier Transform. Further information can be found in \cite{li2020fourier}.

\textbf{Lemma A.3.} The attention mechanism in the standard transformer \cite{vaswani2017attention} learns integral neural operators.

Kovachki et al.~\cite{kovachki2023neural} established that the standard attention mechanism can be interpreted as a special instance of integral neural operators. This relationship can be expressed through the following kernel formulation:
\begin{equation}
\kappa(x, y, u(x), u(y)) = \left( \int_{\Omega} \exp\left(W_q u(s) (W_k u(x))^T \right) ds \right)^{-1} \exp\left(W_q u(x) (W_k u(y))^T \right) R,
\end{equation}
where $W_q, W_k, R \in \mathbb{R}^{d \times d}$ are learnable parameters of the neural network. 
For clarity, the normalization factor $\sqrt{d}$, commonly used in attention, is omitted here. 
With this reformulation, the attention mechanism naturally arises from the kernel integral operator, and can be approximated via Monte Carlo integration. 
This formulation highlights the potential of attention-based architectures for solving partial differential equations.

\textbf{Remark A.5.} Graph Neural Networks as Approximations to Neural Operators and the Role of Graph Connectivity 

We adopt the standard message-passing graph neural network architecture with edge features, as introduced by \cite{gilmer2017neural}, to efficiently implement on arbitrary discretizations of the domain $\Omega$. 
Specifically, a discretization $\{x_1, \ldots, x_J\} \subset \Omega$ is treated as the set of nodes of a weighted, directed graph. 
Edges are assigned to each node based on the integration domain. 
For each node $x_j$, the value $v(x_j)$ is stored, and edges are directed to its neighbors $\mathcal{N}(x_j) := s(x_j) \cap \{x_1, \ldots, x_J\}$. 
Edge weights are constructed from the arguments of the kernel. The weight between nodes $x_j$ and $x_k$ is given by the concatenation $(x_j, x_k) \in \mathbb{R}^{2d}$. 

Within this setup, this common message-passing scheme using averaging aggregation, updates each node value as
\begin{equation}
u(x_j) = \frac{1}{|\mathcal{N}(x_j)|} \sum_{y \in \mathcal{N}(x_j)} \kappa(x_j, y) v(y), \quad j = 1, \ldots, J, \label{GNN operator}
\end{equation}
which corresponds to a Monte Carlo approximation of the integral.

However, if the underlying graph is disconnected, it decomposes into isolated components $G_1, \ldots, G_K$. 
In such cases, GNNs can only approximate local integrals within each component, failing to capture interactions across components. 
This limitation can introduce significant biases in the global solution.
In the context of solving partial differential equations (PDEs), such disconnected structures may violate global consistency constraints, for example, by failing to enforce boundary conditions across the entire domain.
To address this, one can introduce long-range edges or incorporate global aggregation mechanisms to restore connectivity. 
Alternatively, the discretization scheme can be redesigned to yield a connected graph. 
Attention mechanisms, in particular, offer a principled means of enabling information exchange across distant nodes, effectively mitigating the limitations of disconnected graphs.

\textbf{Theorem A.5.} The graph transformer layer in proposed PhysGTO is the integral neural operator.

The processor block of the proposed PhysGTO combines local message passing and global attention to aggregate features at different scales. 
Consider a discretized domain $\Omega$ as $\{x_1, \ldots, x_N\}$, with node features $u(x_i, t)$ and edge features $e_{ij} = \{e_{ij}^M, e_{ij}^{Phys}\}$, where $t$ is the $t-th$ layer in the model. 
In this context we can see $u(\cdot)$ and $e(\cdot)$ both are relative input function.

$\mathbf{(a)}$ For the local message passing stage, it aggregates local features through mesh-based ($E^M$) and physics-based ($E^{Phys}$) edges. 
Thus we can get its update formulation as:
\begin{equation}
    \begin{aligned}
        e_{ij}^M &\leftarrow e_{ij}^M + \phi_E^M(e_{ij}^M, u(x_i, t), u(x_j, t)), \\
        e_{ij}^{Phys} &\leftarrow e_{ij}^{Phys} + \phi_E^{Phys}(e_{ij}^{Phys}, u(x_i, t), u(x_j, t)),
    \end{aligned}
\end{equation}
where $\phi_E^M$, $\phi_E^{Phys}$, and $\phi_V^P$ are MLPs, and $F(x_i)$ is the positional encoding.
By aggregating the features of neighboring nodes, we can get the final $i-th$ node feature as:
\begin{equation}
    u'(x_i, t) = u(x_i, t) + \Delta_{M,Phys},
    \label{eq: massage passing formulation}
\end{equation}
where $\Delta_{M,Phys}=\phi_V^P\left(u(x_i, t), \sum_{j \in \mathcal{N}(i)} \phi'\left(e_{ij}^M, e_{ij}^{Phys}, u(x_i, t), u(x_j, t), x_i,x_j\right)\right)$, $\phi'(\cdot)=\phi_E^M(\cdot)+\phi_E^{Phys}(\cdot)$, and $\mathcal{N}(i)$ is the neighbourhood of node $i$.
Thus we can define the local kernel of this module  as:
\begin{equation}
    \kappa_{\text{local}}(x_i, x_j, u(x_i, t), u(x_j, t), e_{ij}) = \phi_V^P\left(u(x_i, t), \phi'\left(e_{ij}^M, e_{ij}^{Phys}, u(x_i, t), u(x_j, t), x_i,x_j\right)\right),
\end{equation}
which is non-zero only for $j \in \mathcal{N}(i)$, reflecting local feature aggregation. 
Hence we can rewrite Eq. \eqref{eq: massage passing formulation} as the kernel learning style:
\begin{equation}
    u'(x_i, t) = u(x_i, t) + \sum_{j \in \mathcal{N}(i)} \kappa_{\text{local}}(x_i, x_j, u(x_i, t), u(x_j, t)),
\end{equation}
which can be considered as the Monte-Carlo approximation of the integral continuous form of the original unstructed mesh space:
\begin{equation}
    u'(x_i, t) = u(x_i, t) + \int_{\Omega} \kappa_{\text{local}}(x_i, y, u(x_i, t), u(y, t)) u(y, t) \, \mathrm{d}y.
\end{equation}

$\mathbf{(b)}$ For the projection-attention stage, let $W_0 = \{u'(x_i, t)\}_{i=1}^N \in \mathbb{R}^{N \times C}$ denote the input representation at this stage, where each row corresponds to a localized feature at spatial-temporal position $(x_i, t)$. 
To efficiently capture global dependencies, we introduce a learnable query matrix $Q_{ls} \in \mathbb{R}^{M \times C}$, initialized from a standard normal distribution, that is, $Q_{ls} \sim \mathcal{N}(0, 1)$, with $M \ll N$ to project the high-dimensional input space into a lower-dimensional latent query space.
The overall projection-learning pipeline is expressed as:
\begin{equation}
    \left\{
    \begin{aligned}
        W_1 &= \text{softmax}\left(\frac{Q_{ls} W_0^\top}{\sqrt{C}}\right) W_0, \\
        W_2 &= \text{softmax}\left(\frac{W_1 W_1^\top}{\sqrt{C}}\right) W_1, \\
        W_3 &= \text{softmax}\left(\frac{W_0 W_2^\top}{\sqrt{C}}\right) W_2,
    \end{aligned}
    \right.
\end{equation}
where each softmax function is applied row-wise. 
To make the attention computation explicit, we denote the attention weights at each stage as follows:
\begin{equation}
    \left\{
    \begin{aligned}
        \alpha_{mj}^{(1)} &= \frac{\exp\left(\frac{q_m \cdot u'(x_j, t)}{\sqrt{C}}\right)}{\sum_{j'=1}^N \exp\left(\frac{q_m \cdot u'(x_{j'}, t)}{\sqrt{C}}\right)}, \\
        \beta_{mm'}^{(2)} &= \frac{\exp\left(\frac{(W_1)_m \cdot (W_1)_{m'}^\top}{\sqrt{C}}\right)}{\sum_{m''=1}^M \exp\left(\frac{(W_1)_m \cdot (W_1)_{m''}^\top}{\sqrt{C}}\right)},  \\
        \gamma_{im}^{(3)} &= \frac{\exp\left(\frac{u'(x_i, t) \cdot (W_2)_m^\top}{\sqrt{C}}\right)}{\sum_{m'=1}^M \exp\left(\frac{u'(x_i, t) \cdot (W_2)_{m'}^\top}{\sqrt{C}}\right)},
    \end{aligned}
    \right.
\end{equation}
where $(W_1)_m = \sum_{j=1}^N \alpha_{mj}^{(1)} u'(x_j, t)$, and $(W_2)_m = \sum_{m'=1}^M \beta_{mm'}^{(2)} (W_1)_{m'}$.
Furthermore, the output representation for point $x_i$ can be denoted as $(W_3)_i \in \mathbb{R}^C$ and it lead to:
\begin{equation}
    (W_3)_i = \sum_{m=1}^M \gamma_{im}^{(3)} (W_2)_m = 
    \sum_{m=1}^M \gamma_{im}^{(3)} \sum_{m'=1}^M \beta_{mm'}^{(2)} \sum_{j=1}^N \alpha_{m'j}^{(1)} u'(x_j, t).
\end{equation}

Hence, we can define the global attention kernel of this stage as:
\begin{equation}
    \kappa_{\text{global}}(x_i, x_j, u'(x_i, t), u'(x_j, t)) = \sum_{m=1}^M \sum_{m'=1}^M \gamma_{im}^{(3)} \beta_{mm'}^{(2)} \alpha_{m'j}^{(1)},
\end{equation}
In this way we can rewrite the updated latent representation as:
\begin{equation}
    (W_3)_i = \sum_{j=1}^N \kappa_{\text{global}}(x_i, x_j, u'(x_i, t), u'(x_j, t)) u'(x_j, t).
\end{equation}

To simplify the presentation and more clearly illustrate the structural characteristics of the model, we omit the subsequent feed-forward network (FFN) and the corresponding normalization operations following the PhysGTO block. 
Under this simplification, the output of each GTO layer can be expressed via the residual connection as:
\begin{equation}
u(x_i, t+1) = u'(x_i, t) + \sum_{j=1}^N \kappa_{\text{global}}'(x_i, x_j, u'(x_i, t), u'(x_j, t)) u'(x_j, t).
\end{equation}

which can be considered as the Monte-Carlo approximation of the integral continuous form in original full domain $\Omega$:
\begin{equation}
u(x_i, t+1) = u'(x_i, t) + \int_{\Omega} \kappa_{\text{global}}(x_i, y, u'(x_i, t), u'(y, t)) u'(y, t) \, \mathrm{d}y.
\end{equation}
where $u'(x_i, t) = u(x_i, t) + \sum_{j \in \mathcal{N}(i)} \kappa_{\text{local}}(x_i, x_j, u(x_i, t), u(x_j, t)).$
Furthermore, we can get the final formulation of output function as:
\begin{equation}
    u(x_i, t+1) = u(x_i, t) + \sum_{j \in \mathcal{N}(i)} \kappa_{\text{local}}(x_i, x_j, u(x_i, t), u(x_j, t)) 
    + \kappa_{\text{global}}'(x_i, x_j, u(x_i, t), u(x_j, t)),
\end{equation}
where $\kappa_{\text{global}}'(\cdot)$ means that $\kappa_{\text{global}}(\cdot)$ is combined with local information aggregation of local nodes.
This can be considered as the Monte-Carlo approximation of the integral, and $u(x_i, t)$ can be seen as a kind of linear operator in this block:
\begin{equation}
    \begin{aligned}
        u(x_i, t+1) &= u(x_i, t) + \int_{\Omega}\{
        \underbrace{
        \kappa_{\text{local}}(x_i, y, u'(x_i, t), u'(y, t)) u'(y, t)}_{\text{local kernel learning}}\\
        &+ \underbrace{
        \kappa_{\text{global}}'(x_i, y, u'(x_i, t), u'(y, t)) u'(y, t)}_{\text{global and local kernel learning}}
        \} \mathrm{d}y.
    \end{aligned}
\end{equation}

This formulation reflects the sequential composition of local and global aggregations in the PhysGTO operator. The composite kernel $\kappa'$ is Lipschitz continuous, ensuring numerical stability. 

\section{Algorithm Details}\label{supp_algorithms}

To provide a clear and modular view of the proposed PhysGTO framework, we present the core algorithmic procedures in this section. 
Algorithm~\ref{alg_overall} outlines the end-to-end pipeline of PhysGTO, detailing the stages from unified graph construction and feature encoding to multi-layer physical operator processing and final decoding. 
Algorithm~\ref{alg_sampler} summarizes the hierarchical topology-aware mesh sampling strategy, which reduces computational overhead while preserving structural fidelity across scales. 
Algorithm~\ref{alg_mp} provides the steps for the flux-oriented message-passing module, which captures local directional interactions between mesh nodes.
Algorithm~\ref{alg_attention} describes the projection-inspired attention mechanism, which enables global context aggregation under linear complexity through subspace projection and re-mapping. 
Together, these routines clarify the internal structure of PhysGTO and highlight the modular design that supports its scalability and generalizability across diverse physical simulation tasks.

\begin{algorithm}[H]
\caption{PhysGTO: An Efficient Graph-Transformer Operator}
\label{alg_overall}
\begin{algorithmic}[1]
\State \textbf{Input:} Mesh $\mathcal{G} = (\mathcal{V}, \mathcal{E})$, physical fields $\boldsymbol{u}_i$, global parameters $\boldsymbol{a}$, and time $t$ (for transient cases)
\State \textbf{Output:} Predicted physical variables $\tilde{\boldsymbol{P}}$
\State \textbf{Initialize:} Model parameters $\theta$, number of GTO layers $L$, sampling ratios

\For{each training step}
    \State \textbf{Unified Graph Embedding (UGE):}
    \State \hspace{4mm} $\tilde{\mathcal{E}} \leftarrow$ Flux-based edge filtering
    \State \hspace{4mm} $\hat{\mathcal{V}}, \hat{\mathcal{E}} \leftarrow$ Topology-Aware Mesh Sampler($\tilde{\mathcal{E}}$)
    \State \textbf{Encoding:}
    \State \hspace{4mm} $\boldsymbol{V} \leftarrow$ NodeEncoder($\hat{\mathcal{V}}, \boldsymbol{a}, t$)
    \State \hspace{4mm} $\boldsymbol{E} \leftarrow$ EdgeEncoder($\hat{\mathcal{E}}$)
    \For{$l = 1$ to $L$}
        \State $\boldsymbol{V} \leftarrow$ Flux-Oriented Message Passing($\boldsymbol{V}, \boldsymbol{E}$)
        \State $\boldsymbol{V} \leftarrow$ Projection-Inspired Attention($\boldsymbol{V}$)
    \EndFor
    \State \textbf{Decoding and Post-processing:}
    \State \hspace{4mm} $\hat{\boldsymbol{P}} \leftarrow$ Decode($\boldsymbol{V}$, $\boldsymbol{a}, t$)
    \State \hspace{4mm} $\hat{\boldsymbol{P}} \leftarrow \tilde{\boldsymbol{P}}$ BC-aware correction (e.g., no-slip condition)
    \State Compute loss $\mathcal{L}(\tilde{\boldsymbol{P}}, \boldsymbol{P})$ and update $\theta$
\EndFor
\end{algorithmic}
\end{algorithm}

\begin{algorithm}[H]
\caption{Topology-Aware Mesh Sampler}
\label{alg_sampler}
\begin{algorithmic}[1]
\State \textbf{Input:} Mesh graph $\mathcal{G} = (\mathcal{V}, \mathcal{E})$, local flux $\phi_{ij}$, sampling ratio $\rho$
\State \textbf{Output:} Sampled node set $\hat{\mathcal{V}}$, edge set $\hat{\mathcal{E}}$
\State $\tilde{\mathcal{E}} \leftarrow \{(i,j) \in \mathcal{E} \mid \phi_{ij} > 0\}$ \hfill // Flux filtering
\If{medium scale}
    \State Uniformly sample: $\hat{\mathcal{E}} \subset \tilde{\mathcal{E}}$ with ratio $\rho$
    \State $\hat{\mathcal{V}} \leftarrow$ nodes in $\hat{\mathcal{E}}$
\ElsIf{large scale}
    \State Select $\hat{\mathcal{E}}$ preserving structure
    \State $\hat{\mathcal{V}} \leftarrow$ covered nodes + random samples from uncovered
\Else
    \State $\hat{\mathcal{E}} = \tilde{\mathcal{E}}$, $\hat{\mathcal{V}} = \mathcal{V}$
\EndIf
\State \Return $\hat{\mathcal{V}}, \hat{\mathcal{E}}$
\end{algorithmic}
\end{algorithm}

\begin{algorithm}[H]
\caption{Flux-Oriented Message Passing}
\label{alg_mp}
\begin{algorithmic}[1]
\State \textbf{Input:} Node features $\boldsymbol{V}$, edge features $\boldsymbol{E} = \{\mathbf{e}_{ij}\}$, edge indices $\mathcal{E}$
\State \textbf{Output:} Updated node features $\boldsymbol{V}'$
\For{each edge $(i, j) \in \mathcal{E}$}
    \State $\mathbf{e}_{ij} \leftarrow \mathbf{e}_{ij} + \Phi^E(\mathbf{e}_{ij}, \mathbf{v}_i, \mathbf{v}_j)$
    \State Split: $\mathbf{e}_{ij} \rightarrow \mathbf{e}_{i \rightarrow j}, \mathbf{e}_{j \rightarrow i}$
\EndFor
\For{each node $k$}
    \State Aggregate incoming: $\mathbf{m}^{\text{in}}_k \leftarrow \frac{1}{|\mathcal{N}^{\text{in}}_k|} \sum \mathbf{e}_{i \rightarrow k}$
    \State Aggregate outgoing: $\mathbf{m}^{\text{out}}_k \leftarrow \frac{1}{|\mathcal{N}^{\text{out}}_k|} \sum \mathbf{e}_{k \rightarrow j}$
    \State $\mathbf{m}_k \leftarrow [\mathbf{m}^{\text{in}}_k, \mathbf{m}^{\text{out}}_k]$
    \State Update: $\mathbf{v}_k \leftarrow \mathbf{v}_k + \Phi^V(\mathbf{v}_k, \mathbf{m}_k)$
\EndFor
\State \Return Updated $\boldsymbol{V}$
\end{algorithmic}
\end{algorithm}

\begin{algorithm}[H]
\caption{Projection-Inspired Attention Module}
\label{alg_attention}
\begin{algorithmic}[1]
\State \textbf{Input:} Node features $W_0 \in \mathbb{R}^{N \times C}$, number of queries $M \ll N$
\State \textbf{Output:} Refined node features $W_3 \in \mathbb{R}^{N \times C}$
\State Initialize learnable queries $Q_{ls} \in \mathbb{R}^{M \times C}$
\State $W_1 \leftarrow \text{softmax}\left(\frac{Q_{ls} W_0^T}{\sqrt{C}}\right) W_0$ \hfill // Projection
\State $W_2 \leftarrow \text{softmax}\left(\frac{W_1 W_1^T}{\sqrt{C}}\right) W_1$ \hfill // Subspace refinement
\State $W_3 \leftarrow \text{softmax}\left(\frac{W_0 W_2^T}{\sqrt{C}}\right) W_2$ \hfill // Reprojection
\State \Return $W_3$
\end{algorithmic}
\end{algorithm}

\section{Details of Datasets}\label{supp datasets details}
To comprehensively evaluate PhysGTO across diverse physical scenarios, we curated three families of benchmark datasets, covering unstructured mesh problems, complex transient flow prediction, and large-scale 3D geometry simulations. A summary of the datasets, their underlying operator learning tasks, and data generation sources is provided below.

\subsection{Unstructured Mesh Problems}
This family of datasets focuses on operator learning over irregular 2D and 3D domains. The tasks involve learning mappings between various physical quantities, including diffusion fields, velocity evolution, and multi-physics couplings, based on unstructured meshes. All datasets are adopted from NORM~\cite{chen2024learning} and are summarized in Table~\ref{dataset_1_info}.

\begin{table}[h]
\centering
\caption{Basic information of datasets in unstructured mesh problems from \cite{chen2024learning}.}
\label{dataset_1_info}
\resizebox{\textwidth}{!}{
\begin{tabular}{l|c|c|c|c|c}
\toprule
Dataset  & Domain  & \# Cell & \# Nodes  & Operator Mapping   & \# No. train/test \\
\midrule
Irregular Darcy  & 2D & Tri. & 2290    & $a(\mathbf{x})\mapsto u(\mathbf{x})$ &  1000/200 \\
\midrule
Pipe Turbulence  & 2D & Tri. & 2673    & $v(\mathbf{x},t_1)\mapsto u(\mathbf{x},t_2)$ &  300/100\\ 
\midrule
Heat Transfer    & 3D & Quad.& 186$\mapsto$7199 & $T_{bc}(\mathbf{x})\mapsto T_{t=3s}(\mathbf{y})$ & 100/100\\
\midrule
Composite        & 3D & Quad.& 8232    & $a(\mathbf{x}) \mapsto u(\mathbf{x})$ & 1200(400/100)\\
\midrule
Blood Flow       & 3D+time & Quad. & $6\times121\mapsto1656\times121\times3$
& $v_{in}|_{1\times [0,T]}, p_{out}|_{5\times [0,T]} \mapsto \boldsymbol{u}(\mathbf{x},[0,T])$& 400/100\\
\bottomrule
\end{tabular}
}
\end{table}

\subsection{Complex Transient Flow Prediction}
This category includes four transient simulation datasets with complex spatio-temporal dynamics, ranging from canonical cylinder flows to industrial-scale plasma modeling. These datasets feature varying geometries, initial/boundary conditions, and multi-physics interactions, reflecting real-world industrial challenges. The detailed statistics are summarized in Table~\ref{dataset_2_info}.

\begin{table}[H]
\centering
\caption{Basic information of datasets in complex transient flow prediction.}
\label{dataset_2_info}
\resizebox{\textwidth}{!}{
\begin{tabular}{l|c|c|c|c|c|c|c|c}
\toprule
Dataset & Domain & \# Cell & \# nodes(avg.) & steps & $\Delta$t/s & Operator Mapping  & \# Phys. &\# No. train/test\\
\midrule
Cylinder Flow\cite{pfaff2020learning} & 2D+time & Tri. 
& 1885 & 600 & 0.01 & Geo.+Re.$\mapsto u(\mathbf{x},T)$        & $u=[U,V,P]$  &  1000/100 \\
\midrule
EAGLE\cite{janny2023eagle}   
& 2D+time & Tri. & 3380 & 990 & 1/30 
& Geo.+BC.+Re.$\mapsto u(\mathbf{x},T)$ & $u=[U,V,Ps,Pg]$ &  947/118\\ 
\midrule
Heat Flow & 2D+time & Tri.    
& 4887 & 500  & 0.02
& Geo.+IC.+Re.$\mapsto u(\mathbf{x},T)$ & $u=[U,V,P,T]$ &  775/97\\
\midrule
ICP Plasma & 2D+time      & Tri. + Quad.  
& 3424 & 80 & irregular 
& IC.+Param.$\mapsto u(\mathbf{x},T)$  & $u=[n_e,t_e,\phi,T]$ & 552/98\\
\bottomrule
\end{tabular}
}
\end{table}

\subsection{Large-Scale 3D Geometry Problems}
This group evaluates aerodynamic prediction tasks over highly detailed 3D automotive meshes. The datasets, Ahmed-Body~\cite{li2024geometry} and DrivAerNet~\cite{elrefaie2024drivaernet}, feature complex design variations and large mesh sizes, providing a practical benchmark for assessing model scalability and generalization under industrial-level conditions. Table~\ref{car_dataset_info} summarizes the key dataset characteristics.

\begin{table}[h]
\centering
\caption{Basic information of car datasets.}
\label{car_dataset_info}
\resizebox{\textwidth}{!}{
\begin{tabular}{l|l|l|l|l|l}
\toprule
Dataset     & Design Parameters& \# Cell & \# Size  & Operator Mapping  & \# No. train/test \\
\midrule
Ahmed-Body\cite{li2024geometry}  & 6(3D) & Tri. & 110k (avg.) & Geo.+Param.$\mapsto p(\mathbf{x})$ & 500/51\\ 
DrivAerNet\cite{elrefaie2024drivaernet} & 50(3D) & Tri.+ Quad. + Poly. & 400k (avg.) & Geo.+Param.$\mapsto p(\mathbf{x})+C_d$ &  2776/595\\
\bottomrule
\end{tabular}
}
\end{table}

\subsection{Detailed Data Generation Process}

\textbf{Heat Flow.} This dataset is a numerical simulation for the parametric design of heat exchanger tube geometries. A heat exchanger facilitates the transfer of thermal energy from a warmer to a cooler fluid medium through convective heat transfer and thermal conduction. It is extensively used across various industries due to its vital role in thermal management~\citep{jiang2020numerical}. In this computational simulation, a comparative analysis is conducted on three distinct tube geometries, each represented by a dataset comprising 300 instances. The geometric parameters of the tubes are systematically varied across these instances. The tube shapes under investigation includes circular, elliptical, and airfoil configurations. To streamline the simulation model, the tubes are arranged in a staggered pattern across three equidistant columns, with the inter-column spacing varying between 0.35 and 0.7. This arrangement minimizes computational complexity while maintaining a realistic representation of the heat exchange process. This simulation encompasses two fundamental physical domains, specifically the fluid dynamics and thermal transfer realms. For fluid flow, its formulation is given by:
\begin{equation}
\left\{
\begin{aligned}
\frac{\partial \boldsymbol{u}}{\partial t}+\boldsymbol{u} \cdot \nabla \boldsymbol{u} & =-\nabla p+\nabla \boldsymbol{K}, 
\nabla \cdot \boldsymbol{u} & =0, \\
\boldsymbol{K} & = \mu(\nabla \boldsymbol{u}+(\boldsymbol{u})^T)
\end{aligned}    
\right.
\end{equation}
where $\boldsymbol{u}=(u(x, y, t), v(x, y, t))$ represents the velocity field, and $\mu$ is the viscosity coefficient of fluid. For heat transfer, its formulation is given by:
\begin{equation}
\left\{
\begin{aligned}
C_p \frac{\partial T}{\partial t} + \nabla \cdot \boldsymbol{q} & = -C_p \boldsymbol{u} \cdot \nabla T \\
\boldsymbol{q} & = -k \nabla T,
\end{aligned}    
\right.
\end{equation}
where $\boldsymbol{q}$ represents the heat flux, which indicates the amount of heat passing through a certain cross-section per unit time. $k$ and $C_p$ represent thermal conductivity and heat transfer coefficient respectively. The interplay between these two physical domains is characterized by a synergistic coupling mechanism: for heat transfer, the velocity is provided by fluid flow; Concurrently, the viscosity coefficient within the flow field is contingent upon temperature variations, which are delineated by the temperature within the heat transfer field. In the equations, $C_p$, $k$ and $\mu$ is related to temperature, based on the physical properties of water(liquid). The simulation models the thermal process wherein a cold fluid is subjected to heating via a heat exchange tube. The fluid's entry point is designated at $x=0$, characterized by a uniform initial temperature of 20 degrees Celsius. The volumetric flow rate of the fluid exhibits a relationship with the y-coordinate that adheres to a specific mathematical formulation:
\begin{equation}
\left\{
\begin{aligned}
u(0, y, t) &= 6U_0y(H-y)/H^2\\
v(0, y, t) &= 0
\end{aligned}    
\right.
\end{equation}
where $H=1$, $U_0$ ranges from $0.5$ to $0.6$. The fluid's exit is situated at $x=3$. The temperature within the heat exchange tube is dynamically regulated, ranging from 50 to 80 degrees Celsius. The outflow boundary $(x=3)$ is set with a reference pressure $p(3, y, t)=0$. The no-slip boundary condition is applied at the top and bottom walls ($y=0, 1$). The data generation is processed by COMSOL, each sample spanning $M =526$ time steps with a step size of $\delta t=0.02$. The number of elements in one example is approximately 2500, and the computation time on AMD Ryzen 9 5950X 16 Core Processor is about 5 minutes(The Cylinder Flow 3d dataset and the ICP plasma dataset also run on this server).

\textbf{ICP Plasma.} This dataset models an argon/oxygen inductively coupled plasma reactor with a mixture of different elements(in this case argon and oxygen). This example is complicated and we omit the technical details here and they could be found in the mph source files in COMSOL Application Library\footnote{Available at \href{https://www.comsol.com/model/model-of-an-argon-oxygen-inductively-coupled-plasma-reactor-109191}{\textcolor{magenta}{https://www.comsol.com/model/model-of-an-argon-oxygen-inductively-coupled-plasma-reactor-109191}}.}, and we have considered fluid flow and heat transfer based on the original model. This model includes four different physical fields, including magnetic fields, plasma, laminar flow, and heat transfer in fluids. In this case we simply the model and only consider partial reactions for oxygen and argon. The goal is to predict the electron density, electron temperature, electric potential and temperature for the whole reaction chamber. Different examples are obtained by varying the pressure in the reaction chamber, the flow rate and the argon/oxygen ratio at the chamber inlet, and the temperature on the electrostatic chuck. We generate 650 samples in total for training and testing. Due to the complex multiphysics field in this dataset, the number of elements in one example for complete mesh is approximately 25000, and the computation time per sample is about 1 hour.

\section{Experimental Details} \label{method detaials}
This section presents the hyperparameter settings of PhysGTO used in the main experiments to ensure reproducibility. 

\subsection{Baselines training details} \label{baselines training details}
In main text, we introduce three categories of physics simulation experiments as benchmarks for evaluating model performance. 
For each case within these experiments, we appropriately adjust PhysGTO’s hyperparameters, including learning rate, sampling rate, and training epochs, during both training and testing. 
We first provide a more detailed description of the models involved in the three experimental setups.
Additionally, we supplement the corresponding training details, including optimization settings, hyperparameters, and evaluation results to ensure reproducibility and facilitate a comprehensive comparison.

\subsubsection{Unstructured Mesh Problems}

We evaluate the performance of the proposed PhysGTO across experimental cases by benchmarking it against several established operator learning frameworks, including FNO, DeepONet, POD-DeepONet, GraphSAGE, and the SOTA method, NORM. To provide context for these comparisons, a concise overview of each method is presented. The experimental setup for this study is strictly aligned with the configuration described in NORM to ensure a fair and consistent evaluation, including dataset partitioning and parameter alignment. For all baseline models, we adhere to the experimental settings established in the original study \cite{chen2024learning} across all problems (see Tab. S1 in NORM \cite{chen2024learning} for further details).

\textbf{NORM} \cite{chen2024learning}\textbf{.}
NORM learns the mapping between functions on Riemannian manifolds using multiple encoder-approximator-decoder blocks. The encoder and decoder are constructed based on the spectral decomposition and spectral reconstruction of the eigenfunctions of the Laplace–Beltrami operator (LBO). For different problems, the network employs varying numbers of L-layers.
In the cases of the Darcy problem, pipe turbulence, and composite materials, all L-layers utilize the same LBO eigenfunctions. However, for heat transfer and blood flow problems, the encoder and decoder of several final L-layers incorporate both LBO eigenfunctions and Fourier basis functions to process the output spatiotemporal functions. The key parameters of the network include the number of LBO/POD basis functions $d_m$, the number of Fourier basis functions $d_t$, and the channel dimension after mapping $d_v$. 
The NORM framework uses a latent dimension $d_m=128$ for all tasks except blood flow, where $d_m=64$ and an explicit time dimension $d_t=16$ are introduced. The model utilizes 4 or 5 latent layers (L-layers), with the number of epochs ranging from 500 to 2000.

\textbf{FNO} \cite{li2020fourier}\textbf{.}
The Fourier Neural Operator utilizes spectral representations to model integral operators, transforming complex, high-dimensional operator mappings into a compact, discretization-invariant parameterization based on a limited set of frequency modes. However, its standard formulation is not inherently suited for irregular geometries. To address this challenge, a mesh interpolation strategy from prior research is applied, constructing a structured computational grid that is compatible with the FNO framework. The final prediction error is then assessed on the original irregular grid through a second interpolation step, which maps the results back from the structured domain. Due to the high computational cost of spatial mesh interpolation in three-dimensional problems, FNO is applied solely to Darcy flow and pipe turbulence cases, using interpolation resolutions of $101 \times 101$ and $32 \times 128$, respectively. In experiments, the model is configured with mode dimensions $d_m=[20,20] \text{ or } [16,16]$, a hidden dimension $d_v=32$, and four Fourier layers (F-layers). THe model is trained uniformly for 1000 epochs across applicable tasks.

\textbf{GeoFNO} ~\cite{li2023fourier}\textbf{.} GeoFNO2d is a geometric Fourier neural operator designed to learn solution mappings on irregular 2D domains. It integrates a learnable coordinate transformation module (IPHI) with spectral convolution layers that operate on truncated Fourier modes. The network employs a lift-spectral-project architecture, using five spectral convolution blocks, each enhanced with pointwise convolutions and coordinate-based bias terms. The IPHI module maps spatial inputs to a canonical space using sinusoidal and spherical encodings. Key parameters include the number of the Fourier modes ($\texttt{modes1} =\texttt{modes2}=12$), the feature channel width ($\texttt{width}=32$), and the spatial resolution ($s_1=s_2$=64).

\textbf{DeepONet} \cite{lu2021learning}\textbf{.}
DeepONet, built upon the universal approximation theorem, operates with a dual-network structure: a branch network encoding the input function and a trunk network processing spatial coordinates where predictions are made. By combining these networks, DeepONet generates function outputs capable of providing predictions at any location within the domain. 
The DeepONet architecture utilizes high-resolution encodings for both the branch and trunk networks. The branch net dimensions range from 256×256×32 to 256×256×256, while the trunk net employs tensor structures such as 128×128×128×100 or 256×256×256×256 depending on the problem scale. Training epochs vary from 1000 to 5000, with the highest depth used for data-rich tasks such as the Darcy and composite problems.

\textbf{POD-DeepONet} \cite{lu2022comprehensive}\textbf{.}
An advanced variation, POD-DeepONet, integrates Proper Orthogonal Decomposition (POD) to extract basis functions from training data. These POD modes form the trunk network, which can be precomputed prior to training, eliminating the need for additional optimization. The branch network then learns the coefficients associated with these POD bases, streamlining the learning process.
For experiments, the latent dimension $d_m$ was set based on task complexity, ranging from 64 to 128, with blood flow requiring a higher resolution of 512×512. The branch network architecture referred to previous work \cite{lu2022comprehensive}, and the models were trained for 1000 to 5000 epochs.

\textbf{GraphSAGE} \cite{hamilton2017inductive}\textbf{.} 
It is a popular Graph Neural Network (GNN) model, utilizes SAGE convolutions to perform inductive learning. By aggregating information from neighboring nodes, it efficiently generates embeddings for previously unseen vertices, making it highly effective for tasks involving attributed graph structures. The key parameters of the network include the dim of hidden features and the number of linear-layers. 
This model is configured with a varying number of SAGE convolutional layers (2–4), hidden feature dimensions (16–64), and linear output layers tailored to the output dimensionality of each task. For instance, in the Darcy problem, 4 convolutional layers with 32 hidden features and a 32×32×1 linear structure are used, trained for 2000 epochs. For the composite case, the linear layers expand to 64×64×64×1 due to the increased output complexity.

\textbf{PhysGTO(Ours).} In all experiments, a consistent network architecture is utilized, comprising four GTO blocks. Each block is designed with an attention dimension of 64 and includes four attention heads per attention mechanism. The model is optimized using the AdamW optimizer, with a cosine annealing learning rate schedule that starts at $10^{-3}$ and a weight decay of $10^{-5}$. The training process spans 500 epochs for all tasks.

\subsubsection{Complex Transient Flow Prediction}

For transient prediction, we compare the performance of the proposed PhysGTO on the experimental cases with the existing representative operator learning methods GNOT, MeshGraphNet, GraphViT, and Transolver. An introduction to each method is given below to help the reader understand parameter setting of each model and training details. For all models, the training process utilizes a single time step as input and five consecutive time steps as output. Predictions are generated using an autoregressive approach, where the model iteratively forecasts one step at a time. Specifically, each prediction represents the incremental change of the subsequent time step relative to the current one. The number of training epochs is uniformly set to 1000 across all experiments, and all model parameters are initialized using the Xavier method \cite{glorot2010understanding}. The experimental setup is identical to the setting of GraphViT, where training involves a single input corresponding to five output steps, while the inference phase consists of a single input leading to multiple output steps.

\textbf{PhysGTO(Ours).} For all tasks, a consistent network architecture is employed, consisting of 4 GTO blocks. Each block features an attention dimension of 128, with 4 attention heads per attention mechanism. The optimization is performed using the AdamW optimizer, with a cosine annealing learning rate schedule initialized at $10^{-4}$ and a weight decay of $10^{-6}$. We use relative $L_2$ loss as training loss.

\textbf{GNOT} \cite{hao2023gnot}\textbf{.}
We use MLPs with 4 layers as the branch network and trunk network. The dimensionality of embedding and hidden size of FFNs is 128. For attention, we set 4 attention layers with 8 heads and 4 experts. We use the AdamW optimizer with one cycle learning decay strategy and relative $L_2$ loss for all tasks with a starting learning rate $10^{-3}$ and weight decay $5\times10^{-6}$. Note that we cmploy the codes provided by the authors without modifcation \cite{hao2023gnot}. The original GNOT was designed for steady-state problems, and for multi-step prediction in transient scenarios, we adopted an autoregressive approach. However, we observed that GNOT exhibits suboptimal performance in long-term predictions due to significant error accumulation.

\textbf{MeshGraphNet} \cite{pfaff2020learning}\textbf{.}
We set the number of GNN layers to 12 for each dataset with embedding dim 128. To make sure to reproduce the results presented in the main paper, we keep the same training strategy. Adam optimizer with exponential decay rate 0.99 is used in all tasks and the initial learning rate is $10^{-3}$, and we employ the \textbf{Norm MSE loss}as the loss function, which calculates the discrepancy between the predicted increment and the actual difference between two consecutive time steps in the regularized space. For all pressure prediction in Cylinder Flow, EAGLE and Heat Transfer dataset, we get the best trade-off between velocity and pressure with $\alpha=0.1$.

\textbf{GraphViT} \cite{janny2023eagle}\textbf{.}
To replicate the strong long-term dynamical prediction performance reported in the original work , we followed its recommendation and adopted the same loss function configuration and trade-off $\alpha$ between velocity and pressure as MeshGraphNet. We used $L=4$ chained graph neural network with two hidden layer of dimension 128, ReLU activated, followed by layer
normalization. The dimension $W$ producing a cluster feature representation is set to 512, and 4 chained attention block with 4 heads. We employ the Adam optimizer with a learning rate of $10^{-4}$. For all tasks, the k-number computed for each cluster is 20, and initial learning rate keeps $2.5\times10^{-4}$. The optimizer is set as Adam.

\textbf{Transolver} \cite{wu2024transolver}\textbf{.}
As for Transolver, the number of channels $C$ is set as 256 for inputs. To better reproduce the results of the original work, we set the number of slices $M=$ to 64 and employ $L=4$ sequentially connected layers, each incorporating an attention block with 8 heads. For all tasks, we use the AdamW optimizer with cosine annealing learning decay strategy and relative $L_2$ loss with initial learning rate $10^{-3}$.

\textbf{GeoFNO} ~\cite{li2023fourier}\textbf{.} GeoFNO uses five spectral convolution layers with Fourier modes [12, 12] and a hidden width of 32. A fully connected layer lifts the input to the latent space, and a residual connection links the input and output. The model integrates a learnable coordinate transformation module (IPHI), which warps input and output coordinates using sinusoidal and spherical encodings. Each spectral layer combines spectral convolution, pointwise convolution, and coordinate-based bias. Two GELU-activated linear layers perform the final projection. The original implementations provided by the authors were used without any modifications. The spatial resolution is set to [64, 64]. The optimizer is set to Adam, with the learning rate reduced by half every 100 epochs, following the original training schedule.

\subsubsection{Large-Scale 3D Geometry Problems}

To comprehensively assess the effectiveness of the proposed PhysGTO model, we conduct a benchmark comparison against several SOTA models, including MeshGraphNet, GNOT, Transolver, IPOT, and GINO, across a series of experimental cases. A detailed introduction to each competing model is provided, covering their architectural configurations and training methodologies to facilitate a thorough understanding of their respective design principles.

For model training, all experiments employ the AdamW optimizer in conjunction with a cosine annealing decay strategy. The training procedure spans 200 epochs, with an initial learning rate of ${1, 2.5} \times 10^{-4}$, which gradually decays to a final learning rate of ${1, 10} \times 10^{-7}$. Except for predefined parameters, all model weights are initialized using the Xavier method with a scaling factor of $c = 0.01$. To ensure fairness in performance evaluation, a uniform batch size is maintained across all models. Additionally, to mitigate the influence of randomness in data partitioning, each experiment is conducted three times, and the final performance metrics are reported as the average over these independent runs.

\textbf{PhysGTO(Ours).}
A consistent hyperparameter setup is adopted for both datasets. The encoder employs a latent dimension of 128 with the SiLU activation function. The number of GTO block is 4. Each block features an attention dimension of 128, with 8 attention heads per attention mechanism. The decoder is implemented as a three-layer MLP with PReLU activation.

\textbf{MeshGraphNet} \cite{pfaff2020learning}\textbf{.}
It consists of 15 message-passing layers, each with a feature dimension of 128, and employs the ReLU activation function. The encoder comprises two ReLU-activated MLP layers for encoding edges and nodes, which are mirrored in the decoder.

\textbf{GNOT} \cite{hao2023gnot}\textbf{.}
It incorporates a linear attention mechanism with 128 feature dimensions and four attention heads. The mixture-of-experts (MOE) module consists of three experts, each containing three inner layers. The model also utilizes two-layer GELU-based MLPs and integrates four attention layers in its architecture.

\textbf{Transolver} \cite{wu2024transolver}\textbf{.}
It is composed of eight layers with a hidden dimension of 256. The attention module employs eight heads, and the entire architecture is activated using the SiLU function. The mapping ratio is set to 1, while the number of slices is configured to 128. To ensure a fair comparison and reproducibility, we adopt the original codebases provided by the respective authors without making any changes \cite{wu2024transolver}.

\textbf{IPOT} \cite{lee2024inducing}\textbf{.}
It is configured with 128 spectral bands and a maximum resolution of 128. The model utilizes 512 latent vectors, each containing 256 channels. It incorporates four cross-attention layers and four self-attention layers, with each attention module featuring four heads.

\textbf{GINO} \cite{li2024geometry}\textbf{.}
It is characterized by an inner and outer radius of 0.055 and an embedding dimension of 32. The Fourier Neural Operator (FNO) employs mode settings of [32, 32, 32] and utilizes 64 hidden and output channels. The model processes a maximum of 5000 input points and maintains a signed distance field (SDF) spatial resolution of [64, 64, 64]. The model is executed using the official implementations provided by their authors, with no changes made to the original code \cite{li2024geometry}.

\subsection{Comparison with DeepONet-based Methods} \label{supplementary: compare with deeponet}

We interpret the learnable latent space tokens in PhysGTO as a data-driven basis, analogous to handcrafted basis functions used in DeepONet-based methods. 
To further validate this basis-oriented interpretation, we consider three increasingly challenging datasets: \textbf{Laplace}, \textbf{Heat Transfer}, and \textbf{Blood Flow}, as summarized in Tab.~\ref{dataset_1_info_3}. All tasks fall under the general category of operator learning from geometry and boundary conditions to the full-field physical solution. Notably, all three involve \textit{input-output mismatch}, i.e., the input and output point sets are not aligned in space or size. Here, Laplace~\cite{yin2024scalable} is a relatively simple 2D problem with moderate geometric variation and small-scale meshes. Heat Transfer~\cite{chen2024learning} introduces more difficulty: it is defined over 3D geometries with significantly imbalanced input-output sizes (186 input points vs. 7199 output nodes). Blood Flow~\cite{chen2024learning} presents the most complex scenario, involving time-dependent flow in realistic cardiac geometries, strict boundary conditions, and multi-physics field mappings, far beyond DIMON’s original single-physics scope.

\begin{table}[t]
\centering
\caption{Basic information of datasets in unstructured mesh problems from \cite{chen2024learning}.}
\label{dataset_1_info_3}
\resizebox{\textwidth}{!}{
\begin{tabular}{l|c|c|c|c|c}
\toprule
Dataset  & Domain  & \# Cell & \# Nodes  & Operator Mapping   & \# No. train/test \\
\midrule
Laplace \cite{yin2024scalable}  
& 2D & Tri. & 204$\mapsto$2601 & Geo.+$u_{bc}(\mathbf{x})\mapsto u(\mathbf{y})$&  3303/197 \\
\midrule
Heat Transfer \cite{chen2024learning}   & 3D & Quad.& 186$\mapsto$7199 &  Geo.+$T_{bc}(\mathbf{x})\mapsto T_{t=3s}(\mathbf{y})$ & 100/100\\
\midrule
Blood Flow \cite{chen2024learning}      & 3D+time & Quad. & $6\times121\mapsto1656\times121\times3$
& Geo.+$v_{in}|_{1\times [0,T]}, p_{out}|_{5\times [0,T]} \mapsto \boldsymbol{u}(\mathbf{x},[0,T])$& 400/100\\
\bottomrule
\end{tabular}
}
\end{table}

\begin{table}[t]
\centering
\renewcommand{\arraystretch}{1}
\setlength{\tabcolsep}{4pt}
\footnotesize{
\caption{Performance comparison (average relative $L_2$ error, as $\bar{\varepsilon}_{L_2}$, lower is better) on the unstructured mesh problems adopted from \cite{yin2024scalable} and \cite{chen2024learning}. \textbf{Bold} and \underline{underline} numbers indicate the best and second-best performance, respectively. $\Delta$ denotes the relative improvement between PhysGTO and DIMON.
}\label{compare_with_dimon}
\begin{tabular}{l|c|c|c|c}
\toprule
\multirow{2}{*}{Model}&
\multirow{2}{*}{Basis}&
\multicolumn{3}{c}{$\bar{\varepsilon}_{L_2}$ $\downarrow$}\\
\cmidrule(lr){3-5}
&&Laplace \cite{yin2024scalable}& Heat Transfer \cite{chen2024learning}& Blood Flow \cite{chen2024learning}\\
\midrule
DeepONet\cite{lu2021learning}         
& Chebyshev 
& 1.43e-2
& 7.20e-4 
& 8.93e-1\\
POD-DeepONet\cite{lu2022comprehensive}
& POD
& 1.29e-2
& 5.70e-4 
& 3.74e-1\\
DIMON\cite{yin2024scalable}
& PCA
& \underline{8.30e-3}
& \underline{3.01e-4} 
& \underline{1.29e-1}\\
\midrule
\textbf{PhysGTO} (ours)      
& Learnable
& \textbf{7.93e-3}
& \textbf{1.56e-4} 
& \textbf{2.40e-2}\\
\midrule
$\Delta$ & - &\cellcolor{blue!10}{\textbf{4.46\%}}&\cellcolor{blue!10}{\textbf{48.17\%}} &\cellcolor{blue!10}{\textbf{81.40\%}}\\
\bottomrule
\end{tabular}
}
\end{table}

In Tab.~\ref{compare_with_dimon}, we present a systematic comparison between PhysGTO and several DeepONet-based methods, including DeepONet\cite{lu2021learning}, POD-DeepONet\cite{lu2022comprehensive}, and DIMON\cite{yin2024scalable}. DIMON~\cite{yin2024scalable} is a recent variant of DeepONet~\cite{lu2021learning}, which builds upon the operator learning paradigm by introducing a shared reference domain under a homeomorphic mapping assumption. Specifically, for problems with varying geometries, DIMON aligns all domains to a common latent space and encodes geometry and boundary conditions through distinct branch networks, enabling a mapping from input conditions to the solution of the underlying PDE. The model's performance thus relies on the feasibility and quality of this shared reference mapping, which can be limiting for more complex and non-aligned geometries.

To ensure a fair comparison, all DeepONet-based methods—including DeepONet~\cite{lu2021learning}, POD-DeepONet~\cite{lu2022comprehensive}, and DIMON~\cite{yin2024scalable}—are implemented using the official code and hyperparameter settings reported in their original papers. 
The results span multiple basis representations—Chebyshev, POD, and PCA—as well as our proposed learnable basis within PhysGTO. As the geometric and physical complexity increases from Laplace to Blood Flow, our method consistently yields substantial improvements. In particular, PhysGTO achieves a \textbf{4.46\%} error reduction over DIMON on Laplace, a \textbf{48.17\%} reduction on Heat Transfer, and a dramatic \textbf{81.40\%} reduction on Blood Flow. These results validate the effectiveness of our proposed approach that performs \textbf{explicit manifold embedding} in both physical and latent spaces. The improvement trends strongly correlate with increasing geometric/topological complexity and physical fidelity, suggesting that PhysGTO is more robust to nontrivial domain structures, multi-field mappings, and in-out mismatch scenarios.

\section{Visualization of three classes of datasets} \label{supp visulaization}

In this section, we provide a detailed overview and visualization of the three benchmark datasets used to evaluate physical simulation tasks: unstructured mesh problems, complex transient flow prediction problems, and large-scale 3D geometry problems. 

\begin{figure}[]
\centering
\vspace{-10mm}
\includegraphics[width=1\textwidth]{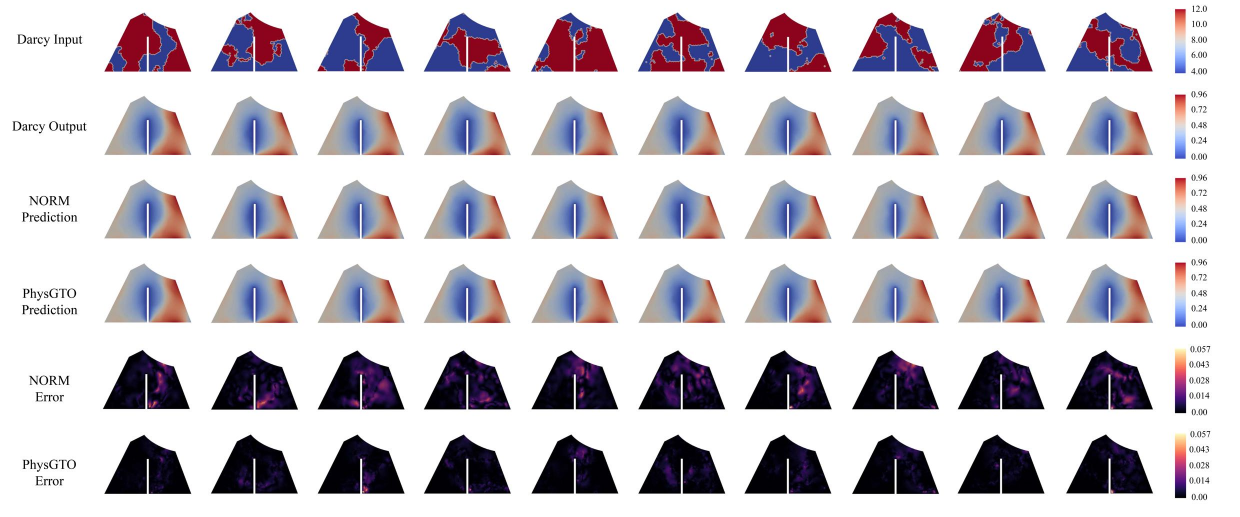}
\vspace{-10mm}
\caption{Visualization of additional representative cases comparing PhysGTO with the SOTA NORM for Darcy problem. Here, \textbf{Error} refers to $|\hat{u}(x,y) - u(x,y)|$, where $\hat{u}(x,y)$ and $u(x,y)$ denote the predicted and ground-truth values, respectively.}
\vspace{-20mm}
\label{sup: task1_darcy}
\end{figure}

\begin{figure}[]
\centering
\vspace{-1mm}
\includegraphics[width=1\textwidth]{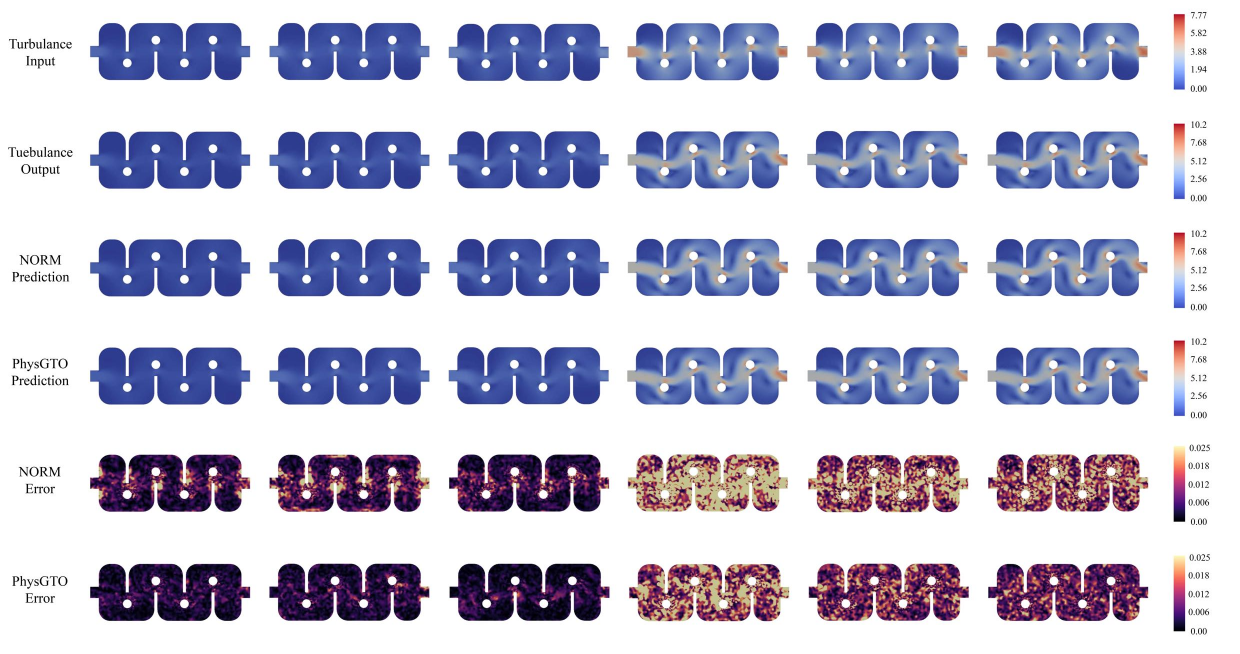}
\vspace{-10mm}
\caption{Visualization of additional representative cases comparing PhysGTO with the SOTA NORM for the prediction in the magnitude of velocity $u(x,y)$ in Pipe Turbulence problem. 
Here, \textbf{Error} refers to $|\hat{u}(x,y) - u(x,y)|$, where $\hat{u}(x,y)$ and $u(x,y)$ denote the predicted and ground-truth values, respectively.}
\label{sup: task1_pipe_turbulence}
\end{figure}

\begin{figure}[]
\centering
\vspace{-6mm}
\includegraphics[width=1\textwidth]{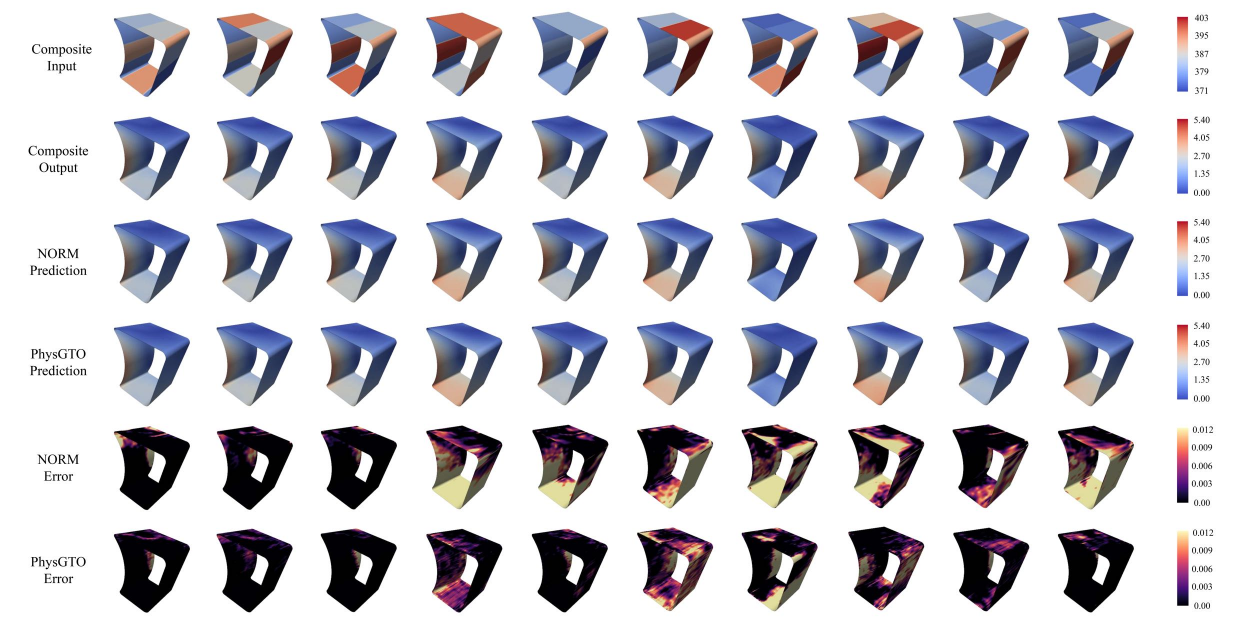}
\vspace{-10mm}
\caption{Visualization of additional representative cases comparing PhysGTO with the SOTA NORM for the deformation field $D(x,y,z)$ in Composite problem. Here, \textbf{Error} refers to $|\hat{D}(x,y,z) - D(x,y,z)|$, where $\hat{D}(x,y,z)$ and $D(x,y,z)$ denote the predicted and ground-truth values, respectively.}
\label{sup: task1_compsite}
\end{figure}

For unstructured mesh problems, we selecte more samples from the test set to compare the prediction performance of PhysGTO with the corresponding SOTA model, NORM~\cite{chen2024learning}. 
The visualizations of the input, ground truth output, and model predictions are provided. 
The visualization results for the Darcy problem, Pipe Turbulence problem, Heat Transfer problem, and Composite problem are shown in Fig.~\ref{sup: task1_darcy}, Fig.~\ref{sup: task1_pipe_turbulence}, Fig.~\ref{sup: task1_heat_transfer} and Fig.~\ref{sup: task1_compsite}.
It is clearly observed that PhysGTO demonstrates superior predictive performance compared to NORM, especially in complex and irregular domain regions, and exhibits strong generalization across different samples.

\begin{figure}[]
\centering
\includegraphics[width=1\textwidth]{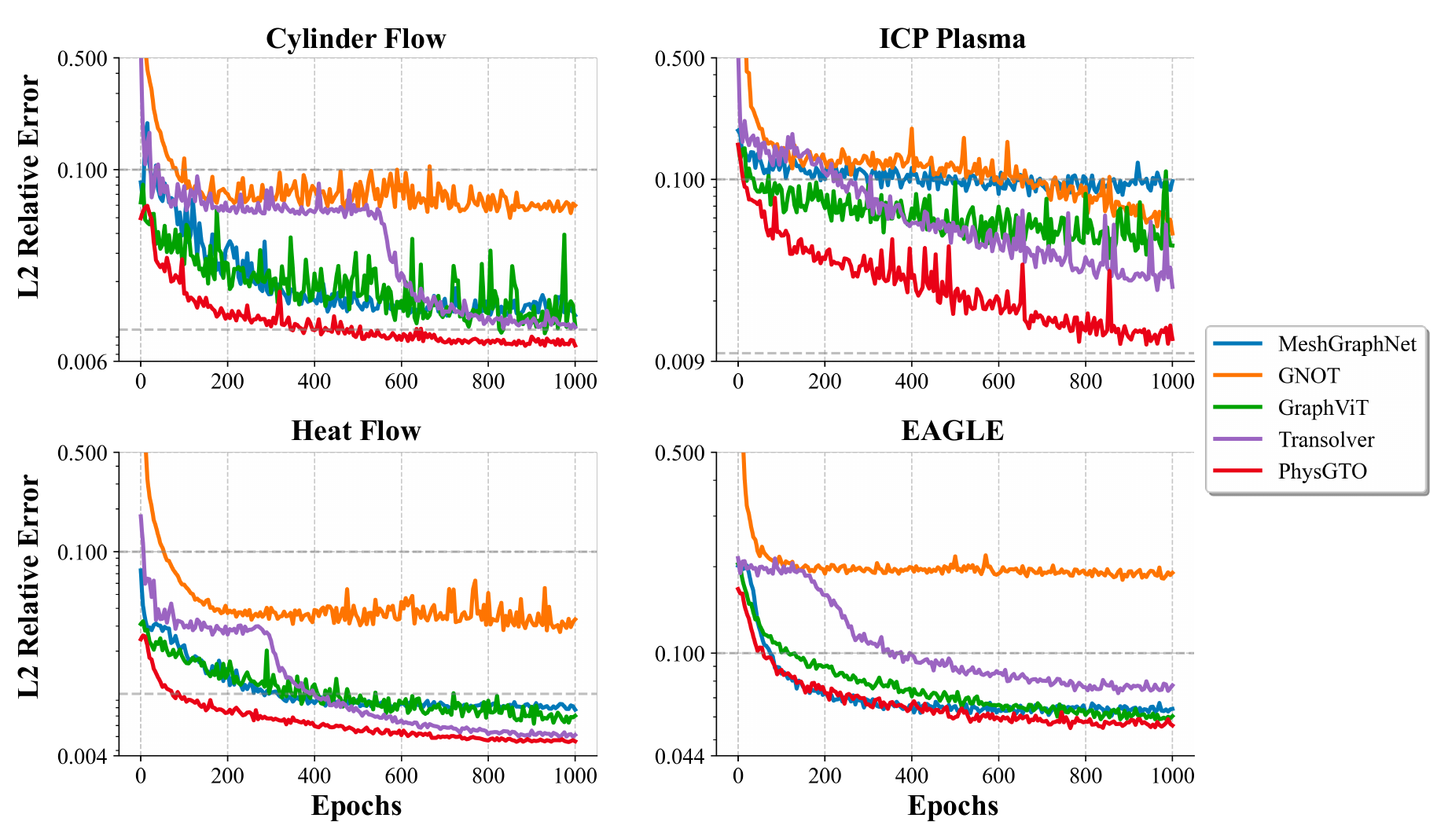}
\vspace{-10mm}
\caption{The l2 error curve of the test dataset 5-step inference loss throughout the training process.}
\label{sup: task2_lr}
\end{figure}

\begin{figure}[]
\centering
\includegraphics[width=1\textwidth]{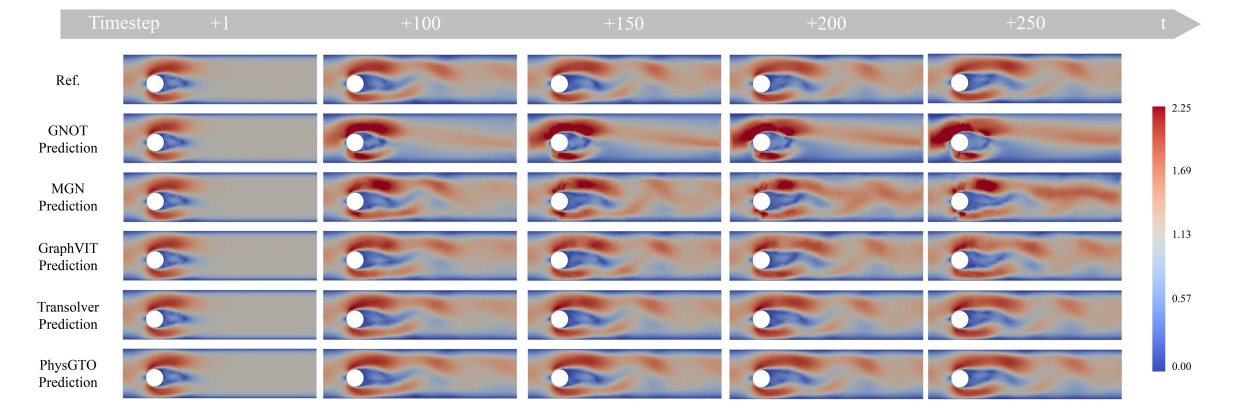}
\vspace{-10mm}
\caption{Long-time velocity $u(x,y)=\sqrt{v_x(x,y)^2+v_y(x,y)^2}$ predictions on Cylinder Flow.
Comparison of PhysGTO with GNOT, MGN, GraphViT, and Transolver at selected future time steps ($t=1$, $t=100$, $t=150$, $t=200$, and $t=250$), against the ground truth. Note that $v_x$ and $v_y$ are predicted separately.}
\label{sup: task2_cylinder}
\end{figure}

\begin{figure}[]
\centering
\includegraphics[width=1\textwidth]{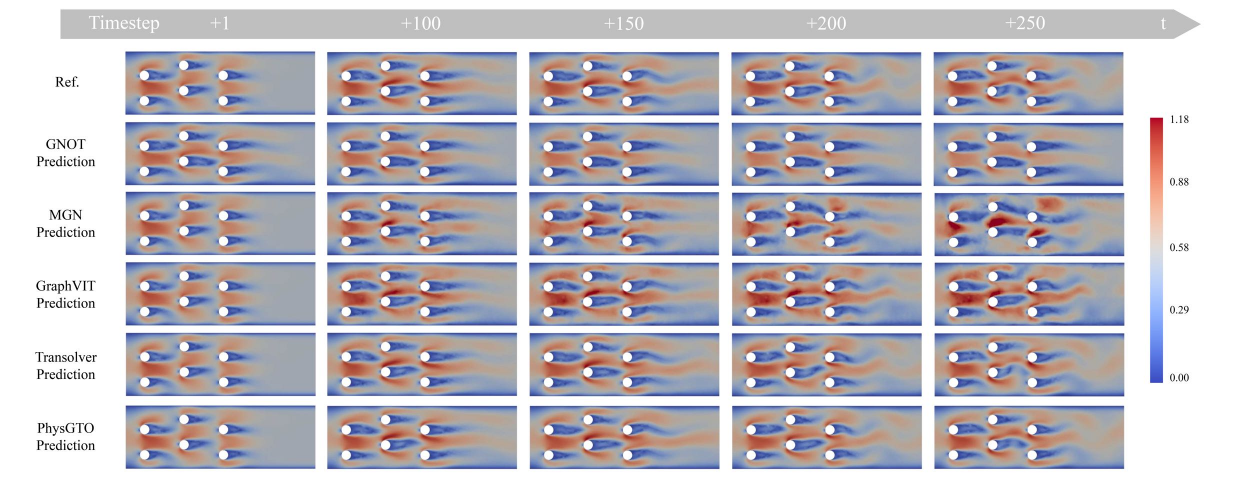}
\vspace{-10mm}
\caption{Long-time velocity $u(x,y)=\sqrt{v_x(x,y)^2+v_y(x,y)^2}$ predictions on Heat Flow.
Comparison of PhysGTO with GNOT, MGN, GraphViT, and Transolver at selected future time steps ($t=1$, $t=100$, $t=150$, $t=200$, and $t=250$), against the ground truth. Note that $v_x$ and $v_y$ are predicted separately.}
\label{sup: task2_heat}
\end{figure}

\begin{figure}[]
\centering
\includegraphics[width=1\textwidth]{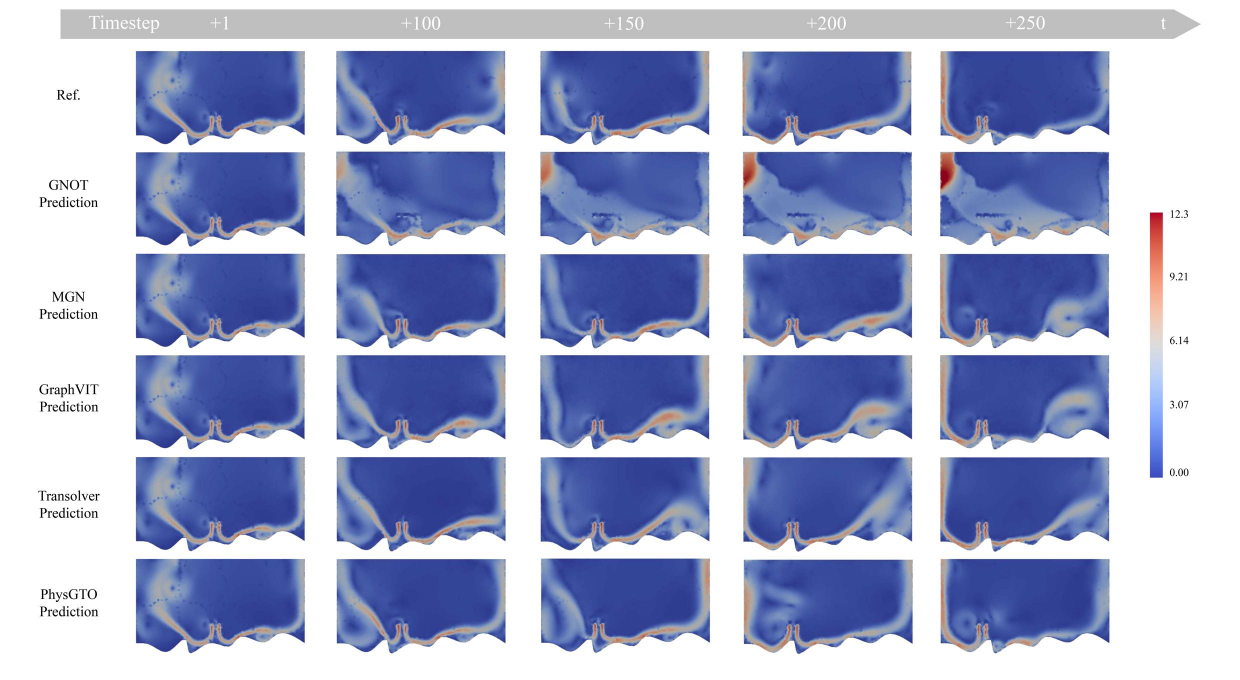}
\vspace{-10mm}
\caption{Long-time velocity $u(x,y)=\sqrt{v_x(x,y)^2+v_y(x,y)^2}$ predictions on EAGLE.
Comparison of PhysGTO with GNOT, MGN, GraphViT, and Transolver at selected future time steps ($t=1$, $t=100$, $t=150$, $t=200$, and $t=250$), against the ground truth. Note that $v_x$ and $v_y$ are predicted separately.}
\label{sup: task2_uav}
\end{figure}

\begin{figure}[]
\centering
\includegraphics[width=1\textwidth]{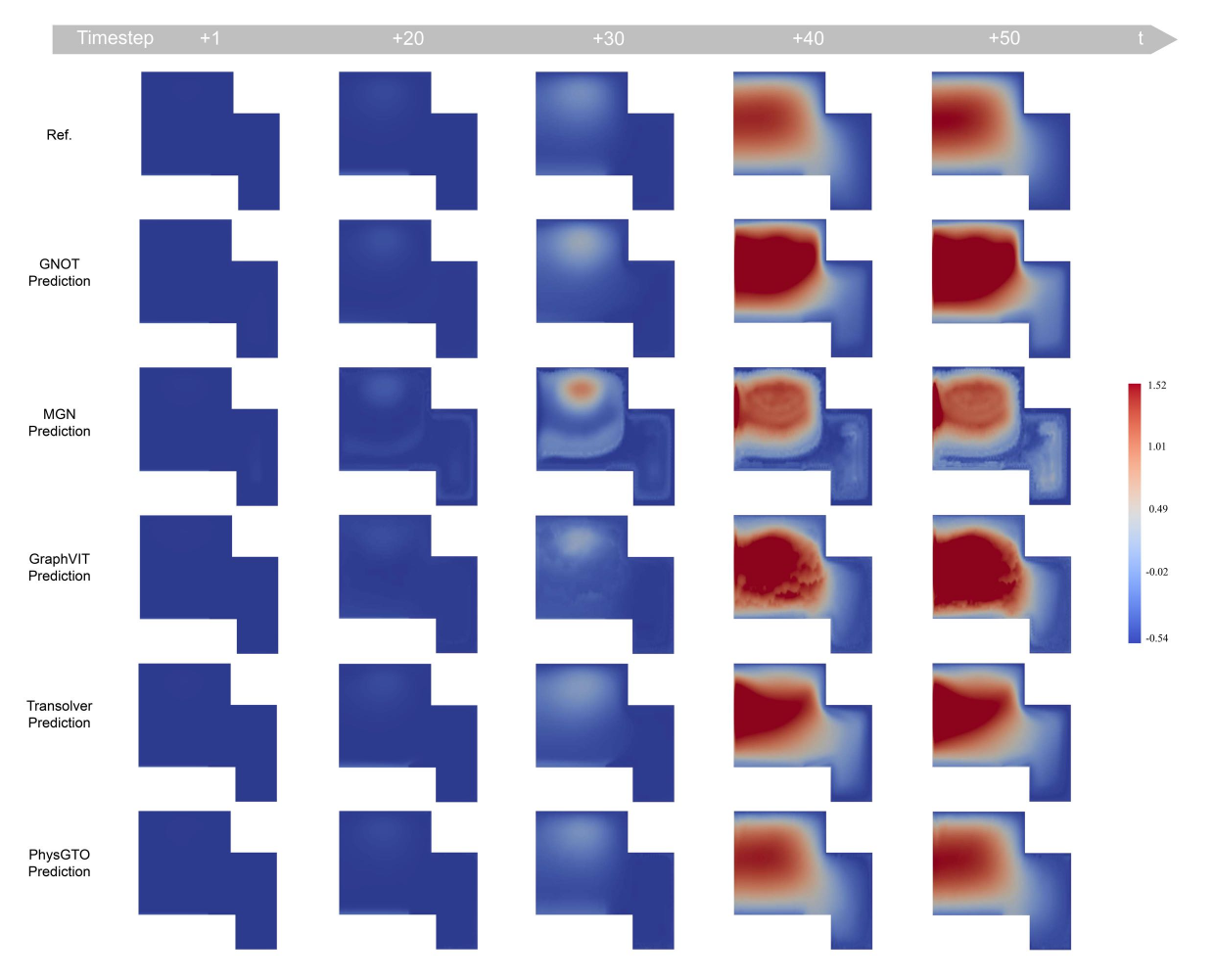}
\vspace{-10mm}
\caption{Long-time electron density $n_e(x,y)$ predictions on ICP Plasma.
Comparison of PhysGTO with GNOT, MGN, GraphViT, and Transolver at selected future time steps ($t=1$, $t=100$, $t=150$, $t=200$, and $t=250$), against the ground truth.}
\label{sup: task2_plasma}
\end{figure}

For complex transient flow prediction problems, we also select more samples to compare the long-term prediction performance of PhysGTO with other baseline models, including GNOT~\cite{hao2023gnot}, MGN~\cite{pfaff2020learning}, GraphViT~\cite{janny2023eagle}, and Transolver~\cite{wu2024transolver}. 
In this evaluation process, we fix several future time steps ($t=1$, $t=100$, $t=150$, $t=200$, and $t=250$) for visualization. 
We record the test $L_2$ error across training epochs for all models on the four datasets, as shown in Fig.~\ref{sup: task2_lr}. 
It is interesting that PhysGTO exhibits the most rapid decline in test error and reaches the lowest error region among all methods. 
This not only reflects the high predictive accuracy of PhysGTO, but also highlights its training efficiency. 
Meanwhile, the comparative visualization results for the Cylinder Flow problem, Heat Flow problem, ICP Plasma problem, and EAGLE problem are shown in Fig.\ref{sup: task2_cylinder}, Fig.\ref{sup: task2_heat}, Fig.\ref{sup: task2_plasma}, and Fig.\ref{sup: task2_uav}, respectively. 
As the prediction horizon extends, error accumulation becomes increasingly evident across all baseline models. 
Nevertheless, PhysGTO consistently yields more accurate predictions compared to the baselines, indicating superior temporal stability in physical simulations. 
This suggests that PhysGTO is more effective in mitigating long-term error propagation, thereby demonstrating enhanced robustness and adaptability in dynamic, time-dependent environments.

\begin{figure}[]
\centering
\vspace{-10mm}
\includegraphics[width=1\textwidth]{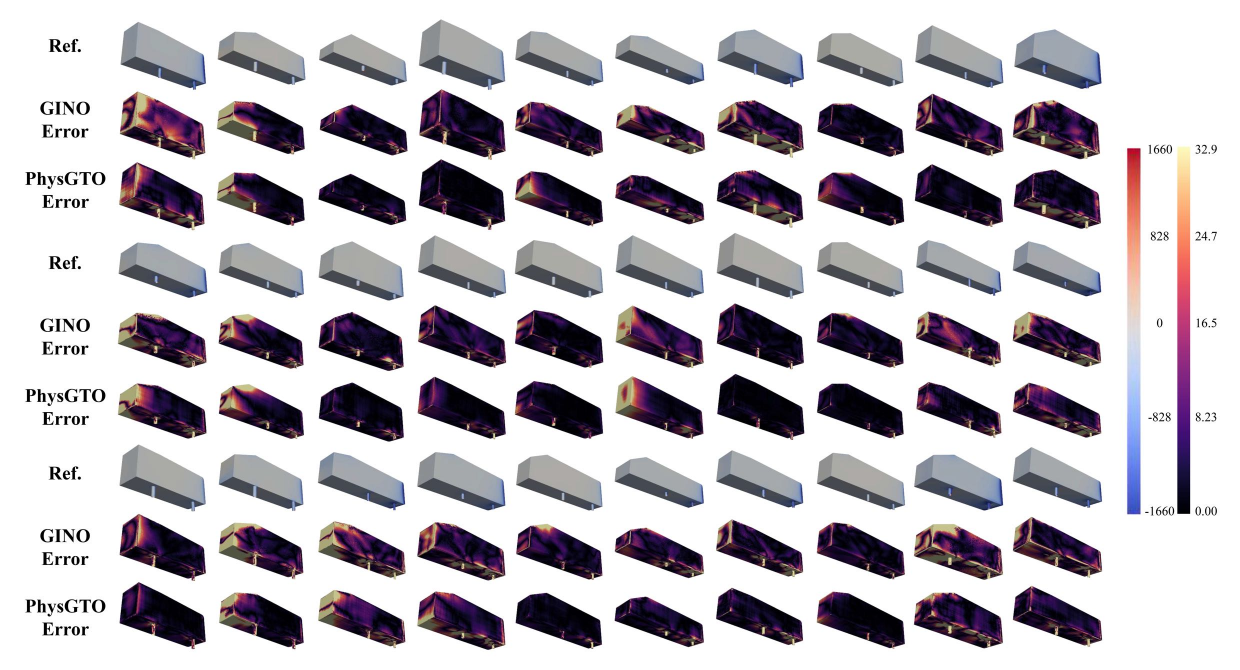}
\vspace{-10mm}
\caption{Large-scale 3D pressure prediction $p(x,y,z)$ on the Ahmed body. Additional visualization of PhysGTO’s prediction errors compared to the current SOTA method, GINO. Here, \textbf{Error} refers to $|\hat{p}(x,y,z) - p(x,y,z)|$, where $\hat{p}(x,y,z)$ and $p(x,y,z)$ denote the predicted and ground-truth values, respectively.}
\label{sup: task3_ahmed}
\end{figure}

\begin{figure}[]
\centering
\includegraphics[width=1\textwidth]{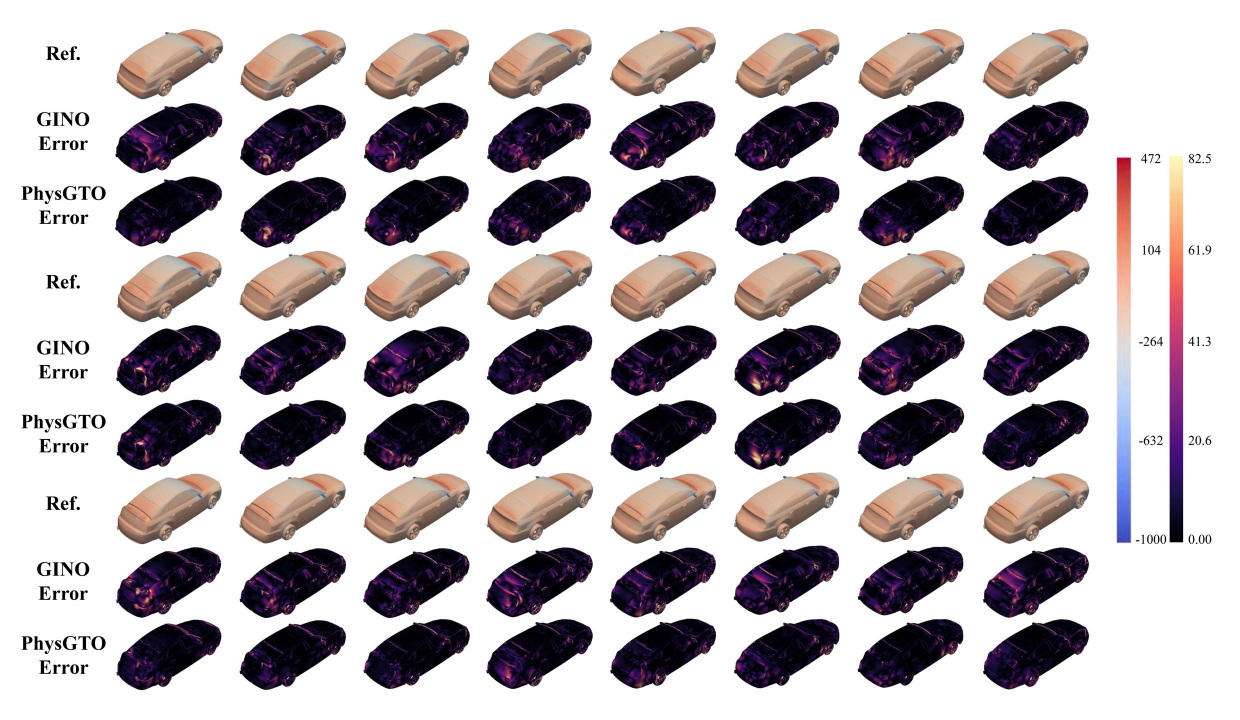}
\vspace{-10mm}
\caption{Large-scale 3D pressure prediction $p(x,y,z)$ on the DrivaerNet. Additional visualization of PhysGTO’s prediction errors compared to the current SOTA method, GINO. Here, \textbf{Error} refers to $|\hat{p}(x,y,z) - p(x,y,z)|$, where $\hat{p}(x,y,z)$ and $u(x,y,z)$ denote the predicted and ground-truth values, respectively.}
\label{sup: task3_drivaer}
\end{figure}

For large-scale 3D geometry problems, we assess an array of vehicle types from the test set to benchmark the predictive performance of PhysGTO against the current state-of-the-art model, GINO~\cite{li2024geometry}. 
Visualization results for the Ahmed Body and DrivAerNet datasets are displayed in Fig.~\ref{sup: task3_ahmed} and Fig.~\ref{sup: task3_drivaer}, respectively. 
PhysGTO consistently demonstrates superior simulation accuracy across these extensive datasets. 
For the Ahmed Body, it accurately delineates the frontal stagnation region and the flow separation zones where the dynamics of the physics change abruptly. 
In the DrivAerNet dataset, PhysGTO offers more precise modeling of intricate automotive parts—such as windows and wheel hubs—producing pressure field predictions more closely aligned with the actual data. 
These results demonstrate that PhysGTO achieves higher modeling accuracy than existing neural operator approaches on standard benchmarks, while also generalizing reliably to complex, real-world scenarios.

Notably, complete examples for all datasets—including the corresponding data and visualization code—are provided in the supplementary materials and are publicly accessible via our repository.

\clearpage
\newpage

\begin{figure}[]
\centering
\includegraphics[width=1\textwidth]{fig_new/fig_task1_heat_transfer.pdf}
\vspace{-10mm}
\caption{Visualization of additional representative cases comparing PhysGTO with the SOTA NORM for the Temperature $T(x,y)$ in Heat Transfer problem. Here, \textbf{Error} refers to $|\hat{T}(x,y) - T(x,y)|$, where $\hat{T}(x,y)$ and $T(x,y)$ denote the predicted and ground-truth values, respectively.}
\label{sup: task1_heat_transfer}
\end{figure}

\end{document}